\documentclass[twocolumn,prd,superscriptaddress,nofootinbib]{revtex4-1}

\usepackage{graphicx}
\usepackage{amsmath,bm}
\usepackage[usenames,dvipsnames]{xcolor}
\usepackage{ulem}
\usepackage{url}
\usepackage{multirow}

\usepackage[colorlinks  = true,
            linkcolor   = NavyBlue,
            urlcolor    = NavyBlue,
            citecolor   = NavyBlue,
            anchorcolor = NavyBlue]{hyperref}

\newcommand{\sS}{\sqrt{s}}
\newcommand{\xR}{x_R}

\newcommand{\pT}{p_T}

\newcommand{\sigmaInv}{\sigma_{\rm inv}}

\def \aap  {A\&A}

\def \apj  {ApJ}

\def \nat {Nature}

\begin{document}

\widetext

\title{New determination of the production cross section for secondary positrons and electrons in the Galaxy}

\author{Luca Orusa}
\affiliation{Department of Physics, University of Torino, via P. Giuria, 1, 10125 Torino, Italy}
\affiliation{Istituto Nazionale di Fisica Nucleare, via P. Giuria, 1, 10125 Torino, Italy}
\author{Mattia Di Mauro}
\affiliation{Istituto Nazionale di Fisica Nucleare, via P. Giuria, 1, 10125 Torino, Italy}
\author{Fiorenza Donato}
\affiliation{Department of Physics, University of Torino, via P. Giuria, 1, 10125 Torino, Italy}
\affiliation{Istituto Nazionale di Fisica Nucleare, via P. Giuria, 1, 10125 Torino, Italy}
\author{Michael Korsmeier}
\affiliation{The Oskar Klein Centre for Cosmoparticle Physics, Department of Physics, Stockholm University, Alba Nova, 10691 Stockholm, Sweden}

\date{\today}

\begin{abstract}
The cosmic-ray fluxes of electrons and positrons ($e^{\pm}$) are measured with high precision by the space-borne particle spectrometer AMS-02. To infer a precise interpretation of the production processes for $e^{\pm}$ in our Galaxy, it is necessary to have an accurate description of the secondary component, produced by the interaction of cosmic-ray proton and helium with the interstellar medium atoms.
We determine new analytical functions of the Lorentz invariant cross section for the production of $\pi^\pm$ and $K^\pm$ by fitting data from collider experiments. We also evaluate the invariant
cross sections for several other channels, involving for example hyperon decays, contributing at the few \% level on the total cross section. 
For all these particles, the relevant 2 and 3 body decay channels are implemented, with the polarized $\mu^\pm$ decay computed with next-to-leading order corrections. 
The cross section for scattering of nuclei heavier than protons is modeled by fitting data on $p+C$ collisions.
The total differential cross section  $d\sigma/dT_{e^\pm}(p+p\rightarrow e^\pm+X)$ is predicted from 10 MeV  up to 10 TeV of $e^\pm$ energy with an uncertainty of about 5-7\% in the energies relevant for AMS-02 positron flux, thus dramatically reducing the precision of the theoretical model with respect to the state of the art.
Finally, we provide a prediction for the secondary Galactic $e^\pm$ source spectrum with an uncertainty of the same level.
As a service for the scientific community, we provide numerical tables and a script to calculate energy-differential cross sections.

\end{abstract}

\maketitle

\section{Introduction}

During the last decades, the space-based experiments PAMELA, AMS-02, DAMPE and CALET have performed unprecedented precise measurements of the cosmic-ray (CR) fluxes with uncertainties at the few percent level in an energy range from 1 GeV to tens of TeV, making the physics of charged CRs a precision discipline. 
These experiments have measured the CR nuclear \cite{Adriani_2011,Adriani_2014,Aguilar:2015ooa,article,Aguilar:2016vqr,2019,Adriani_2019} and leptonic (positron and electron, $e^{\pm}$) \cite{Adriani2011,Aguilar:2019ksn,2017Natur.552...63D,Adriani_2018,PhysRevLett.122.041102} components, as well as cosmic antiprotons \cite{Aguilar:2016kjl,Adriani_2010}. The most recent positron flux measurement by AMS-02 extends from 0.5 to 1000 GeV with an uncertainty $< 5\%$ for almost the whole energy range. 
The new precise flux data have stimulated numerous analyses on Galactic CR propagation \cite{Korsmeier:2016kha,Tomassetti:2017hbe,Liu:2018ujp,Genolini:2019ewc,Weinrich:2020cmw,Weinrich:2020ftb,Evoli:2019wwu,Evoli:2019iih,Boschini:2018baj,Boschini:2019gow,Boschini:2020jty,Di_Mauro_2021,Luque:2021nxb,DeLaTorreLuque:2021yfq,Schroer:2021ojh,Korsmeier:2021brc,Korsmeier:2021bkw}, lepton production from astrophysical sources like pulsars and supernova remnants
\cite{Hooper:2008kg,Ahlers:2009ae,Boudaud:2014dta,Boudaud:2016jvj,Manconi_2017,Manconi:2018azw,Fornieri_2020,Manconi:2020ipm,DiMauro:2019yvh,Orusa_2021,Evoli_2021,Diesing:2020jtm,Cholis_2018,Cholis:2021kqk}, and particle dark matter annihilation or decay into antimatter \cite{Cirelli:2008id,Bergstrom:2013jra,DiMauro:2015jxa,Di_Mauro_2021}. 

It is generally established that the so-called secondary production, \textit{i.e.}~production by the interaction of CRs with the interstellar medium (ISM) atoms, contributes to $e^{\pm}$ flux in our Galaxy (see, e.g., \cite{Delahaye_2009}).
In particular, the flux of cosmic $e^+$ is dominated by this process at energies below 10 GeV.
Instead, above 10 GeV the data (see, e.g., \cite{PhysRevLett.122.041102}) are higher than the predictions for the secondary production. This is called the positron excess and its origin remains unresolved. To infer reliable conclusions on the possible contribution of primary sources, such as pulsars or dark matter, to the positron excess, an accurate description of the secondary production is necessary. 

The dominant production of secondary flux comes from the proton-proton ($p+p$) channel, namely CR protons interacting on ISM hydrogen atoms. 
Other relevant contributions involve CR projectile or ISM target atoms given by helium (He$+p$, $p+$He, and He$+$He). Following the results obtained with secondary antiprotons for which the calculation involves the same CRs and ISM atoms (see, e.g., \cite{Korsmeier_2018}), channels involving heavier CR species and atoms can contribute at the few percent level to secondary $e^{\pm}$. Secondary $e^{\pm}$ are mainly produced by spallation processes between CRs and ISM atoms producing pions ($\pi^{\pm}$) and kaons ($K^{\pm}$), which subsequently decay into $e^{\pm}$. Therefore, the cross sections for the production of $\pi^{\pm}$ and $K^{\pm}$ are key elements for the calculation of secondary $e^{\pm}$.

There are two different strategies to parametrize the $e^{\pm}$ production cross sections. The first possibility is to find an analytic description of the double differential and Lorentz invariant cross section for the production of $\pi^{\pm}$ and $K^{\pm}$, performing a fit to cross section data. 
This strategy was first pursued by \cite{Tan:1984ha} and then repeated with new data by \cite{Blattnig:2000zf}. The other option is to use predictions from Monte Carlo event generators \cite{Sjostrand:2014zea,Kelner:2006tc,Koldobskiy:2021nld}. The authors of \cite{Kamae:2006bf} used this strategy to extract the required cross sections.

Both methods have advantages and drawbacks. Analytic functions permit to calibrate cross sections very precisely on existing data, but they imply large extrapolations in the parameter space where measurements are not available. Moreover, it is hard to use this method on production channels for which data are scarce or not available, as for example for $p+$He.
Monte Carlo generators can be used to derive the cross sections for all the possible channels of production, \textit{i.e.}~also for nuclei or hyperon contributions, but they typically do not fully reproduce the available data which is relevant for CRs at low energies (see, e.g., \cite{Kachelriess:2015wpa,Kachelriess:2019ifk} for antiprotons). In fact, codes like Pythia or QGSJET are mainly tuned to high-energy data (with center of mass energy of the order of TeV). As outlined in Ref.~\cite{Delahaye_2009}, the adoption of the predictions from different cross section models \cite{PhysRevD.15.820,Tan:1984ha,Kamae:2006bf} produces a variation in the normalization of the secondary $e^{\pm}$ flux up to a factor of 2. Instead, in Ref.~\cite{Koldobskiy:2021nld} the authors have shown that the differences in the source term obtained by using the results in \cite{Kamae:2006bf} and different event generators can reach $30\%$ in the relevant energies for $e^{\pm}$ CR physics. However, Ref.~\cite{Koldobskiy:2021nld} does not consider the models from Refs. \cite{PhysRevD.15.820,Tan:1984ha}, so the reported uncertainty could be underestimated. 

The \textsc{Galprop} code \cite{strong2009galprop}, widely used in the community for calculating the propagation of CRs, implements for the $e^{\pm}$ production cross sections the pion production in $p+p$ collisions developed by \cite{1986ApJ...307...47D,1986A&A...157..223D}. The $e^{\pm}$ distributions from the muon decay are computed following \cite{Kelner:2006tc}.
On the other hand, \textsc{Dragon} \cite{Evoli:2016xgn,Evoli:2017vim} and \textsc{Usine} \cite{Maurin:2018rmm} codes employ the \cite{Kamae:2006bf} $e^{\pm}$ production cross sections, as well many others (see, e.g., \cite{Delahaye_2009,DiMauro_2014,Weinrich:2020ftb,Di_Mauro_2021}).

The production cross section of $e^{\pm}$ from Kamae \textit{et al.}~\cite{Kamae:2006bf} are largely used by the community, despite being tuned on at least 20-year old data. The analysis by Ref.~\cite{Kamae:2006bf} carefully checks the total $p+p$ cross sections and the separate contribution from nondiffractive, diffractive, and resonance-excitation processes. However, this does not guarantee that the cross sections catch the correct dependence in the relevant kinematic phase space (e.g.~in the transverse momentum and rapidity).
The reason is that, until recently, the available dataset was limited to data collected from the sixties to the eighties. In the last decades, however, new experimental datasets have become available. For example, the NA49 and NA61/SHINE collaborations at the CERN Super Proton Synchrotron (SPS) \cite{2005_NA49,Aduszkiewicz:2017sei} provide important information for the energies of interest for AMS-02 and a wide range of the double differential cross section. Moreover, high-energy data at center-of-mass-energy (CM) $\sqrt{s}>200$ GeV have been collected from different experiments \cite{2007_BRAHMS,2011_PHENIX,Adam:2015qaa,2017_CMS}. These data permit to calibrate precisely the dependence with the $\sqrt{s}$.
Given the importance of these data in astroparticle physics, a reevaluation of the leptonic production cross sections is mandatory for $p+p$, He$+p$, $p+$He, and He$+$He collisions. In this paper, we engage ourselves in this task, in order to provide an updated parametrization of the inclusive $e^{\pm}$ production cross section.

The paper is structured as follows. In Sec.~\ref{sec:cross_sections} we report the model for the calculation of the source term from the double differential cross section of pions and kaons. In Sec.~ \ref{sec:cross_section_eplus}, we provided a detailed discussion of the pion channel for positron production in proton-proton collisions. Then, in Sec.~\ref{sec:other_channels} we discuss all the other channels from proton-proton collisions as shown in Fig.~\ref{fig:positron_production_channels}. Sec.~\ref{sec:nuclei} is dedicated to nuclei collision and we discuss how to scale the cross sections from proton-proton to proton-nuclei collisions. 
Our results for the total positron and electron production cross section as well as for the source spectrum are presented in Sec.~\ref{sec:source_term_eplus} and Sec.~\ref{sec:electrons}, respectively.
Finally, we conclude in Sec.~\ref{sec:conclusions}.

\section{From cross sections to the source term}
\label{sec:cross_sections}

\begin{figure*}[t]
  \centering {
    \includegraphics[width=0.80\textwidth]{"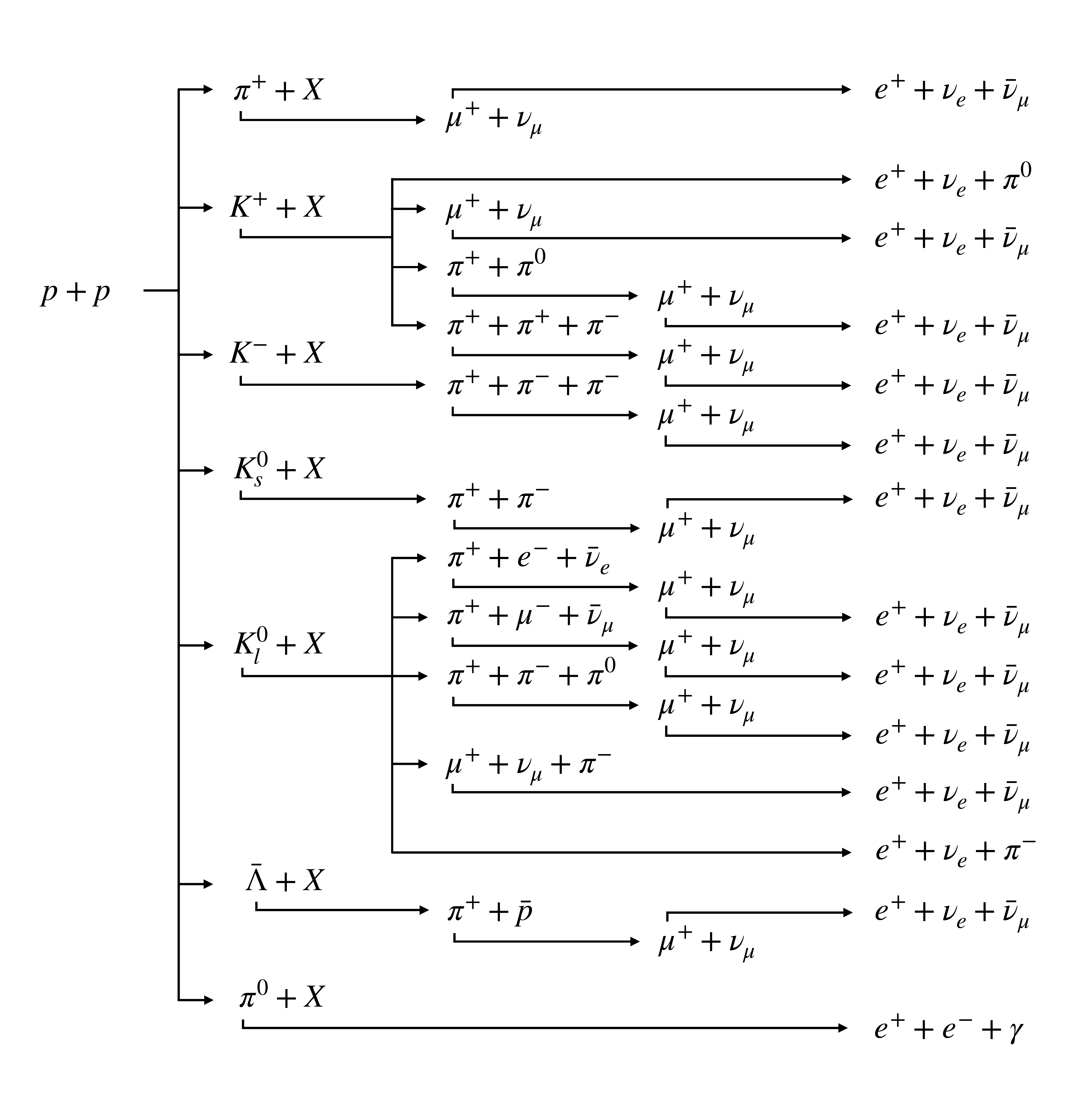"}
  }
  \caption{
             This diagram represents the $e^+$ production channels from a $p+p$ collision considered in our analysis.
             The same scheme holds for $e^-$ production under charge conjugation (except for the initial $p+p$ state). We report here only the channels that produce at least 0.5\% of the total yield (see the main text for further details).
             \label{fig:positron_production_channels}
  }
\end{figure*}

The source term is computed as the convolution between the primary CR flux ($\phi$), the density of the ISM ($n_{\mathrm{ISM}}$) and the energy-differential cross section for $e^{\pm}$ production ($d\sigma/dT_{e^{\pm}}$).
In particular, the total source term is calculated as the sum of all the possible combinations of the $i$-th CR species with the $j$-th ISM components as: 

\begin{eqnarray}    
    \label{eq:source_term}
    q(T_{e^{\pm}})&=& \sum_{i,j} 4 \pi\, n_{\mathrm{ISM},j}
    \int dT_i \, \phi_i(T_i)\frac{d\sigma_{ij}}{d T_{e^{\pm}}}(T_i,T_{e^{\pm}}) \,,
\end{eqnarray}
where $T_{e^{\pm}}$ is the $e^{\pm}$ kinetic energy, $\phi_i$ is the CR flux at the kinetic energy $T_i$, $n_{\mathrm{ISM},j}$ is the number density of the ISM $j$-th atom, and $d\sigma_{ij}/d T_{e^{\pm}}$ is the energy-differential production cross section for the reaction $i+j\rightarrow e^{\pm} + X$.
The factor $4\pi$ corresponds to the angular integration of the isotropic CR flux.
We note that, in general, the source term depends on the position in the Galaxy because both the CR gas density and the CR flux are a function of the position.
Almost the entire ISM ($99\%$) consists of hydrogen and helium atoms \cite{Ferriere:2001rg}. CRs share the same hierarchy with most of the flux given by protons and helium nuclei. Therefore, the main channels for the production of secondary $e^{\pm}$ are $p+p$, $p+$He, He$+p$ and He+He.

Secondary positrons and electrons are not produced directly in the proton-proton (or nuclei) collisions but rather by the decay of intermediate mesons and hadrons. In Fig.~\ref{fig:positron_production_channels}, we show a sketch of all the production channels for $e^+$ that are considered in this analysis. 
The channels that produce $e^-$ are the same as in Fig.~\ref{fig:positron_production_channels}, but all particles have to be replaced by their antiparticles (e.g. $\pi^+ \rightarrow \pi^-$ and $\mu^+ \rightarrow \mu^-$). We neglect production or decay channels that contribute less than $0.5\%$ to the total positron production. One example is the production of positrons (electrons) from the decay of antineutrons (neutrons). This channel is suppressed because in the decay almost all of the energy is carried away by the antiproton (proton) and positrons (electrons) are only produced at very small energies \cite{Kamae:2006bf}.
We will discuss other channels that we neglect or that we include with a simple rescaling of other contributions in Sec.~\ref{sec:eplus_otherchannels} .

We provide now the calculations to find the source term starting from the production cross sections of pions and kaons. We focus on $e^+$ and consider the dominant channel which involves intermediate $\pi^+$ and gives a contribution of about 80-90 \% to the final positron yield. After production, pions first decay into muons with a branching ratio of 99.99\ \%, and then the muons decay into positrons. This discussion shows that the derivation of the differential cross section for the production of positrons is split into two steps. First, we must model the pion production cross section and then the decays of the pion to the positron. 
 
The positron production cross section is calculated from the pion production cross section as follow:
\begin{equation}    
    \label{eq:convolution}
    \frac{d\sigma_{ij}}{d T_{e^{\pm}}}(T_i,T_{e^{\pm}})= 
     \int  d T_{\pi^{\pm}} \, \frac{d\sigma_{ij}}{dT_{\pi^{\pm}}}(T_i,T_{\pi^{\pm}}) \; P(T_{\pi^{\pm}}, T_{e^{\pm}}) \,
\end{equation}
where $T_{\pi^{\pm}}$ is the kinetic energy of the pion that decays into a $e^{\pm}$ with kinetic energy $T_{e^{\pm}}$. $P(T_{\pi^{\pm}}, T_{e^{\pm}})$ is the probability density function of the process which can be computed analytically. In Sec.~\ref{sec:piplus_decay} we detail how we obtain $P$.

In contrast to the pion decay, the pion production cross section cannot be derived from first principles. It rather has to be modeled and fitted to experimental data. High-energy experiments provide measurements of the fully differential production cross section usually stated in the Lorentz invariant form: 
\begin{equation}
    \sigma^{(ij)}_{ {\rm inv}}= E_{\pi^{\pm}} \frac{d^3 \sigma_{ij}}{dp_{\pi^{\pm}}^3}.
    \label{eq:invariant}
\end{equation}
Here $E_{\pi^{\pm}}$ is the total $\pi^{\pm}$ energy and $p_{\pi^{\pm}}$ its momentum. The fully differential cross section is a function of three kinematic variables. We choose them to be the center of mass energy $\sqrt{s}$, the transverse momentum of the pion $p_T$, and the radial scaling $x_R$. The latter is defined as the pion energy divided by the maximal pion energy in the center of mass frame, $x_R = E_{\pi^{\pm}}^\ast/E_{\pi^{\pm}}^{\max\ast}$, where the asterisk denotes the center of mass reference frame.

After modeling the Lorentz invariant cross section, the energy-differential cross section for pion production as required in Eq.~\eqref{eq:convolution} is obtained by first transforming the kinetic variables into the fix-target frame, \textit{i.e.} the frame where the ISM target atom is at rest, and then by integrating over the solid angle $\Omega$: 
\begin{eqnarray}
    \frac{d\sigma_{ij}}{d T_{\pi^{\pm}}}(T_i,T_{\pi^{\pm}}) &=& p_{\pi^{\pm}}\int d \Omega \; \sigma_{ {\rm inv}}^{(ij)}(T_i,T_{\pi^{\pm}},\theta) \\
    &=& 2\pi p_{\pi^{\pm}} \int^{+1}_{-1} d \cos{\theta} \; \sigma_{ {\rm inv}}^{(ij)}(T_i,T_{\pi^{\pm}},\theta), \nonumber
    \label{eq:solid_int}
\end{eqnarray}
where $\theta$ is the angle between the incident projectile and the produced $\pi^{\pm}$ in the LAB frame.
The derivation of the other channels works in analogy to the pion channel, namely, we first model the production and then the decay. We will first concentrate on the $e^+$ production cross sections, then we provide parallel results for secondary $e^-$. 
The channels and cross sections are very similar, but not identical. In fact, charge conservation implies that the production of $e^+$ is enhanced with respect to $e^-$ since both the target and the projectiles involved in the production process are positively charged particles.

\subsection{Computation of the $\pi^{\pm}$ to $e^{\pm}$ decay rates}
\label{sec:piplus_decay}

\begin{figure}[b]
    \centering {
    \includegraphics[width=0.49\textwidth]{"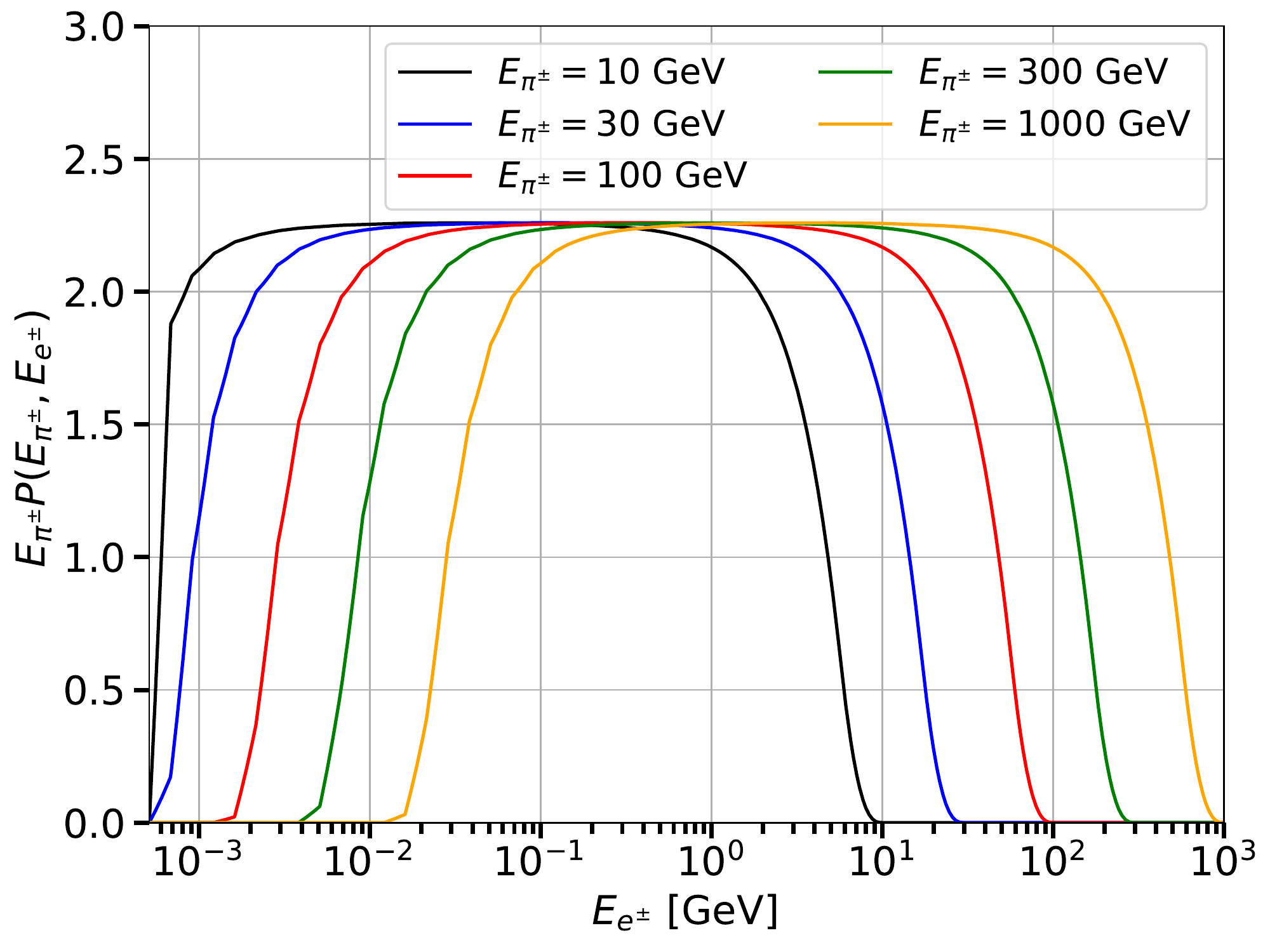"}
     }
     \caption{$E_{\pi^{\pm}}P(E_{e^{\pm}},E_{\pi^{\pm}})$ computed from the $\pi^\pm$ and subsequent $\mu^\pm$ decays for $\pi^\pm$ LAB energies of 10, 30, 100, 300 and 1000 GeV from left to right.
     }
     \label{Fig R1}
\end{figure}

The largest fraction of $e^\pm$ produced in $p+p$ collisions comes from the $\pi^\pm$ and subsequent  $\mu^\pm$ decays, as illustrated in Fig.~\ref{fig:positron_production_channels}. Therefore, we need the probability distribution, $P(E_{\pi^{\pm}}, E_{e^{\pm}})$, for obtaining an $e^\pm$ with energy $E_{e^{\pm}}$ from a $\pi^{\pm}$ with energy $E_{\pi^{\pm}}$. The $\pi^\pm$ decay is entirely determined from kinematics, namely, in the $\pi^\pm$ rest frame, the energy of the $\mu^\pm$ is determined by energy and momentum conservation. In contrast, the $\mu^\pm$ decay goes into three final states and has to be computed in Fermi theory. The $\mu^\pm$ are fully polarized into their direction of motion after the $\pi^\pm$ decays. We implement the polarized $\mu^\pm$ decay rate including the next to leading order (NLO) corrections \cite{Arbuzov_2002}. In the rest frame of the $\mu^\pm$, the decay rate is given by:
\begin{equation}
    \frac{d\Gamma}{dE_{e^{\pm}}'\,d\cos\theta'}=C[f(E_{e^{\pm}}')\pm g(E_{e^{\pm}}') \cos \theta']\, ,
    \label{eq:spectrum}
\end{equation}
where $C$ is a normalization factor, $E_{e^{\pm}}'$ is the energy of the $e^{\pm}$, and $\theta'$ is the angle between the direction of polarization of the $\mu^\pm$ and the direction of motion of the $e^{\pm}$. The apostrophe denotes that quantities are computed in the rest frame of the $\mu^\pm$. We extract the functions $f(E_{e^{\pm}}')$ and $g(E_{e^{\pm}}')$ at NLO from Ref.\ \cite{Arbuzov_2002}.   

Then, we follow the steps of Ref.~\cite{1965ApJ...141..718S} to obtain $P(E_{\pi^{\pm}}, E_{e^{\pm}})$. In short, we perform two Lorentz transformations, first from the $\mu^\pm$ rest frame to the $\pi^\pm$ rest frame and then from the $\pi^\pm$ rest frame to the LAB frame (\textit{i.e.} the rest frame of the Galaxy). Finally, we integrate over all the possible directions of the $\mu^{\pm}$ and all the directions of the $e^{\pm}$. Figure~\ref{Fig R1} shows our result for $P(E_{e^{\pm}},E_{\pi^{\pm}})$ as a function of $E_{e^\pm}$ for a few different values of $E_{\pi^{\pm}}$. 
We note that our calculations are an improvement over the standard treatment in CR propagation codes. For example, in \textsc{Galprop} \cite{strong2009galprop} the $\pi^\pm$ decay rate is computed according to Ref.~\cite{Kelner:2006tc}, not containing NLO correction and assuming $m_e=0$.

\section{Positrons from $p+p \rightarrow \pi^+ + X$ collisions}
\label{sec:cross_section_eplus}

\begin{table*}[t]
 \caption{ 
           Summary of all $p+p$ datasets used for $\pi^{\pm}$ and/or $K^{\pm}$ fits, their center of mass energies, and references. $\sigma_{inv}$ is the fully differential production cross section usually stated in the Lorentz invariant form  and $n$ is the total multiplicity of a particle. With $\surd$ we indicate when the quantity is considered in the analysis.
         }
 \label{tab::pp_data}
 \begin{tabular}{l c c c c c}
   \\ \hline \hline
   Experiment   & $\sS\;\mathrm{[GeV]}$        &                         & $\sigma_{\rm inv}$ & $n$      & Ref.  \\ \hline
   NA49         & 17.3                         & ($\pi^{\pm},K^{\pm}$)   & $\surd$            & -       & \cite{2005_NA49,NA49_2010} \\
   ALICE        & 900                          & ($\pi^+,K^{\pm})$       & $\surd$            & -       & \cite{2011_ALICE} \\ 
   CMS          & 900, 2760, 7000, 13000       & ($\pi^{\pm},K^{\pm}$)   & $\surd$            & -       & \cite{2012_CMS, 2017_CMS} \\ 
   Antinucci    & 3.0, 3.5, 4.9, 5.0, 6.1, 6.8 & ($\pi^{\pm}$)           & -                  & $\surd$ & \cite{osti_4593576} \\
                & 2.8, 3.0,3.2, 5.3, 6.1, 6.8 & ($K^+$)                 & -                  & $\surd$ & \cite{osti_4593576} \\
                & 4.9, 5.0, 6.1, 6.8           & ($K^-$)                 & -                  & $\surd$ & \cite{osti_4593576} \\
   NA61/SHINE   & 6.3, 7.7, 8.8, 12.3, 17.3    & ($\pi^{\pm},K^{\pm}$)   & -                  & $\surd$ & \cite{Aduszkiewicz:2017sei} \\
  \hline \hline
 \end{tabular}
 \end{table*}
 
In this section, we focus on the $\pi^+$ production channel which is responsible for almost 80 \% of the $e^+$ and, therefore, deserves the most careful discussion. We introduce our strategy and the most important concepts for the modeling of the production cross section and subsequent decays. Many concepts from this section will be applied analogously to the other channels discussed in Sec.~\ref{sec:other_channels}. So, this section also serves as an important reference for the following.

As outlined in Sec.~\ref{sec:cross_sections}, secondary $e^+$  are produced via various different channels. The common scheme is that the $e^+$  are produced indirectly, \textit{i.e.} they come from the decay of one or more intermediate mesons or hadrons. Some channels involve an additional $\mu^+$ decay.

The secondary production gives most of its contribution to AMS-02 positron data in the range between 0.5 and 10 GeV.
These positrons are mostly produced from CR protons with energies between 5 GeV and 200 GeV, which corresponds to center of mass energies between 3.6 and 20 GeV. The measurement of pion production in this energy range and with the widest coverage of the kinetic parameter space is provided by the NA49 experiment \cite{2005_NA49} at $\sqrt{s}=17.3$ GeV. Therefore, we decided to gauge our modeling of the $e^+$ invariant cross section on NA49.

To good approximation the Lorentz invariant production cross section is scaling invariant:
\begin{equation}
   \sigma_{\rm inv}(s,x_R,p_T) \approx
   \sigma_{\rm inv}(s_0,x_R,p_T). 
   \label{eq:main_equation_inv}
\end{equation}
However, two ingredients are violating this approximate invariance: first, the rise of the inelastic cross section for $p+p$ collisions (see Sec.~\ref{sec:piplus}) and, second, the softening of the $p_T$ shape at large center of mass energies (see Sec.~\ref{sec:piplus_differentCME}).

Guided by the above considerations, our strategy is as follows: in the first step, we fix the kinematic shape of the $\pi^+$ production cross section using only the NA49 data. In the second step, we combine measurements of the multiplicity at different $\sS$ down to 3 GeV, and measurements of the multiplicity and the $p_T$ shape by CMS \cite{2012_CMS,2017_CMS} and ALICE \cite{2011_ALICE} to calibrate our model over a large range of energies. 
A summary of the included datasets is provided in Tab.~ \ref{tab::pp_data}. 

We detail the analytic model for the pion production in Sec.~\ref{sec:piplus}. In Sec.~\ref{sec:piplus_fitNA49} and \ref{sec:piplus_differentCME} we discuss the fit to NA49 and other center of mass energies, respectively and show the first results in Sec.~\ref{sec:piplus_results}.

\subsection{Model for the invariant production cross section} \label{sec:piplus}

In this section, we specify the analytical model of the invariant cross section for the inclusive production of $\pi^+$ in $p + p$ collisions. In the past, several empirical parametrizations were proposed and compared to existing data at that time \cite{PhysRevD.15.820,RevModPhys.49.753,ALPER1975237,Tan:1984ha}.
In the meantime, NA49 data \cite{2005_NA49} became available and Ref.~\cite{Norbury_2007} identified three of these parametrizations that provide a moderately good agreement with the new $p+p$ data. However, the parametrizations were not refitted to the new NA49 data and indeed the agreement was not very precise. Finally, the best two of those three parametrizations were also scaled to the $p+$C data, again showing a moderately good agreement. 

Instead of adopting one of these old parametrizations, we propose a new parametrization of $\sigma_{\rm inv}$ which can fit a large number of datasets of the inclusive production of $\pi^+$ in $p + p$ collisions, with $\sqrt{s}$ ranging from few GeV up to LHC energies. 
As outlined in Ref.~\cite{2005_NA49}, the $\pi^+$ are produced by a combination of prompt emission, emerging from the hadronization chains, and the decay of hadronic resonances, in particular from $\rho$ and $\Delta$. Inspired by this idea, we write $\sigma_{\rm inv}$ as the sum of two terms, called $F_p$ and $F_r$, which should roughly follow the prompt and resonance components. However, we emphasize that the individual terms do not have precise physical meaning. It is neither our aim nor do we have the data to precisely distinguish the physically prompt and resonant production. Our only aim is to describe the total cross section which corresponds to the sum of the $F_p$ and $F_r$ terms. 
The Lorenz invariant cross section is given by:
\begin{equation}
   \sigma_{\rm inv}= \sigma_0 (s) \,  c_1 \,   \Big[F_p(s, p_T, x_R) + F_r(p_T, x_R)\Big]
    \, A(s),
   \label{eq:main_equation}
\end{equation}
where $\sigma_0 (s)$ is the total inelastic $p+p$ cross section. The derivation of $\sigma_0 (s)$ along with its uncertainty is discussed in App.~\ref{app:sigma_0}. 
The functional form of $F_p(p_T, x_R)$ is partially inspired by the parametrizations from Ref.~\cite{Norbury_2007} (and Refs. therein). Specifically, we use:
\begin{eqnarray}     
     \label{eq:function_prompt}
     F_p&&(s, p_T, x_R) = (1-x_R)^{c_2} 
        \exp(-c_3 \, x_R) \, p_T^{c_4}  \\ \nonumber
      &&\times \exp\left[ -c_5 \sqrt{s/s_0}^{\;c_6} 
      \left(\sqrt{p_T^2 + m_\pi^2}-m_{\pi}\right)^{c_7 \sqrt{s/s_0}^{\;c_6}} \right] \;, 
\end{eqnarray}
where $\sqrt{s_0}=17.3$ GeV is the energy of NA49 data. The model parameters $c_i$ will be fitted to the available cross section data, as explained in the following of this Section.

\begin{figure*}[t]
    \includegraphics[width=0.49\textwidth]{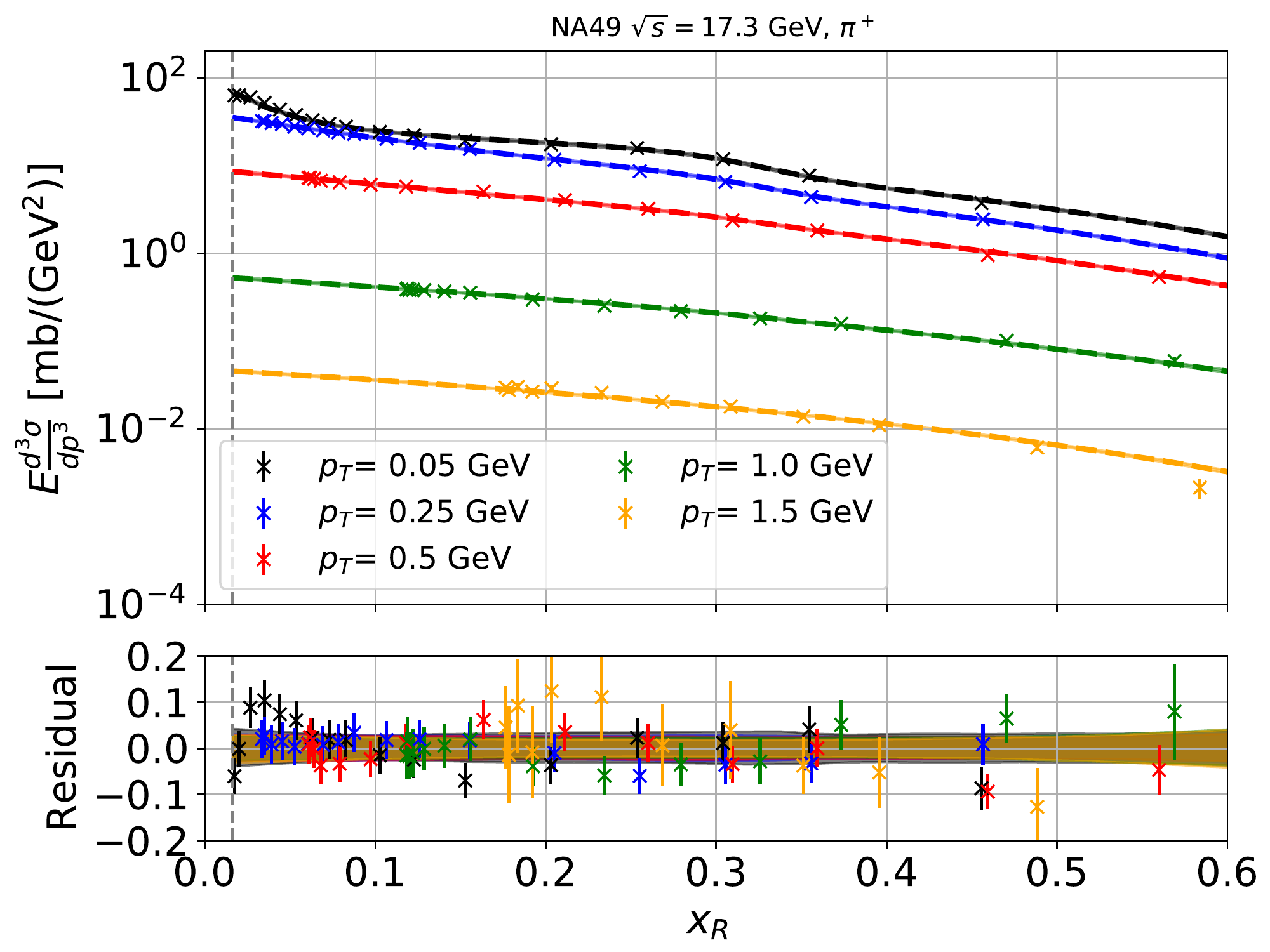}
    \includegraphics[width=0.49\textwidth]{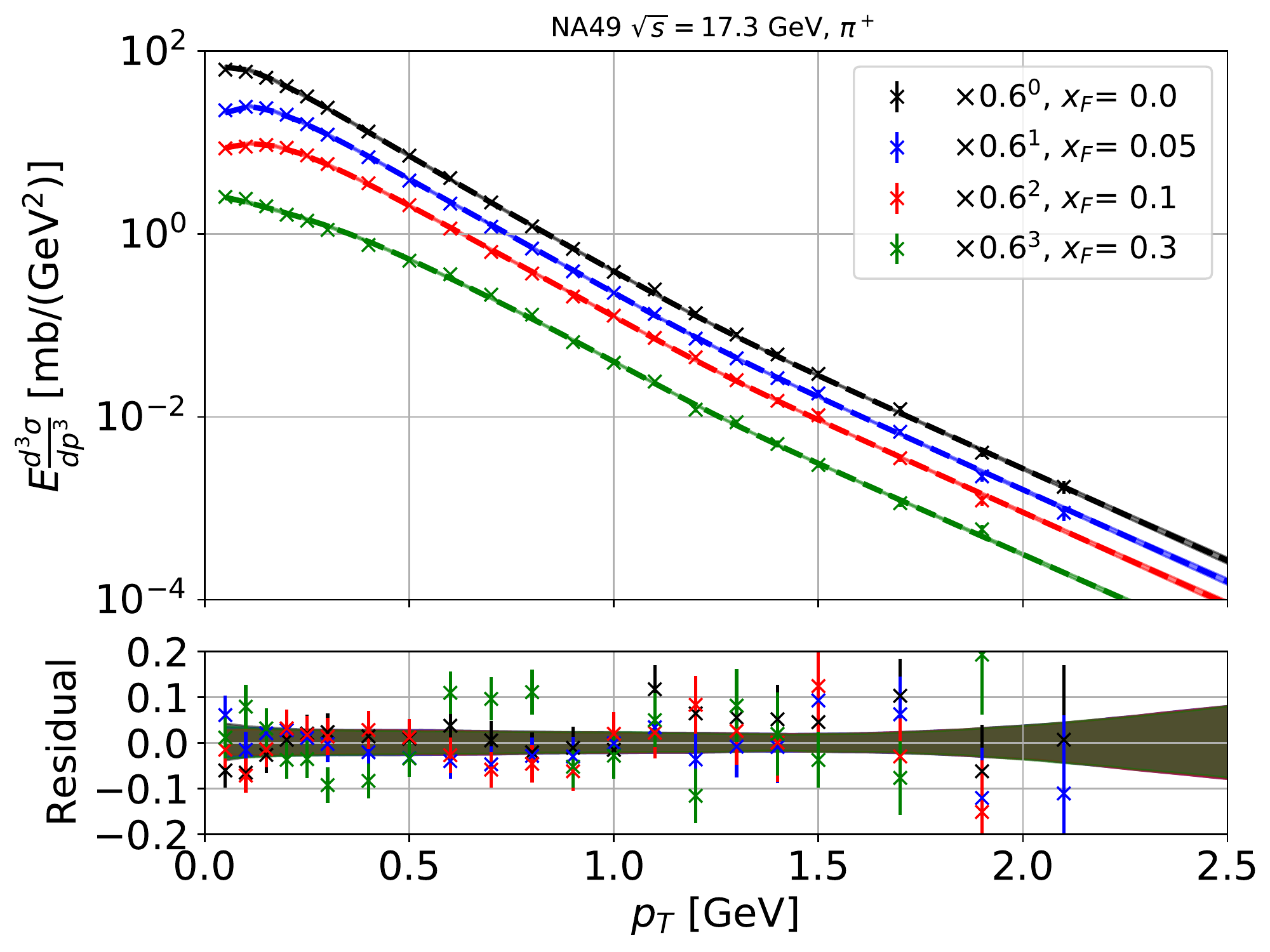}
    \caption{
             Results of the fit on the NA49 data \cite{2005_NA49} invariant cross section for the inclusive $\pi^+$ production in $p+p$ collisions. The left (right) panel shows the NA49 data along with our fit results for representative  $p_T$ ($x_F$) values, as a function of $x_R$ ($p_T$). Each curve is plotted along with its $1\sigma$ uncertainty band. In the bottom part of each panel we plot the residuals, which are defined as (data-model)/model, and the width of the $1\sigma$ uncertainty band on the model. 
    } 
    \label{Fig:pi-plus-NA49}
\end{figure*}

On the other hand, the empirical expression for $F_r$ is motivated by the contributions from resonances, as simulated in Ref.~\cite{2005_NA49} (see their Fig. 54). 
The functional form of $F_r(p_T, x_R)$ reads:
\begin{eqnarray}
  \label{eq:fr}
  F_r&&(p_T, x_R) = (1-x_R)^{c_{8}}  \\ \nonumber
  &&\times \exp\left[ -c_{9}\,p_T - \left(\frac{|p_T-c_{10}|}{c_{11}}\right)^{c_{12}}\right]  \\ \nonumber
  &&\times \biggl[c_{13}\exp(-c_{14}\,p_T^{c_{15}} x_R) +  \\ \nonumber
  && \qquad + c_{16} \exp \left(-\left(\frac{|x_R- c_{17}|}{c_{18}}\right)^{c_{19}}\right) \biggr].
\end{eqnarray}
Finally, we allow for an additional scaling with $\sqrt{s}$, which is required to obtain the correct $\pi^+$ multiplicity at different energies. The functional form is given by
\begin{eqnarray}
  A(s) &=&  \frac{ 1+\left(\sqrt{s/c_{20}}\right)^{c_{21}-c_{22}}}{ 1+\left(\sqrt{s_0/c_{20}}\right)^{\,c_{21}-c_{22}}}\,
             \left(\sqrt{\frac{s}{s_0}}\right)^{c_{22}}, 
  \label{eq:As}
\end{eqnarray}
which represents a smoothly broken power law as function of $\sqrt{s}$ with slopes  $c_{21}$ and  $c_{22}$ above and below the break position at  $\sqrt{s}=c_{20}$, respectively. 
In all the formulas reported in the paper, $p_T, \sqrt{s}$, the mass of the particles,
$\sqrt{s_0}$ and energies are intrinsically normalized to 1 GeV, in order to have dimensionless parameters. 

\subsection{Fit of the $\pi^+$ production to NA49 data}\label{sec:piplus_fitNA49}
The NA49 experiment at the CERN SPS performed precise measurements of $\pi^{+}$ inclusive cross sections of $p+p$ interaction. Data are collected at $\sqrt{s} = 17.3$\ GeV and over a large range of $x_F$ and $p_T$, where $x_F=2 p_L/\sS$ is the reduced longitudinal momentum. In the first step, we fix the shape of the Lorenz invariant cross section as a function of $x_R$ and $p_T$ at the NA49 center of mass energy. To a first approximation, the shape of the cross section is invariant and does not change when going to different values for $\sqrt{s}$. This approximation works very well for $\sqrt{s}$ values below 50 GeV. At higher energies this scaling invariance is broken (see also \ref{sec:piplus_differentCME}). The parametrizations of Eqs.~\eqref{eq:function_prompt}-\eqref{eq:As} contain a few parameters that change the behavior of the invariant cross section as function of $\sqrt{s}$, \textit{i.e.} they break the scaling invariance. More specifically, those parameters are $c_6$, $c_{20}$, $c_{21}$ and $c_{22}$. However, the parametrization is chosen such that the cross section at the center of mass energy of NA49 is independent of those parameters. Hence we can use the NA49 data to fix all the other parameters of our model that do not depend on $\sqrt{s}$. 
We perform a $\chi^2$-fit using the \textsc{Multinest} package \cite{Multinest_2009} to minimize the $\chi^2$, with statistical and systematic uncertainties added in quadrature. We note that there is also a normalization uncertainty of 1.5\%. This normalization uncertainty is not included in the fit but taken into account separately (see below). We use \textsc{Multinest} with 1000 live points, an enlargement factor of \texttt{eft}$=0.7$ and a stopping parameter of \texttt{tol}$=0.1$.

Our results are summarized in Fig.~\ref{Fig:pi-plus-NA49}, where we plot the invariant cross section for the inclusive $\pi^+$ production in $p+p$ collisions as a function of $x_R$ (left) and $p_T$ (right). The data are displayed along with our best fit results and the $1\sigma$ uncertainty for a few representative values at fixed $p_T$ and $x_F$, respectively. The residuals of the data and the width of the theoretical uncertainty band are displayed in the bottom panels. The fit converges to a total $\chi^2_{\rm NA49}= 338$ with 263 degrees of freedom (d.o.f.), meaning that we obtain a very good fit  with $\chi^2_{\rm NA49}$/d.o.f.$=1.29$. The data are well described at all $p_T$ and $x_F$ values. The structures in the low $p_T$ data are very well followed by our parametric formulae, Eqs.~\eqref{eq:function_prompt} and \eqref{eq:fr}. 

Finally, we derive the uncertainties on our cross section fit. To this end, we extract the covariance matrix and the mean parameter values from the \textsc{MultiNest} fit. The covariance matrix $C$ contains the uncertainties and correlations of all the fit parameters. At this point, we account for the previously neglected normalization uncertainty of the NA49 data. The overall normalization of the cross section is dictated by the $c_1$ parameter. So, an additional 1.5\% uncertainty on the normalization can be accounted by resetting the corresponding diagonal entry of the covariance matrix: $C_{1,1} \rightarrow C_{1,1} + 0.015^2\,c_1^2$. Then, we sample 500 parameter realizations using a multivariate Gaussian distribution. Figure~\ref{Fig:pi-plus-NA49} shows the uncertainty band at the 68\% confidence level, which spans about 5\% over all the kinematic range explored by the data. For  $p_T > 2$ GeV it increases to almost 10\%. However, we note that high $p_T$ values are suppressed after the angular integration (see Eq.~\eqref{eq:convolution}).

\subsection{Fit to different center of mass energies}\label{sec:piplus_differentCME}

The general kinematic shape of the invariant $\pi^+$ production cross section has been fixed in the previous section. Here we focus on the scaling of the cross section at different $\sqrt{s}$. Our parametrization introduces two physically different dependencies on $\sqrt{s}$. On the one hand, the parameter $c_6$ in Eq.~ \eqref{eq:function_prompt} allows a softening of the $p_T$ shape as observed at high energies, while on the other hand the factor $A(s)$ and the parameters $c_{20}$ to $c_{22}$ introduce an overall renormalization. In this section, we proceed with the determination of the parameters $c_6$, $c_{20}, c_{21}$,  and $c_{22}$. All the other parameters are fixed to the values of the fit to the NA49 data, as described above in Sec.~\ref{sec:piplus_fitNA49}.

To extend to $\sqrt{s}$ below NA49 measurement we use the multiplicity measurements of NA61/SHINE \cite{Aduszkiewicz:2017sei} as well as a collection of data points provided in Ref.~\cite{osti_4593576} (in the following also called Antinucci). At larger $\sqrt{s}$ we use the $p_T$ dependent data provided by CMS \cite{2012_CMS,2017_CMS} and ALICE \cite{2011_ALICE} at central rapidity. All datasets and their $\sqrt{s}$ are summarized in Tab.~\ref{tab::pp_data}. As in the previous section, we perform a $\chi^2$ fit and use the \textsc{Multinest} \cite{Multinest_2009} package to scan over the parameter space. 

Typically, each cross section measurement contains a statistical, a systematic, and a scale uncertainty. In the last section, we only used a single dataset, the one from NA49, which allowed us to use a simplified treatment where we ignore the scale uncertainty of 1.5\% at first and then added it in a post-procedure. Here we combine datasets from different experiments and, thus, the scaling uncertainty has to be included from the beginning. For datasets with only a single data point, this is straightforward and we can simply add all the individual uncertainties in quadrature. In practice, those are the multiplicity measurements taken by NA61/SHINE and Antinucci.
We note that the Antinucci data points are a collection from different experiments and therefore have independent uncertainties, and the NA61/SHINE are taken at different $\sqrt{s}$. 
On the other hand, at higher energies, we use the measurements of the invariant cross section by ALICE and CMS at central rapidity. The cross section is provided for values of the transverse momentum between 0.1 and 2.5\ GeV. For those data points the scaling uncertainty is fully correlated so we cannot simply add them in quadrature in the definition of the total $\chi^2$. Instead, we follow Ref.~\cite{Korsmeier_2018} and introduce nuisance parameters allowing for an overall renormalization of each dataset from ALICE and CMS. Then, the total $\chi^2$ is defined as the sum of two parts: 
\begin{eqnarray}
  \label{eqn::chiSquare}
  \chi^2 &=& \chi_\mathrm{stat}^2 + \chi_\mathrm{scale}^2.
\end{eqnarray}
Here the first term accounts for the statistical and systematic uncertainty, while the second term constrains the nuisance parameters according to the scale uncertainties. Explicitly, $\chi_\mathrm{stat}^2$ is given by the sum over all data points $i_k$ and all datasets $k$:
\begin{eqnarray}
    \label{eqn::chiSquare_stat}
    \chi_\mathrm{stat}^2 &=& \sum \limits_{k} \sum\limits_{i_k} 
    \frac{\left( \omega_k {\sigmaInv}_{i_k} - {\sigmaInv}(  {\sS}_{i_k}, {x_R}_{i_k}, {p_T}_{i_k}   )\right)^2}{\omega_k^2 \sigma_{i_k}^2}, \quad\quad
\end{eqnarray}
where ${\sigmaInv}_{i_k}$ is the measured cross sections and ${\sigmaInv}(  {\sS}_{i_k}, {x_R}_{i_k}, {p_T}_{i_k})$ is the evaluation of our cross section parametrization at the corresponding kinematic variables. The nuisance parameters $\omega_k$ rescale both the cross section measurement and the uncertainties $\sigma_{i_k}^2$. 
Then, the second term of the Eq.~\eqref{eqn::chiSquare} is given by 
\begin{eqnarray}
  \label{eqn::chiSquare_scale}                                       
  \chi_\mathrm{scale}^2(\omega ) &=& \sum \limits_{k}
                                           \frac{\left( \omega_k - 1 \right)^2}{ \sigma_{\mathrm{scale},k}^2},
\end{eqnarray}
where $\sigma_{\mathrm{scale},k}$ is the scale uncertainty for each data set. 
We stress that the sum in Eq.~\eqref{eqn::chiSquare_stat} runs over every single data point, while the sum in Eq.~\eqref{eqn::chiSquare_scale} only runs over datasets. So, moving up or down all the data points of a dataset by the same factor is only penalized once and not for each data point. 

Finally, we address two more subtleties. 
First, the ALICE and CMS experiments provide $d^2n/(dp_T dy)$ data (that we convert in Lorentz invariant cross section) averaged in relatively large rapidity bins of $|y|<0.5$ and $|y|<1$, respectively. In order to take this into account, we also average our model evaluation over those rapidity ranges.
\begin{figure*}[t]
\setlength{\unitlength}{1\textwidth}
\begin{picture}(1,0.4)
  \put(-0.01 , 0.0){\includegraphics[width=0.505\textwidth]{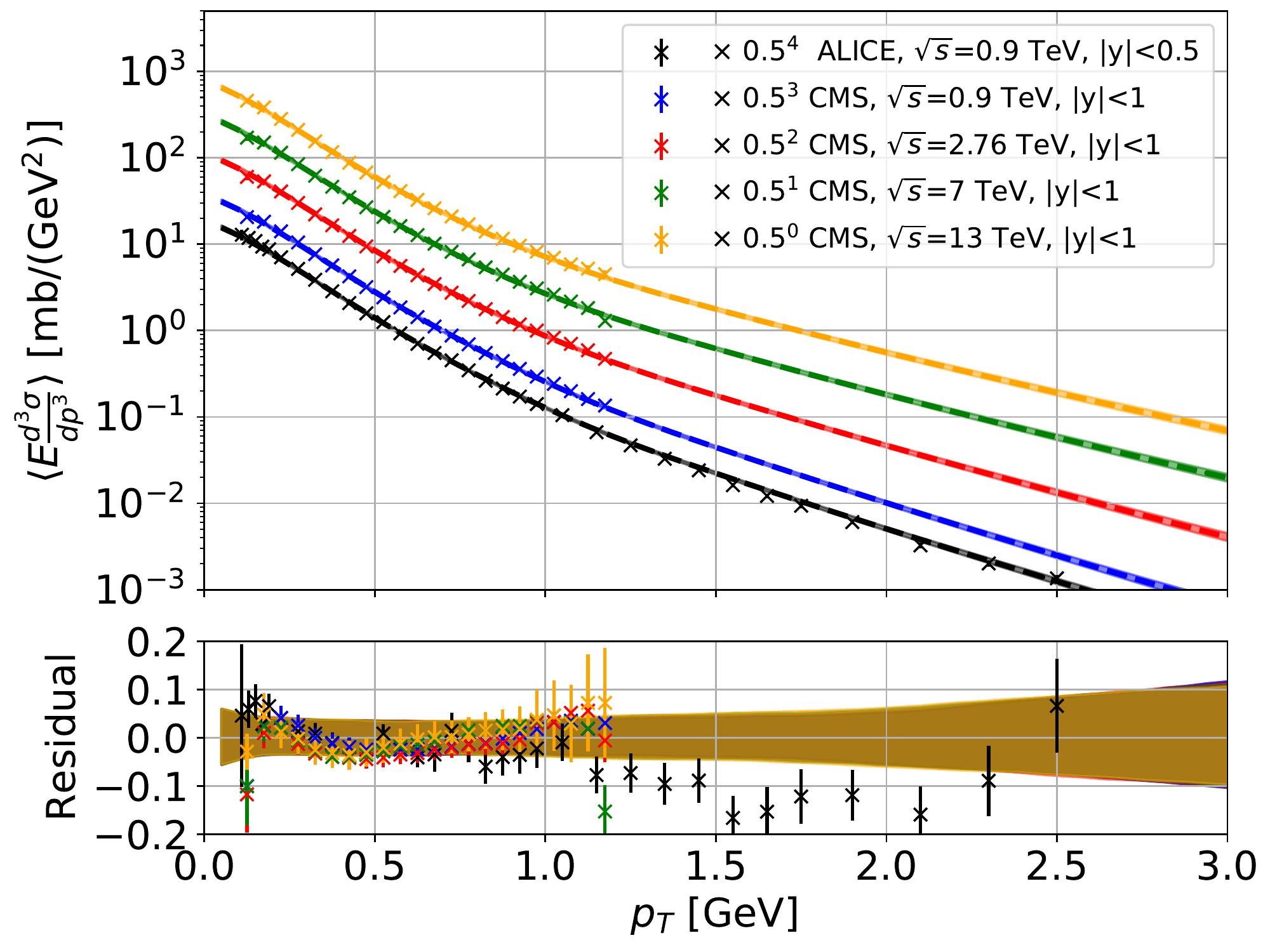}}
  \put( 0.50,-0.004){\includegraphics[width=0.25\textwidth  ]{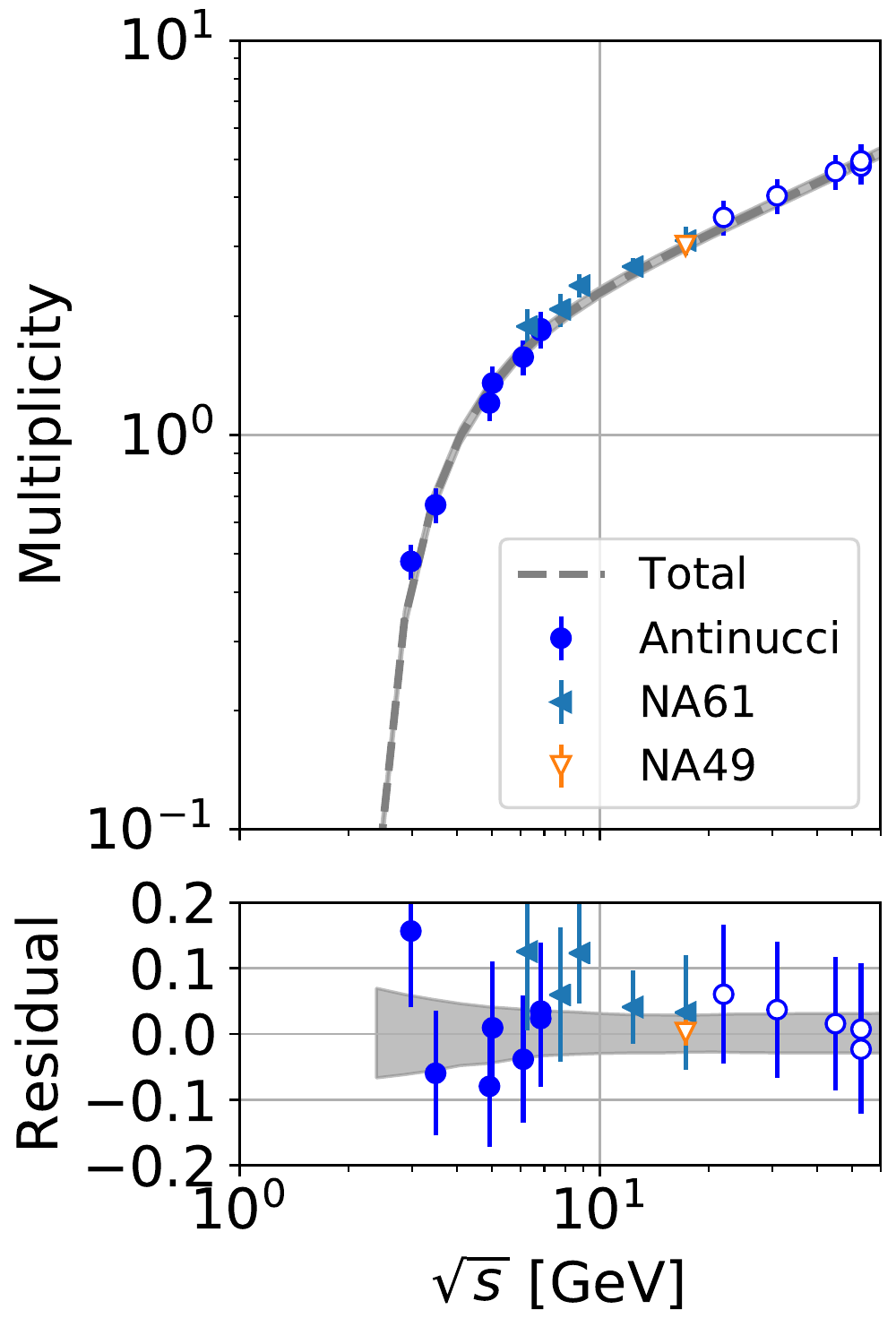}}
  \put( 0.75,-0.004){\includegraphics[width=0.25\textwidth  ]{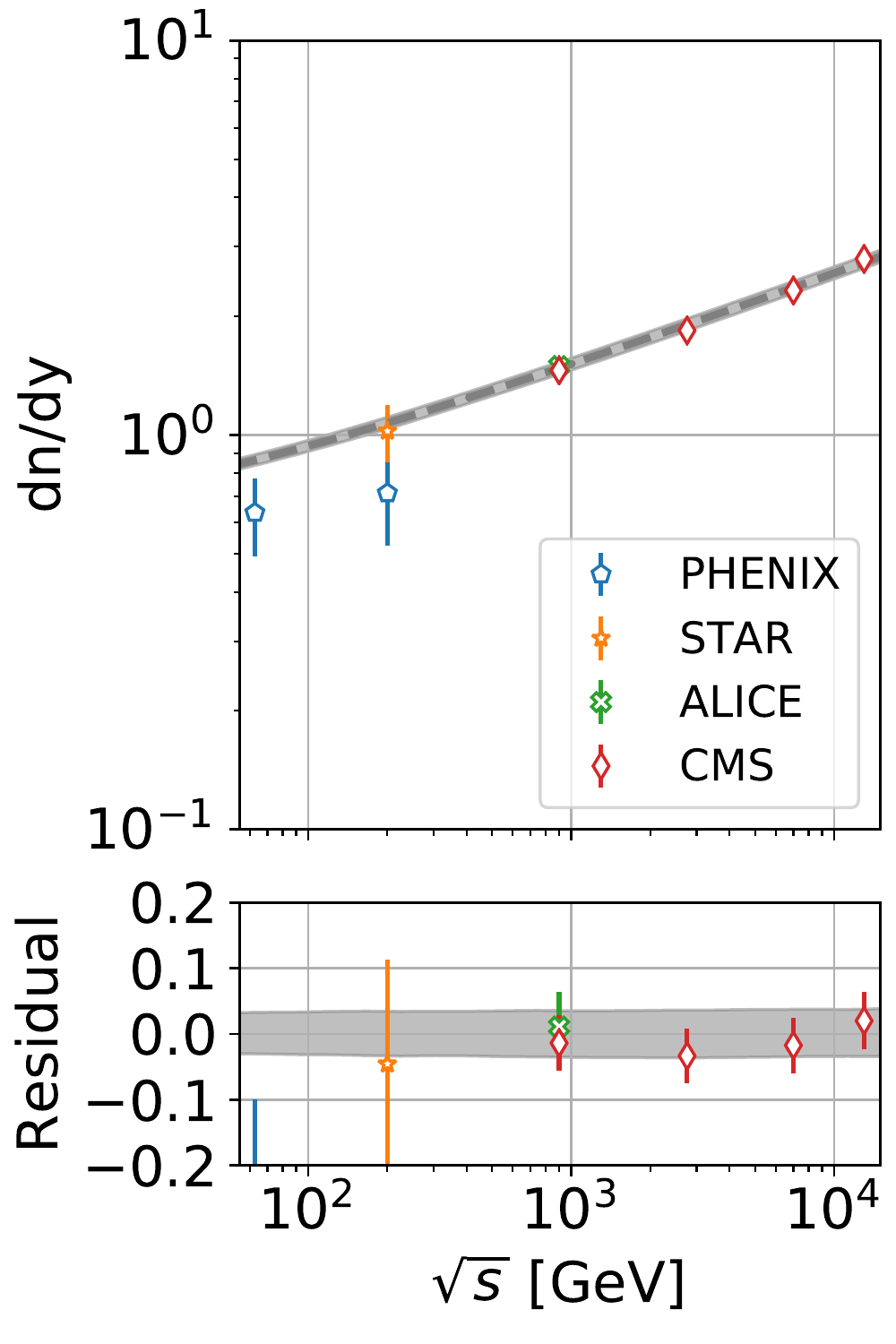}}
\end{picture}
  \caption{
        Left panel: invariant cross section of inclusive $\pi^+$ production in $p+p$ collisions at large $\sqrt{s}$ as measured by ALICE and CMS. The dashed lines represent the best fit parametrization and the shaded bands show the uncertainty at the $1\sigma$ level.
        Right panel: multiplicity (left sub-panel) and $dn/dy$ (right sub-panel) of $\pi^+$ production in $p+p$ collisions measured at different $\sqrt{s}$. The solid lines represent the best fit parametrization and the grey shaded bands show the uncertainty of our fit at the $1\sigma$ level. Filled data points are included in the fit while open data points are only plotted for comparison. The bottom panels shows the residuals defined as (data-model)/model.
  }
  \label{fig:sqrt_s_sigma}
\end{figure*}
Second, the recent experiments (NA61/SHINE, ALICE, and CMS) perform feed-down corrections, namely they subtract the $\pi^+$ production from the weak decay of strange particles which are mainly $K_0^S$, but also $\bar{\Lambda}$ and $\Sigma^+$. In contrast, the collection of multiplicity measurements from Antinucci is not corrected for this feed-down. So, we correct those data points by subtracting the contributions of $K_0^S$ using our estimation from Sec.~\ref{sec:other_channels}. This contributions to the total multiplicity vary from 0.4\% for $\sqrt{s}$=3 GeV to 1.7\% for $\sqrt{s}$=6.8 GeV. To be conservative we add this correction at each data point to the measurement uncertainty in quadrature.

Figure~\ref{fig:sqrt_s_sigma} shows the results of the fit at high energies. The invariant cross section is plotted as a function of $p_T$ and at different energies of the corresponding ALICE and CMS data. The ALICE data at $\sqrt{s}=0.9$\ TeV cover a wide range of $p_T$ from 0.1 GeV up to 2.5 GeV, while the CMS data span a smaller range in $p_T$, only up to 1.2 GeV, but they extend the center of mass energies up to 13 TeV. The fitted function provides a good agreement with the data. The uncertainty on $\sigmaInv$ is about 5\% at the lowest $p_T$ values and increases to 10\% for $p_T>2$ GeV.  

In Fig.~\ref{fig:sqrt_s_sigma} (right panel), we compare the multiplicity from our parametrization with the available data as a function of $\sqrt{s}$. The plot is divided into two energy regimes: at lower energies, experiments determine the total multiplicity which is integrated over the whole kinematic parameter space, while the collider experiments only determine the multiplicity at central rapidity, often expressed as an average $dn/dy$. The fit includes the data points with the filled symbols from Antinucci and NA61/SHINE, while the open data points are only plotted for comparison. At high energies, we show next to the ALICE and CMS data also data from PHENIX \cite{2011_PHENIX} and STAR \cite{STAR:2006xud}. The $dn/dy$ data points at high energies are actually averaged over different rapidity ranges, namely PHENIX: $|\eta|<0.35$, STAR: $|y|<0.5$, CMS: $|y|<1$, ALICE: $|y|<0.5$, while our model is plotted for $|y|<0.5$. However, $dn/dy$ is fairly flat at high energies and midrapidity such that the impact on the model (grey line) is negligible. ALICE and CMS provide measurements which are feed-down corrected, while we perform the feed-down correction for PHENIX and STAR ourselves by subtracting the contributions of $K_0^S$%
\footnote{
    Actually, STAR \cite{STAR:2006xud} measured only the average of $\pi^+$ and $\pi^-$ production, which becomes symmetric at high energies.
}.

In general, both the fitted data and the ones plotted for comparison are in good agreement with our best-fit parametrization. For the data points from NA49, ALICE, and CMS this is expected, since their data in the $x_R - p_T$ plane have been included in the fits to Eqs.~\eqref{eq:function_prompt}-\eqref{eq:As}. Instead, the comparison of the data from STAR and PHENIX provides an independent cross-check. The STAR is in very good agreement with our parametrization, while the PHENIX data lie systematically below our multiplicity line. We note that something similar was also observed for antiprotons \cite{winkler_2017}, potentially pointing to a more general unaccounted systematic. 
\begin{figure*}[t]
    \includegraphics[width=0.49\textwidth]{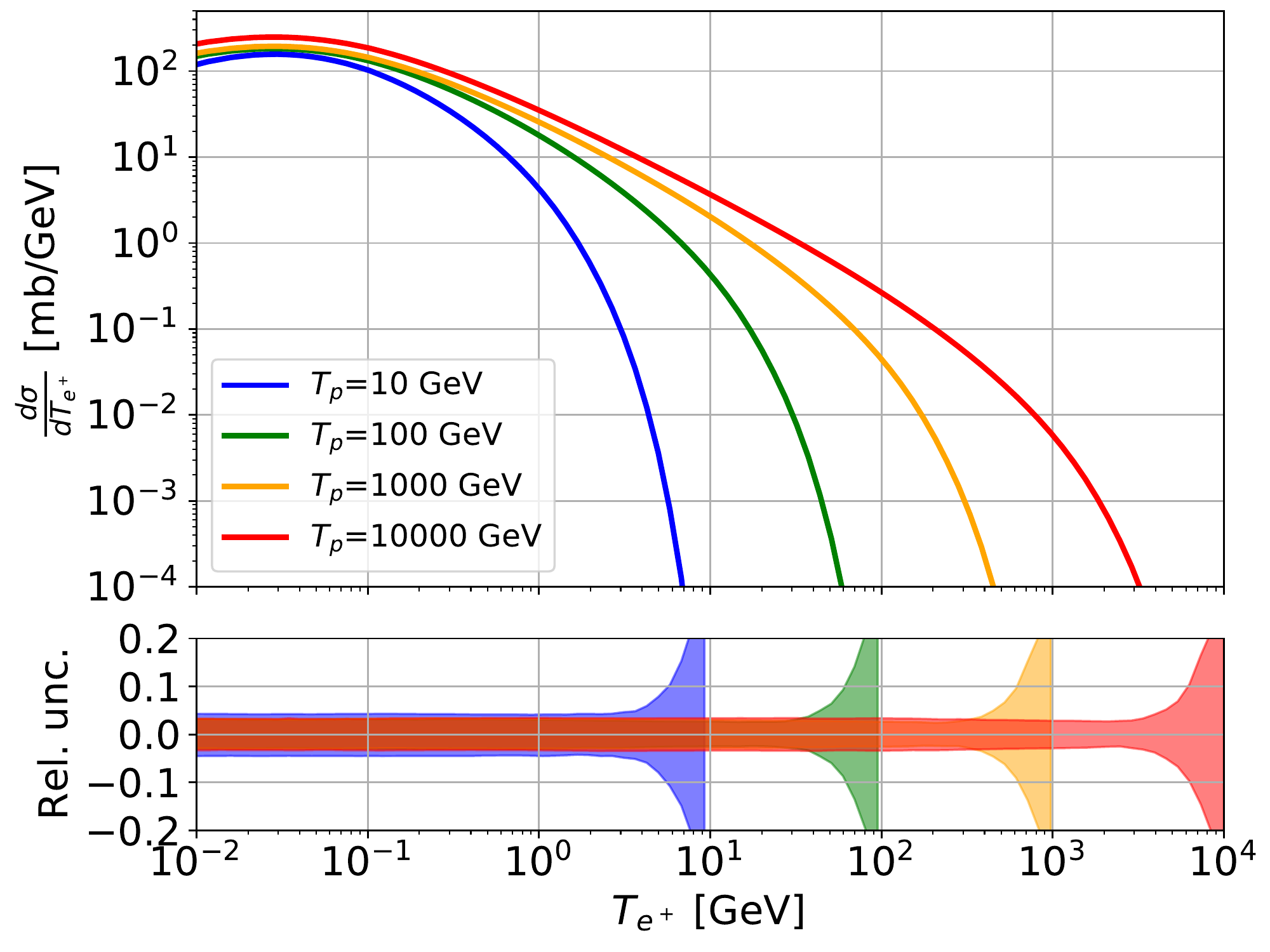}
    \includegraphics[width=0.49\textwidth]{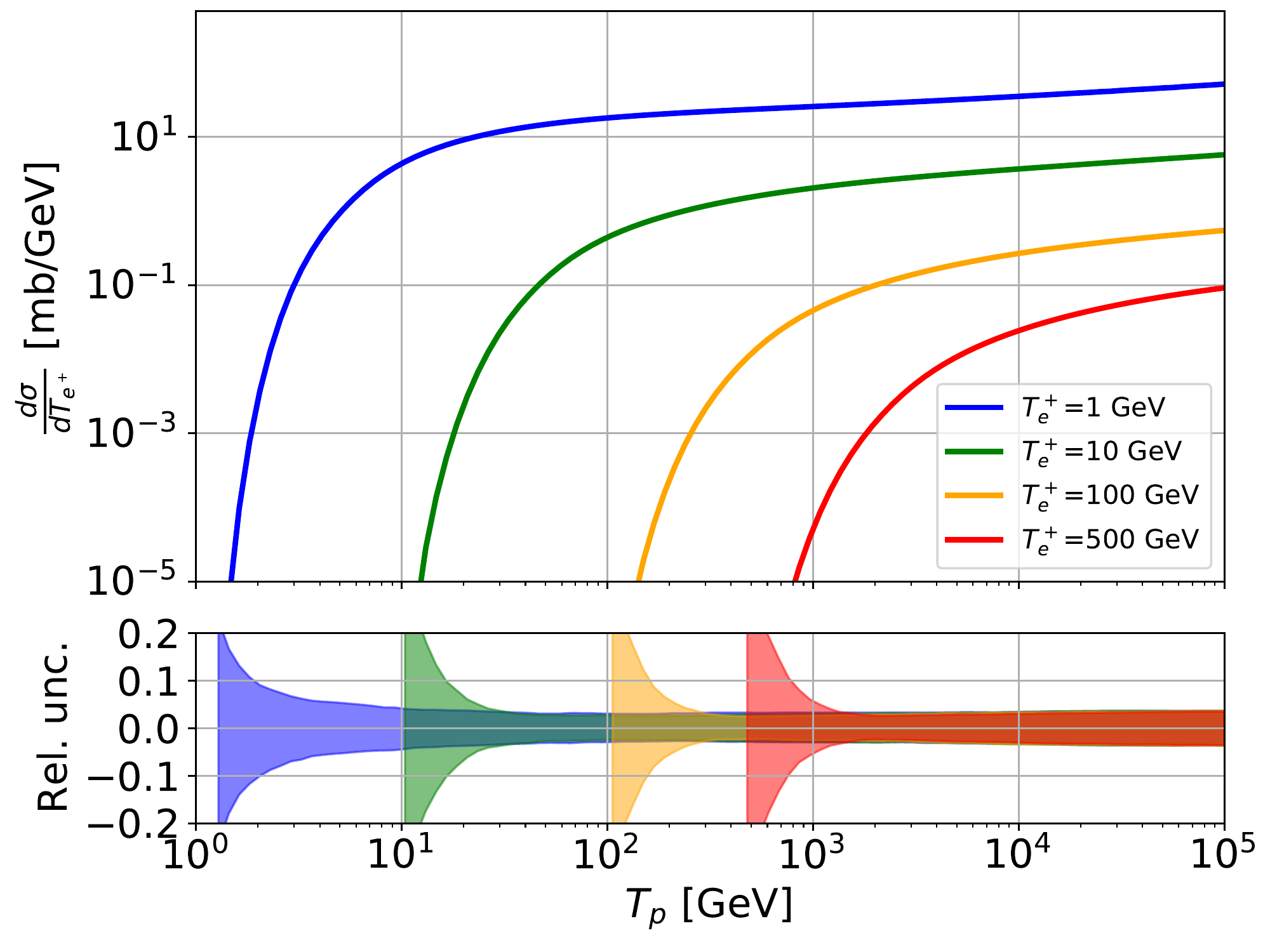}
    \caption{Differential cross section for the production of $e^+$ from $\pi^+$ in $p+p$ collisions, computed for different incident kinetic proton energies as a function of $e^+$ kinetic energy (left) and different $e^+$ kinetic energies as a function of $p$ kinetic energy (right).
    } 
    \label{Fig:pi-plus-final}
\end{figure*}
Overall, our parametrization provides a good fit to the datasets at different $\sqrt{s}$. The $\chi_{n}^2/$d.o.f. of the best fit converges to 189/129. More details and individual contributions are provided in Tab.~\ref{tab::chi2_pp}. Furthermore, the parameters $c_6, c_{21}, c_{22}$, and $c_{23}$ are all well constrained by the fit and their values are summarized in Tab.~\ref{tab::Fit_results_pp_pion}. Within our parametrization, the multiplicity is determined with a precision of about 3\% above $\sqrt{s}$ of 10 GeV, increasing to 5\% at the lowest $\sqrt{s}$.
At high energies, the radial scaling invariance is not only broken by the general increase of the cross section with $\sqrt{s}$ but also because the $p_T$ shape hardens. ALICE and CMS measure the cross section as a function of $p_T$ only at midrapidity. 
  
NA61/SHINE also provided data in the $x_R - p_T$ plane which are however not included in our fit to Eqs.~\eqref{eq:function_prompt}-\eqref{eq:As}. As discussed above we use a phenomenologically motivated function and fix the kinematic shape of the cross section with the most reliable data from NA49 data assuming radial scaling invariance. An additional dataset would require a more careful assessment of systematics to avoid over-constraining the fit parameters, and thus underestimating uncertainties. 
Moreover, we observed some inconsistencies in the tables provided by Ref.~\cite{Aduszkiewicz:2017sei}%
\footnote{
    In the database \url{https://www.hepdata.net/record/ins1598505} referred \cite{Aduszkiewicz:2017sei} for some data points the systematic errors are written to zero or set equal to the central data point.
}. 
We decided therefore not to include this data in the fit. Nevertheless, we have checked that the NA61/SHINE data  are generally consistent with our parametrization also in the $x_R - p_T$ plane. We provide more information in App.~\ref{app:comparison_NA61}.

In our parametrization, we assume that there is no similar violation of scaling in $x_R$. While the bulk of pions (and thus finally also positrons) are produced at midrapidity, the steeply falling CR projectile flux in the source term enhances pions produced in forward direction~\cite{Donato:2017ywo}. The enhancement is supposed to become less important at very high energies, but it might be important at intermediate energies, \textit{i.e.} between NA49 and ALICE/CMS. In the future, more experimental data might help to solve the issue.

\begin{table}[b!]
\caption{Results from the best fit and the 1$\sigma$ error for the parameters in Eqs.~\eqref{eq:main_equation}, \eqref{eq:function_prompt}, \eqref{eq:fr} and \eqref{eq:As}. 
$c_1$ is in units of GeV$^{-2}$. }
\label{tab::Fit_results_pp_pion}
\begin{tabular}{ l  c  c }
 \hline \hline
                             &   $\pi^+$                         &   $\pi^-$                       \\ \hline
$c_1$                        &   $1.05\pm 0.14$                  & $0.85\pm 0.15$                  \\
$c_2$                        &   $3.62\pm 0.24$                  & $5.37\pm 0.30$                  \\
$c_3$                        &   $-1.05\pm 0.36$                 & $-2.25\pm 0.46$                 \\
$c_4$                        &   $0.10\pm 0.04$                  & $0.55\pm 0.11$                  \\
$c_5$                        &   $4.96\pm 0.14$                  & $4.83\pm 0.20$                  \\
$c_6$                        &   $(-3.81\pm 0.04)\cdot 10^{-2}$ & $(-4.45\pm 0.06)\cdot 10^{-2}$ \\
$c_7$                        &   $0.91\pm 0.02$                  & $1.01\pm 0.03$                  \\
$c_8$                        &   $0.11\pm 0.09$                  & $1.04\pm 0.26$                  \\
$c_9$                         &   $6.91\pm 0.10$                  & $7.09\pm 0.19$                  \\
$c_{10}$                     &   $0.54\pm 0.03$                  & $0.60\pm 0.06$                  \\
$c_{11}$                     &   $0.67\pm 0.03$                  & $0.68\pm 0.05$                  \\
$c_{12}$                     &   $3.67\pm 0.41$                  & $2.67\pm 0.25$                  \\
$c_{13}$                     &   $4.68\pm 0.73$                  & $5.80\pm 1.04$                  \\
$c_{14}$                     &   $3.10\pm 0.16$                  & $3.87\pm 0.42$                  \\
$c_{15}$                     &   $-0.84\pm 0.03$                 & $-0.86\pm 0.05$                 \\
$c_{16}$                     &   $0.34\pm 0.07$                  & $3.15\pm 0.03$                  \\
$c_{17}$                     &   $0.14\pm 0.01$                  & $(1.67\pm 0.92)\cdot 10^{-2}$  \\
$c_{18}$                     &   $0.18\pm 0.01$                  & $0.12\pm 0.01$                  \\
$c_{19}$                     &   $5.35\pm 0.73$                  & $0.83\pm 0.06$                  \\
$c_{20}$                     &   $9.79\pm 0.70$                  & $9.61\pm 0.72$                  \\
$c_{21}$                     &   $-0.79\pm 0.05$                 & $-0.90\pm 0.07$                 \\
$c_{22}$                     &   $(2.06\pm 0.04)\cdot 10^{-1}$  & $(2.09\pm 0.04)\cdot 10^{-1}$  \\
 \hline \hline
\end{tabular}
\end{table}

\subsection{Results on the $e^+$ production cross section }\label{sec:piplus_results}

Now we have all the ingredients to compute the differential cross section for the production of $e^+$ as a function of the incident proton energy, $T_p$, and the positron energy, $T_{e^+}$, using Eq.~\eqref{eq:convolution} and Eq.~\eqref{eq:solid_int}. As a matter of fact, it means that we have to perform a double integration in the solid angle and in the $\pi^+$ energy. 
In Fig.~\ref{Fig:pi-plus-final}, we present the result for the cross section ${d\sigma_{pp \rightarrow \pi^+ + X}}/{d T_{e^{+}}}$ as function of $T_{e^+}$ (left panel) and $T_p$ (right) panel for a few representative values of $T_p$ and $T_{e^+}$, respectively. 
The cross section peaks at positron energies below 100 MeV at about 100 to 300 mb/GeV, almost independently of $T_p$, and decreases rapidly to zero for $T_{e^{+}}$ close to the threshold, \textit{i.e.} at $T_{e^{+}}=T_p$. 
The uncertainties are about 5\% for almost all $T_{e^{+}}$, which is in agreement with the results from in Secs.~\ref{sec:piplus_fitNA49} and \ref{sec:piplus_differentCME}. 
The relative uncertainty increases above 20\% when approaching the threshold. We note, however, that this kinematic range is suppressed in the positron source term and has a negligible impact on the final uncertainty.

\begin{figure*}[t]
\setlength{\unitlength}{1\textwidth}
\begin{picture}(1,0.4)
  \put(-0.01 , 0.0){\includegraphics[width=0.505\textwidth]{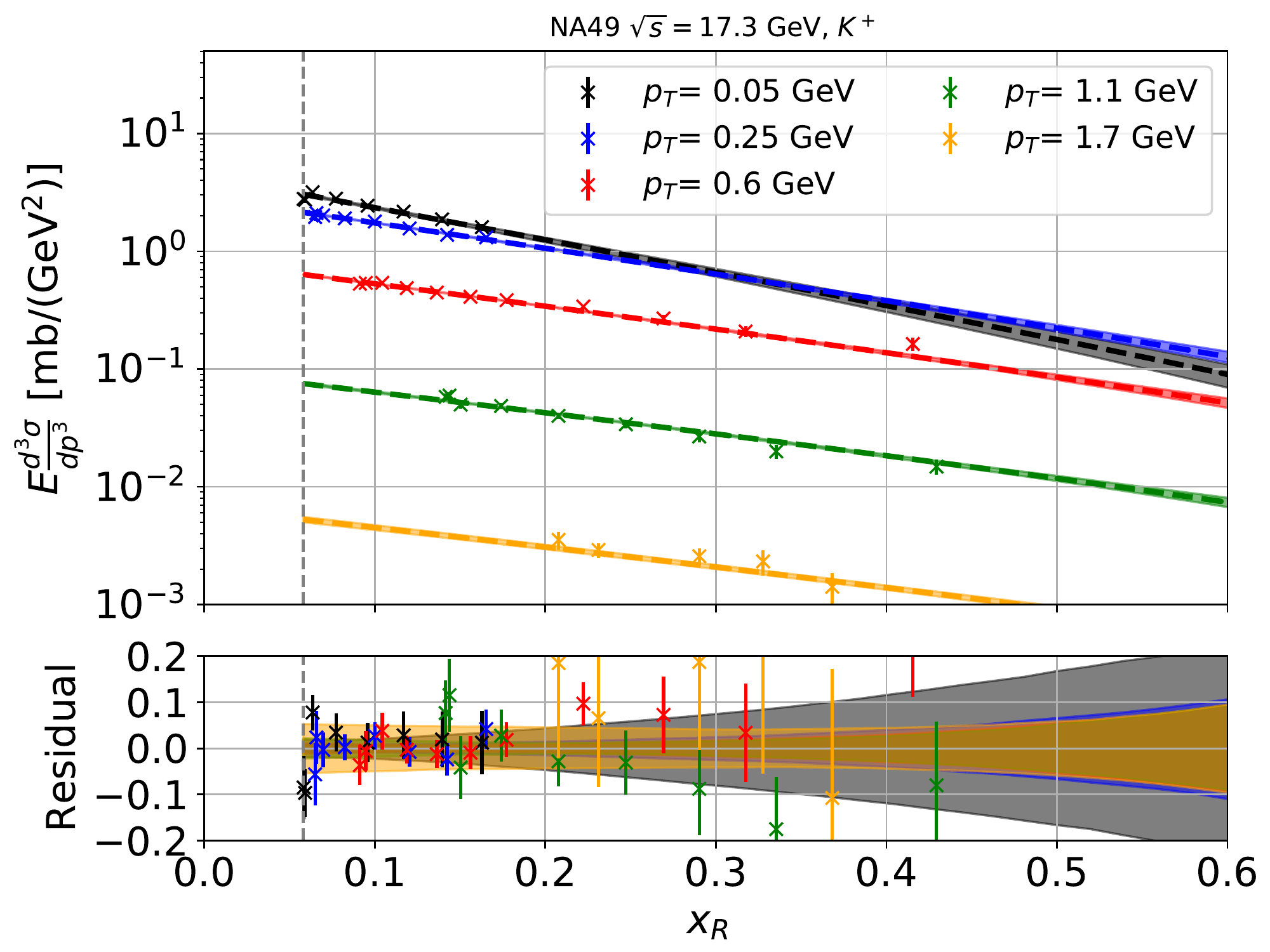}}
  \put( 0.50,-0.004){\includegraphics[width=0.25\textwidth  ]{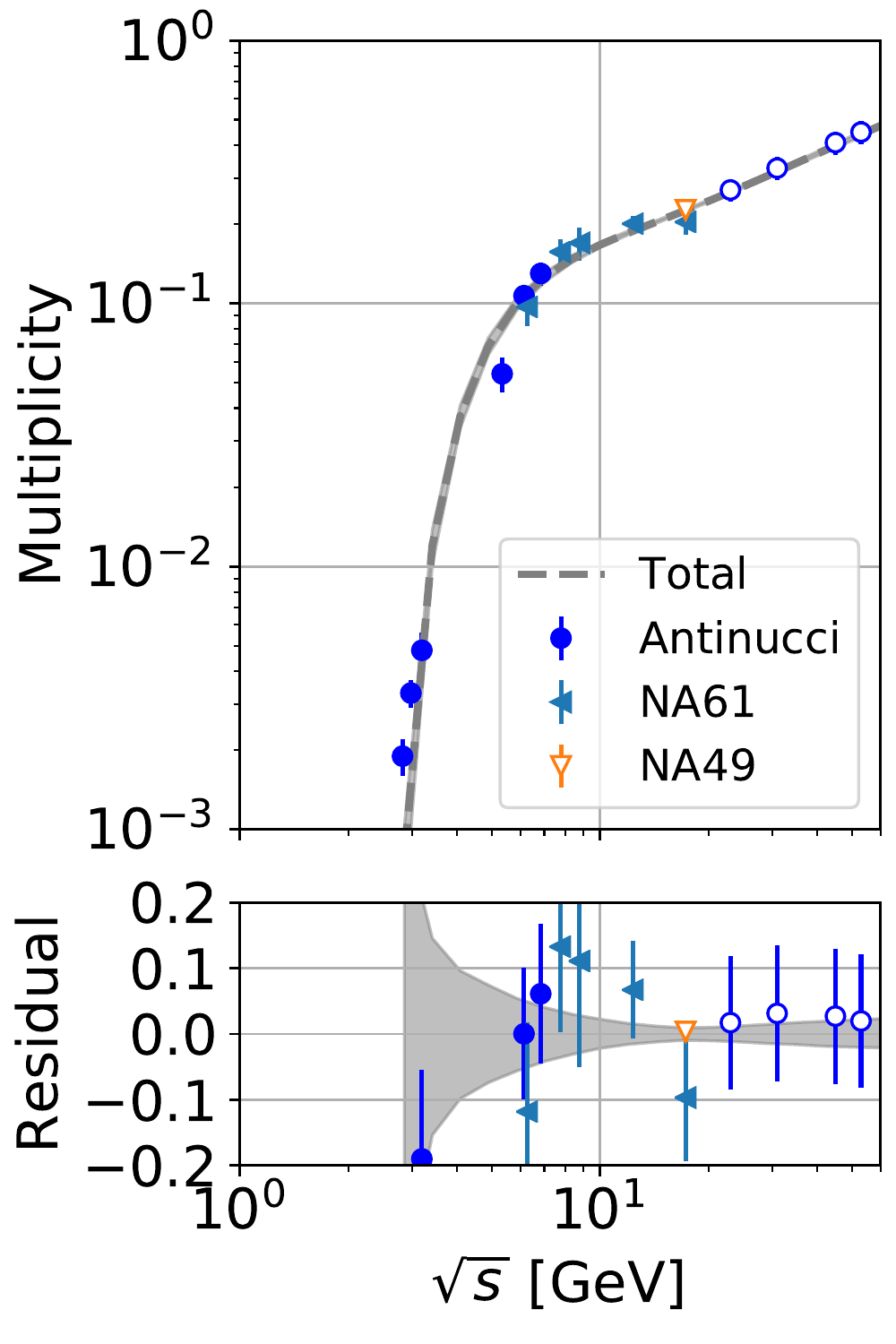}}
  \put( 0.75,-0.004){\includegraphics[width=0.25\textwidth  ]{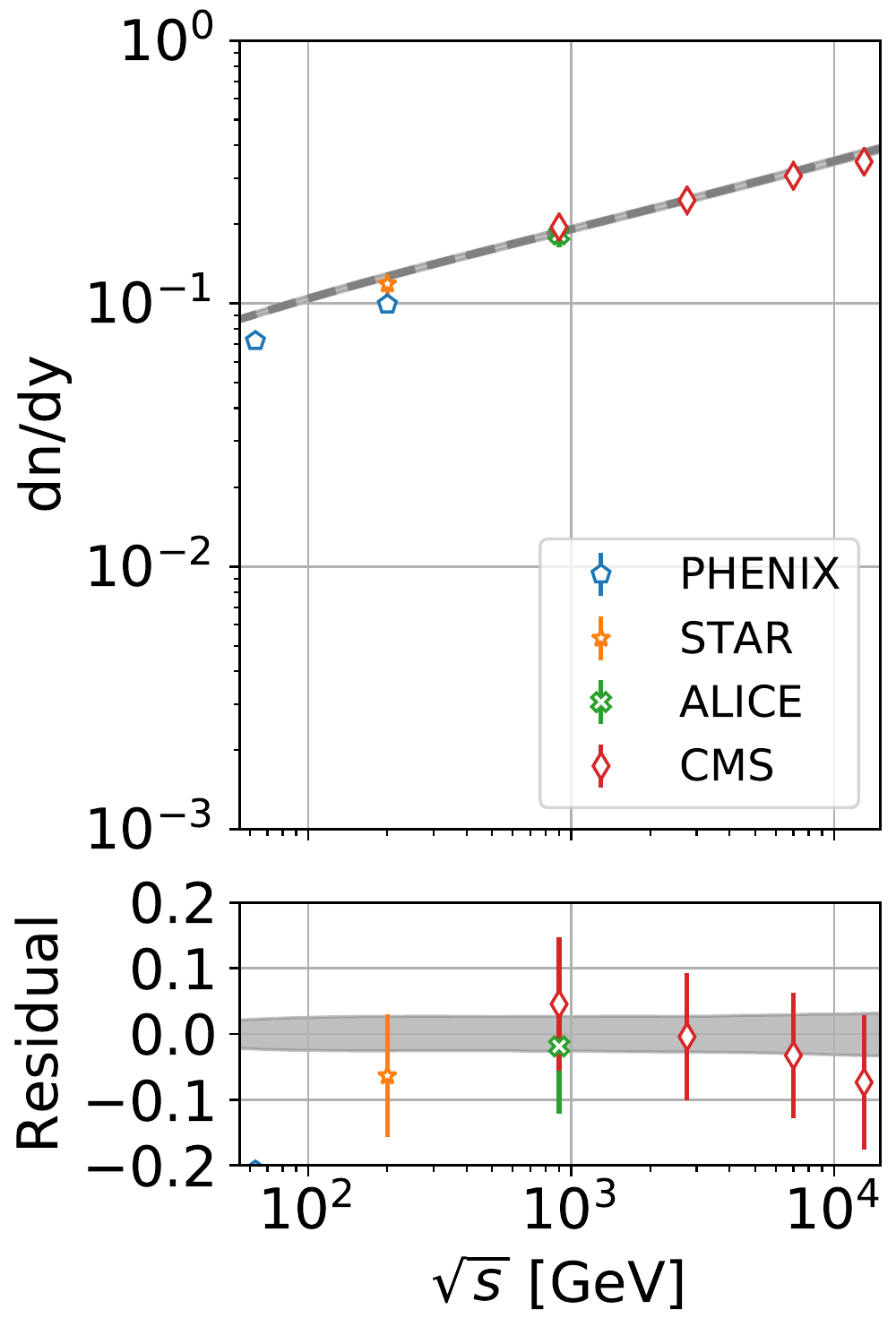}}
\end{picture}
  \caption{
    Comparison of the best-fit cross section parametrization for the inclusive $K^+$ production in $p+p$ collisions with NA49 data (left panel) and multiplicity measurements at different center of mass energies by various experiments (right panel). Right panel: filled data points are included in the fit while open data points are only plotted for comparison. The plots are similar to Figs. \ref{Fig:pi-plus-NA49} and \ref{fig:sqrt_s_sigma}(right). 
  }
  \label{fig:kaon_fit}
\end{figure*}

The projection of the cross section on $T_p$ for fixed values of $T_{e^{+}}$ shows a rapid increase above the threshold which continues for about one order of magnitude in $T_p$. Afterward the cross section keeps rising very slowly with energy. As before, the relative uncertainty is large close to the threshold.
The results of this Section already hint at the final result. The by far dominant contribution of $e^+$ production in $p+p$ collisions comes from $\pi^+$. So, even after adding the contributions from smaller channels, both the general behavior of the cross section and the relative uncertainty will follow the trends in Fig.~\ref{Fig:pi-plus-final}.

\section{Contribution from other channels}\label{sec:other_channels}

\subsection{Contribution from $K^+$}
\label{sec:eplus_kplus}

About 10\% of the positrons produced in $p+p$ collisions come from the decays of charged kaons. As sketched in Fig.~\ref{fig:positron_production_channels}, the main different decay channels considered in this work (branching fraction in brackets) are
\begin{itemize}
    \item $K^+\rightarrow\mu^+ \nu_{\mu}$ (63.6\%), 
    \item $K^+\rightarrow\pi^+ \pi^0$ (20.7\%), 
    \item $K^+\rightarrow\pi^+ \pi^+ \pi^-$ (5.6\%), 
    \item $K^+\rightarrow\pi^0 e^+ \nu_e$ (5.1\%).
\end{itemize}
To obtain the decay spectrum from kaons we proceed in this way: for the $K^+\rightarrow\mu^+ \nu_{\mu}$ channel we follow the same method reported in Sec.~\ref{sec:piplus_decay}, but adapted to $K^+$; for $K^+\rightarrow\pi^+ \pi^0$ we have to add one step to the $\pi^+$ decay, considering all the possible energies of the $\pi^+$ produced from this process; for the last two and less important three-body decay channels we adopt a simplified treatment, assuming that the three particles take $1/3$ of the $K^+$ energy.
To obtain the total positron yield we closely follow the steps from $\pi^+$ as detailed in Sec.~ \ref{sec:cross_section_eplus}, namely, we fit an analytical formula for the Lorentz invariant cross section of the inclusive $K^+$ production in $p+p$ collisions. In contrast to pions, kaons do not contain strong resonant production. So, we can use a simplified version of Eq.~\eqref{eq:main_equation} and Eq.~\eqref{eq:function_prompt} and define the  Lorenz invariant cross section by:
\begin{equation}
   \sigma_{\rm inv}= \sigma_0 (s) \,  d_1 \,F_K(s, p_T, x_R) \, A_K(s) 
   \label{eq:main_equation_K}
\end{equation}
with 
\begin{eqnarray}     
     \label{eq:function_prompt_K}
      &&\!\!F_K(s, p_T, x_R) = (1-x_R)^{d_2} 
        \exp(-d_3 \,p_T^{d_{4}} x_R) \, p_T^{d_5}  \\ \nonumber
      &&\;\;\times \exp\left[ -d_6 \sqrt{s/s_0}^{\;d_7} 
      \left(\sqrt{p_T^2 + m_K^2}-m_{K}\right)^{d_8 \sqrt{s/s_0}^{\;d_7}} \right] \;, 
\end{eqnarray}
where $m_{K}$ is the mass of the kaon, $d_i$ are the fit parameters and $\sqrt{s_0} $ is set to 17.3 GeV. The energy dependent normalization $A_K(s)$ is taken to be:
\begin{eqnarray}
  A_K(s) &=&  A_{K}^0\! \left(1-\frac{\sqrt{s_{\rm th}}}{\sqrt{s}} \right)\! \left(1+\sqrt{\frac{s}{d_9}}^{d_{10}-d_{11}} \right) \!
             \sqrt{s}^{d_{11}} \;\;\;
  \label{eq:As_K}
\end{eqnarray}
where $s_{\rm th}$ is the threshold energy for $K^+$ production and  $A_K^0$ is determined by the condition $A_K(s_0)=1$. 

We follow the two-step procedure previous used for $\pi^+$ (see Secs.~\ref{sec:piplus_fitNA49} and \ref{sec:piplus_differentCME}), fixing first the $x_R$--$p_T$ shape with NA49 data \cite{NA49_2010}, and then adjusting the $\sqrt{s}$ behavior with the multiplicity measurements from Antinucci, NA61/SHINE, ALICE and CMS \cite{osti_4593576,Aduszkiewicz:2017sei,2011_ALICE,2012_CMS,2017_CMS}. In this way, we fit the parameters $d_1$ to $d_6$ and $d_8$ with NA49 data, while the remaining parameters are fixed in a second fit keeping the first set of parameters fixed and using the multiplicity data at smaller and larger $\sqrt{s}$. For ALICE and CMS we use the $p_T$-dependent multiplicity measurements at midrapidity. A summary of the datasets is provided in Tab.~\ref{tab::pp_data}.

The $\chi^2/$d.o.f. converges to 306/253 with the individual contribution $\chi^2_{\rm NA49}/{\rm d.o.f.}=146/151$ from the first fit and the $\chi^2_n/{\rm d.o.f.}=160/102$ from the second fit. The best-fit parameters are reported in Tab.~\ref{tab::Fit_results_pp_kaon}. In Fig.~\ref{fig:kaon_fit}, we compare our best fit parametrization with the experimental measurement. In the left panel, the NA49 data of the invariant cross-section is shown as a function of $x_R$ and for a few representative values of $p_T$, while the right panel shows the comparison with various multiplicity measurements as a function of $\sqrt{s}$. All in all, our parametrization provides a very good description of the available data. The shaded bands mark the $1\sigma$ uncertainty at fixed $p_T$, which is below 5\% at smallest $x_R$ and increases to 15\% at $x_R=0.45$ for the smallest $p_T$. 
Whereas the uncertainties can be larger than the ones in the $\pi^+$ channel, their impact on the final positron yield, $d\sigma/d T_e^+$, is suppressed by the smaller production rate of kaons with respect to pions. A comparison of Fig.~\ref{fig:kaon_fit} with Fig.~\ref{Fig:pi-plus-NA49} shows that the $K^+$ production is suppressed by about one order of magnitude. 
Finally, we also compute the positron cross section from the decay of $K^-$ into $\pi^+\pi^-\pi^-$ and the subsequent decay of the $\pi^+$ into $e^+$. For this, we use the fit of the inclusive $K^-$ production in $p+p$ collisions, which is performed in analogy to the fit of $K^+$.

\begin{table}[b!]
\caption{ Results from the best fit and the 1$\sigma$ error for the parameters in Eqs.~\eqref{eq:main_equation_K}, \eqref{eq:function_prompt_K}) and \eqref{eq:As_K}. $d_1$ is in units of GeV$^{-2}$.}
\label{tab::Fit_results_pp_kaon}
\begin{tabular}{ l  c  c }
 \hline \hline
              &   $K^+$                         &   $K^-$                           \\ \hline
$d_1$         & $(1.22\pm 0.07)\cdot 10^{-1}$  & $(1.20\pm 0.07)\cdot 10^{-1}$    \\
$d_2$         & $0.63\pm 0.45$                  & $1.12\pm 0.52$                    \\
$d_3$         & $3.35\pm 0.59$                  & $6.29\pm 0.71$                    \\
$d_4$         & $-0.17\pm 0.04$                & $-0.09\pm 0.02$                    \\
$d_5$         & $(-4.6\pm 2.4) \cdot 10^{-2}$  &   $(-8.1\pm 20.45)\cdot 10^{-3}$ \\
$d_6$         & $5.08\pm 0.05$                  & $5.13\pm 0.05$                    \\
$d_7$         & $(-5.0\pm 0.1) \cdot 10^{-2}$  & $(-4.8\pm 0.1)\cdot 10^{-2}$    \\
$d_8$         & $0.92\pm 0.01$                  & $0.93\pm 0.02$                    \\
$d_9$         & $11.61\pm 0.50$                 & $10.85\pm 0.56$                   \\
$d_{10}$      & $-1.72\pm 0.08$                 & $-1.34\pm 0.07$                   \\
$d_{11}$      & $(2.02\pm 0.05) \cdot 10^{-1}$ & $(2.06\pm 0.05) \cdot 10^{-1}$   \\
 \hline \hline
\end{tabular}
\end{table}

\subsection{Contribution from $K^0_S$}
\label{sec:eplus_K0s}

$K^0_S$ hadronically decay into neutral or charged pions:  
\begin{itemize}
    \item $K^0_S \rightarrow \pi^0 \pi^0$ ($B_r  = 30.7\%$)
    \item $K^0_S \rightarrow \pi^+ \pi^-$ ($B_r = 69.2\%$),
\end{itemize}
thus contributing to the final positron and electrons cross sections with the same amount.
The first decay channel is negligible because almost all $\pi^0$ decays into two photons and only 1.2\% into $e^+e^-\gamma$. In fact, $K^0_S$ makes between $1-5\%$ of the total yield (see Fig.~\ref{Fig:final_cross_electrons} and Fig.~\ref{Fig:final_cross_positron}) so contribution from the decay into neutral pions will be below the per-mille level. We only consider the second channel. 

The NA61/SHINE experiment recently measured the spectra for the production of $K^0_S$ from $p+p$ collisions with a beam momentum of 158 GeV ($\sqrt{s}=17.3$ GeV) \cite{NA61SHINE:2021iay}. Double differential distributions were obtained in $p_T$ from 0 to 1.5 GeV and in $y$ from -1.75 to 2.25.

\begin{figure*}[t]
  \centering {
    \includegraphics[width=0.5\textwidth]{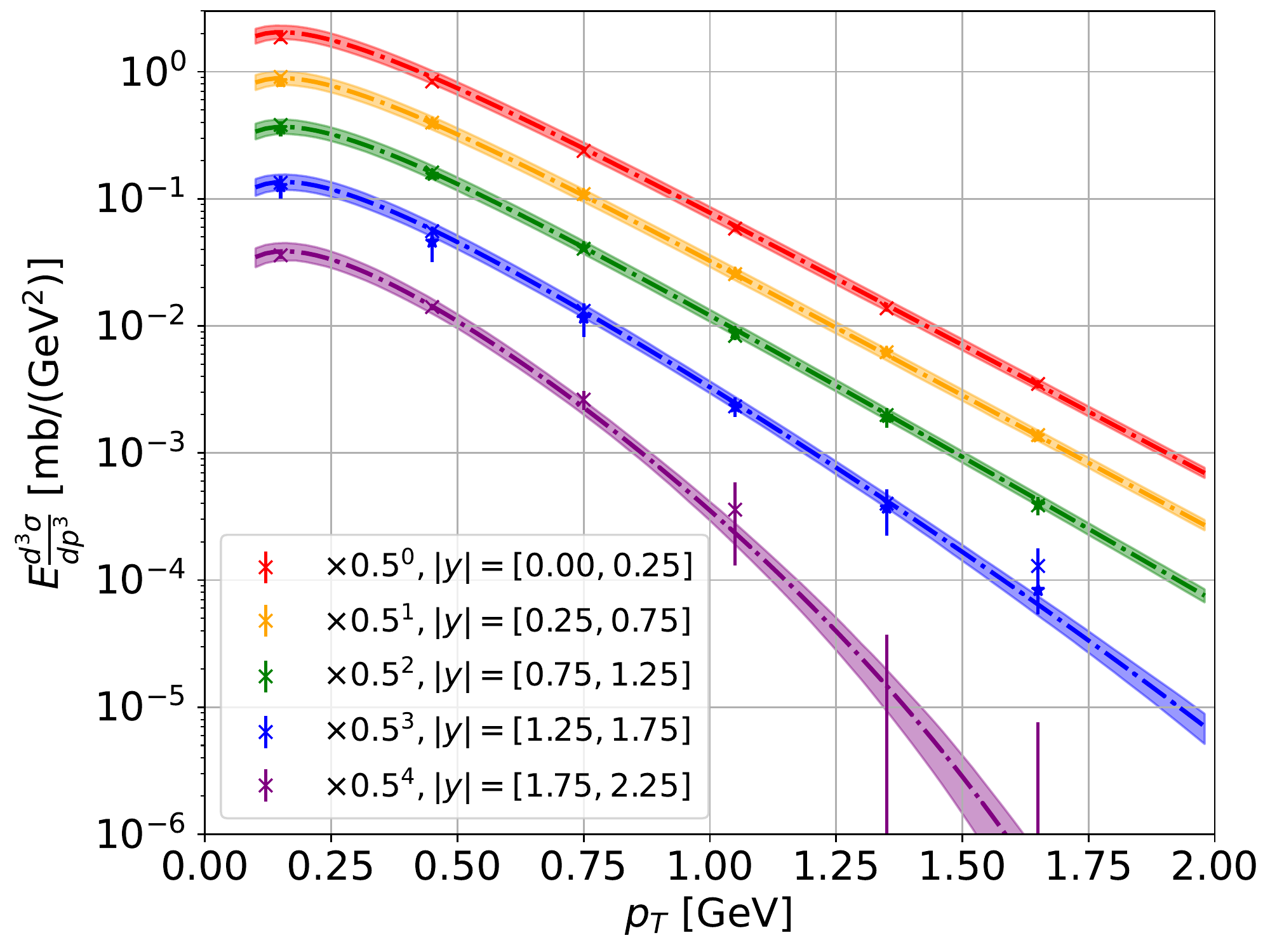}\includegraphics[width=0.5\textwidth]{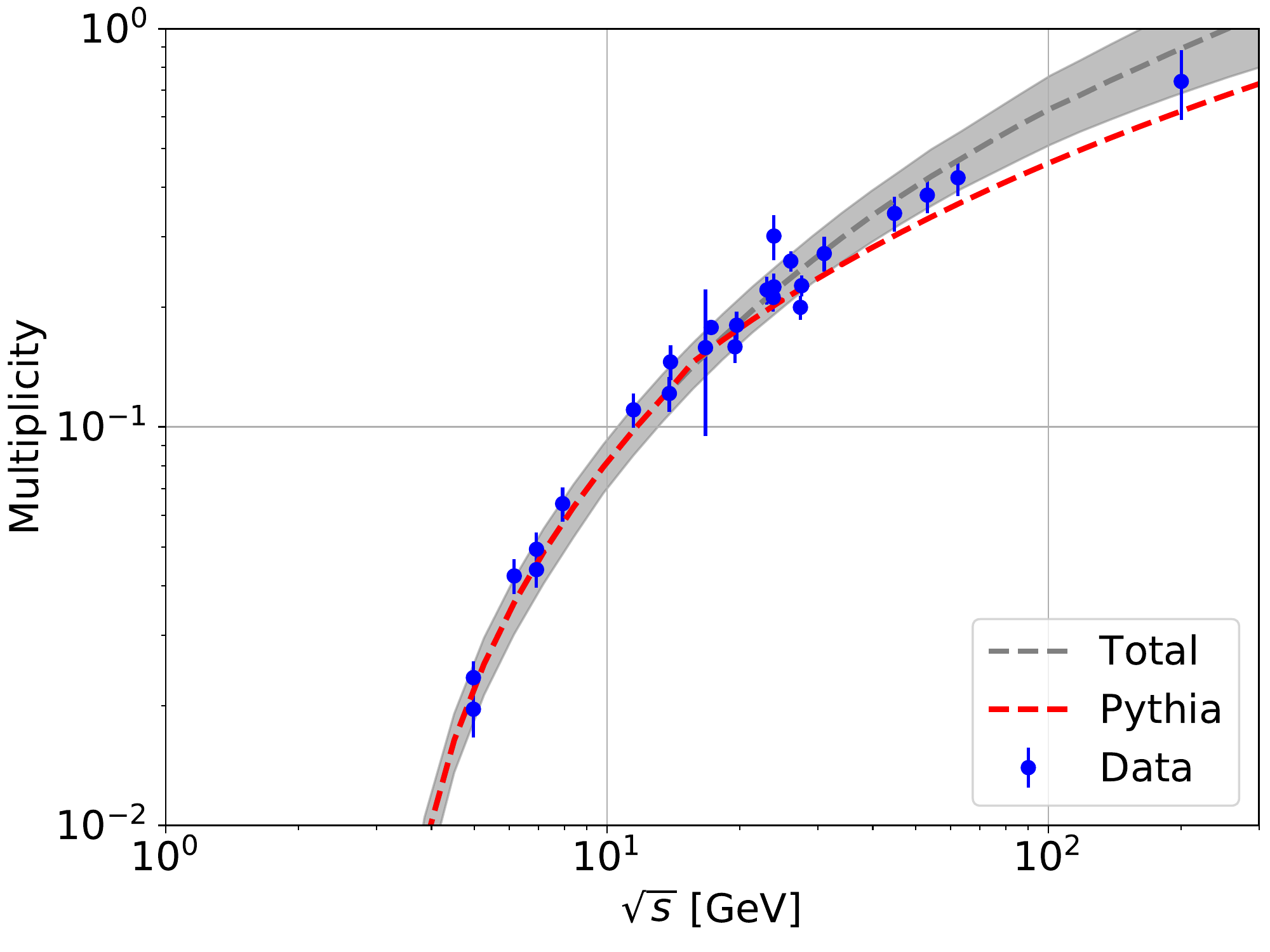}
  }
  \caption{
    Left panel: comparison of the $K^0_S$ production cross section measured by the NA61/SHINE experiment and the best-fit of Eq.~\eqref{eq:k0s}. Right panel: multiplicity data for the production of $K^0_S$ measured by different experiments (blue data points) collected and reported in Fig.~120 of Ref.~\cite{NA49_2010} and the best-fit obtained with our model (grey dashed line) and Pythia 8.3 (red dashed line). 
  }
  \label{fig:k0s}
\end{figure*}

Following a similar strategy as for $\pi^+$ and $K^+$, we first fix the $p_T$ and $x_F$ dependence of the cross section by fitting the data of NA61/SHINE at $\sqrt{s}=17.3$ GeV\footnote{
  The NA61/SHINE data are given in $d^2n/(dydp_T)$, which we transform to $\sigmaInv$ to perform the fit. 
}.
In more detail, we define the  Lorenz invariant cross section by:
\begin{equation}
   \sigma_{\rm inv}= \sigma_0 (s) \,  k_1 \,F_{K^0_S}(p_T, x_F) \, A_{K^0_S}(s),
   \label{eq:main_equation_k0s}
\end{equation}
with 
\begin{eqnarray}     
     \label{eq:k0s}
      &&\!\!F_{K^0_S}(p_T, x_F) = (1-|x_F|)^{k_2} 
        \times   \\ \nonumber
      &&\times \exp(-k_3 \,p_T^{k_{4}} |x_F|) \;\;p_T^{k_5} \exp\left[ -k_6 \;  p_T^{k_7} \right] \;, 
\end{eqnarray}
where $k_i$ are the fit parameters. The energy dependent normalization $A_{K^0_S}(s)$ is taken to be:
\begin{eqnarray}
  A_{K^0_S} (s) &=&  A_{K^0_S,0} \left(1-\sqrt{\frac{k_8}{s}}^{k_{9}-k_{10}} \right) \sqrt{s}^{k_{10}},
  \label{eq:As_K0s}
\end{eqnarray}
where the $A_{K^0_S,0}$ is determined by the condition $A_{K^0_S}(\sqrt{s_0}=17.3\rm{\,GeV})=1$ and the best-fit parameters $k_8$, $k_9$, $k_{10}$ are determined by a second fit to the multiplicities at different $\sqrt{s}$. 
We extract the multiplicity data from Ref.~\cite{NA49_2010} (reported in their Fig.~120) and fit with the function in Eq.~\eqref{eq:main_equation_k0s}.
We obtain a good result for both fits. The $\chi^2$/d.o.f. converge to 20/41 and 42/24 for the fit to NA61/SHINE data and the multiplicity data, respectively. All the values of the best-fit parameters are reported in Tab.~\ref{tab::Fit_results_hyperons}. 
Fig.~\ref{fig:k0s} shows that our parametrization provides a good description of the data. In the left panel, we compare the NA61/SHINE data with the result of Eq.~\eqref{eq:main_equation_k0s}, while in the right panel we show the multiplicity as a function of $\sqrt{s}$ together with the best-fit of our parametrization in Eq.~\eqref{eq:main_equation_k0s}.

For comparison, we checked the predictions of the multiplicity using the Pythia event generators. We employ the Pythia version 8.3 \cite{Sjostrand:2014zea}. Pythia produces predictions for the multiplicity that are close to the data with a shape only slightly different from the best fit obtained with Eq.~\eqref{eq:main_equation_k0s}.

\subsection{Contribution from $K^0_L$}
\label{sec:eplus_K0l}

The decay time of the $K^0_L$ meson is  $5.1\times 10^{-8}$~s which is a factor of about 600 larger than the one of $K^0_S$, making it very difficult to detect $K^0_L$ particles at accelerator experiments.
Moreover, the $K^0_L$ has different decay channels and branching ratios than $K^0_S$:
\begin{itemize}
    \item $K^0_L \rightarrow \pi^{\pm} e^{\mp} \nu_e$ ($B_r=40.6\%$),
    \item $K^0_L \rightarrow \pi^{\pm} \mu^{\mp} \nu_{\mu}$ ($B_r=27.0\%$),
    \item $K^0_L \rightarrow \pi^{0}\pi^{0}\pi^{0}$ ($B_r=19.5\%$), 
    \item $K^0_L \rightarrow \pi^{+} \pi^{-} \pi^{0}$ ($B_r=12.5\%$).
\end{itemize}

The lack of experimental data makes it impossible to determine an independent parametrization of the production cross section. Therefore, we employ the Pythia event generator to compare the $p_T$ and $x_F$ dependence of the final $e^{+}$ spectra from $K^0_S$ and $K^0_L$. 
 
We find that the $p_T$ and $x_F$ shapes for the production of $e^{+}$ is very similar for the $K^0_L$ and $K^0_S$ particles. The difference is simply a normalization factor (for more details see App.~\ref{app:details_K0_Lambda}). The $K^0_L$ meson produces about a factor of 1.16 more $e^{+}$ than $K^0_S$ which can be explained by different decay modes of the two kaons. In particular, it is mainly due to the branching ratio of $K^0_S$ into $2\pi^0$ ($B_r=30.7\%$) which is larger than for $K^0_L$ ($B_r=19.5\%$) suppressing positron production from $K^0_S$ ($(1-0.195)/(1-0.307)=1.16$).

\begin{figure*}[t]
  \centering {
    \includegraphics[width=0.49\textwidth]{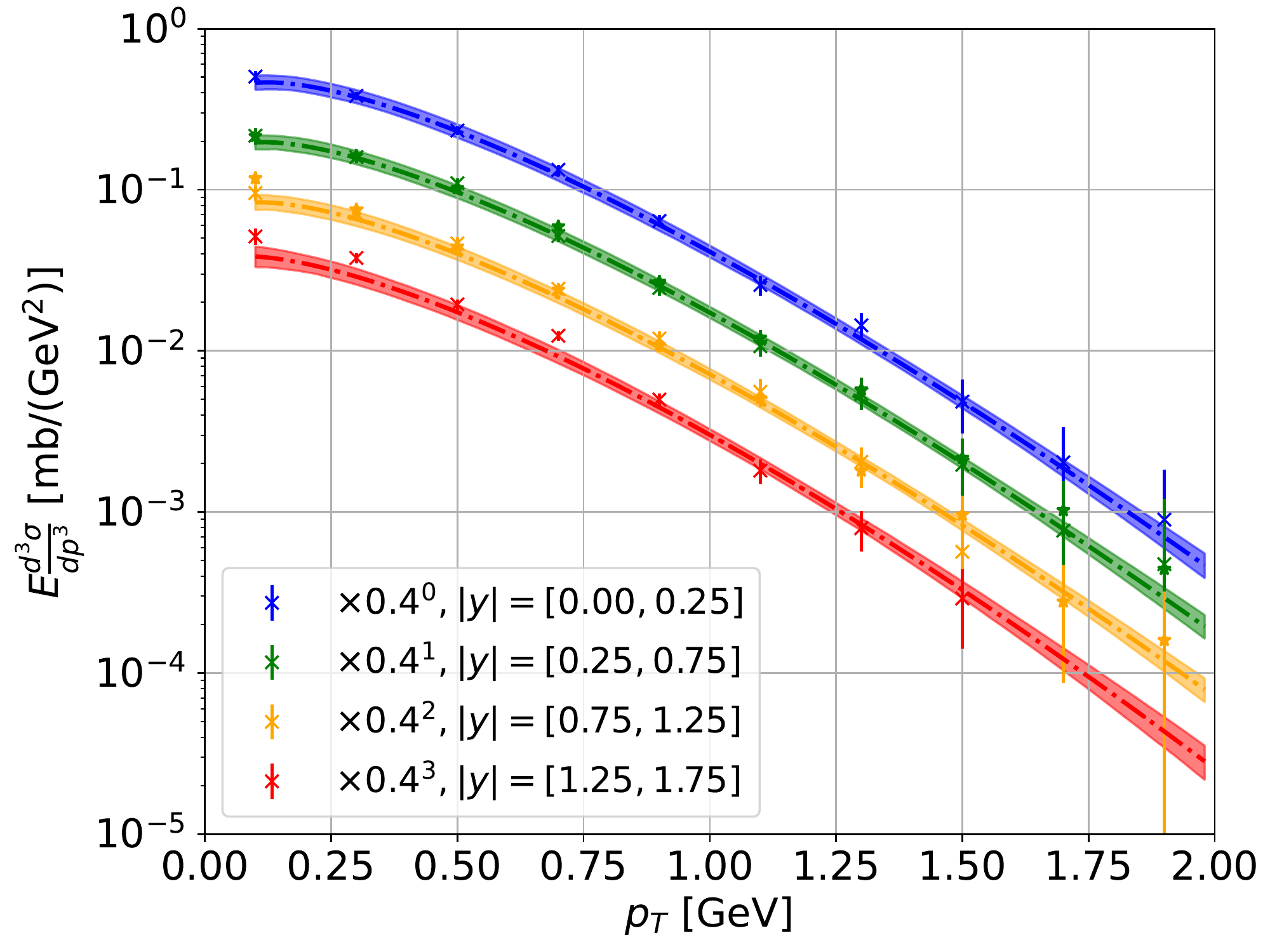}
  \includegraphics[width=0.49\textwidth]{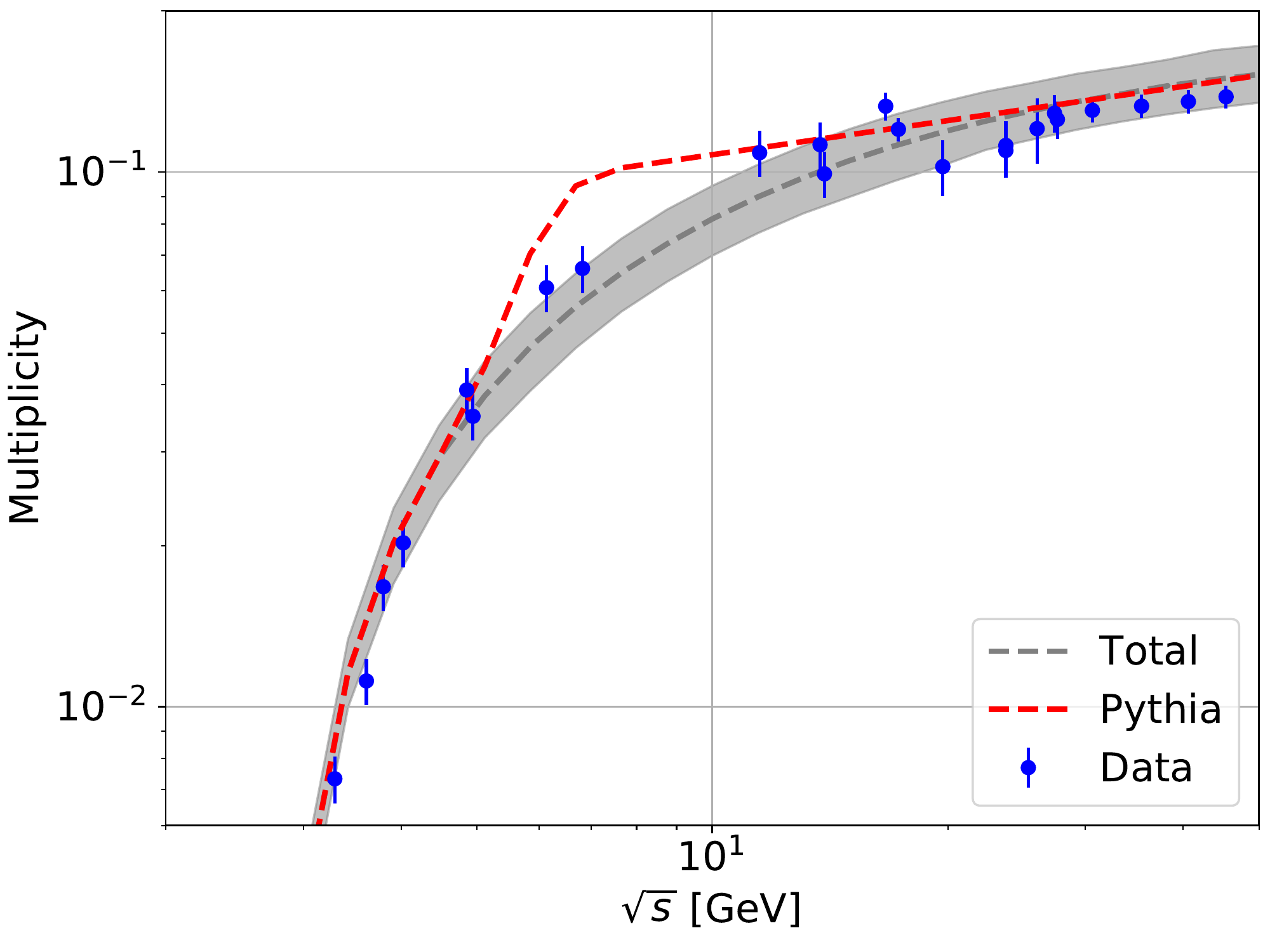}
  }
  \caption{
    Left panel: comparison of the $\Lambda$ production cross section measured by the NA61/SHINE experiment and the best-fit of Eq.~\eqref{eq:main_equation_lambda}. Right panel: multiplicity data for the production of $\Lambda$ measured by different experiments (blue data points) collected and reported in Fig.~16 of \cite{NA61SHINE:2015haq} and the best-fit obtained with our model (grey dashed line) and Pythia 8.3 (red dashed line). 
  }
  \label{fig:lambda}
\end{figure*}

So, in the following we assume that the production cross section of positrons from $K^0_L$ can be obtained from $K^0_S$ by rescaling with a factor $1.16$. Because of charge symmetry, we apply the same results for $e^-$ production.  In particular, we do not add any uncertainty related to the factor $1.16$ used to rescale the results of $K^0_S$ since this comes from the different $B_r$ of $K^0_{L}$ and $K^0_{S}$ decay into pions that are very well measured. We apply the same uncertainty of the $K^0_S$ to the $K^0_L$ channel.

\subsection{Contribution from $\Lambda$}
\label{sec:eminus_lambda}
The $\Lambda$ hyperon decays mainly in:

\begin{itemize}
    \item $\Lambda \rightarrow p \pi^-$ ($B_r = 63.9\%$) 
    \item $\Lambda \rightarrow n\pi^0$ ($B_r= 35.8\%$).
\end{itemize}

The former contributes only to the $e^-$ through the decay of the $\pi^-$ while the latter would contribute to both the $e^{\pm}$ with a negligible contribution through the $\pi^0$ decay (see Sec.~\ref{sec:eplus_pi0}). Instead, the part related to the neutron decay would contribute only at energies below 100 MeV \cite{Kamae:2006bf}.
Given the decay channels reported above, the $\Lambda$ particle contributes mainly to the $e^-$ secondary part. However, the $\Lambda$ production cross section helps to gauge some of the other subdominant channels for $e^+$ production, namely, we will obtain their contribution by a rescaling, as explained in Sec. \ref{sec:eplus_otherchannels}.

The NA61/SHINE experiment recently measured the spectra for the production of $\Lambda$ from $p+p$ collisions with a beam momentum of 158 GeV  ($\sqrt{s}=17.3$ GeV) and for $p_T=[0.,1.9]$ GeV/c and $y=[-1.75,1.25]$ \cite{NA61SHINE:2015haq}.
Following a similar strategy as for $K^0_{S}$, we first fix the $p_T$ and $x_F$ dependence of the cross section by fitting the data of NA61/SHINE at $\sqrt{s_0}=17.3$ GeV 
\footnote{
  The NA61/SHINE data are given in $d^2n/(dydp_T)$, which we transform to $\sigmaInv$ to perform the fit. 
}.
In more detail, we define the  Lorenz invariant cross section by:
\begin{equation}
   \sigma_{\rm inv}= \sigma_0 (s) \,  l_1 \,F_{\Lambda}(p_T, x_F) \, A_{\Lambda}(s) 
   \label{eq:main_equation_lambda}
\end{equation}
with 
\begin{eqnarray}     
     \label{eq:lambda}
      &&\!\!F_{\Lambda}(p_T, x_F) = (1-|x_F|)^{l_2} 
        \times   \\ \nonumber
      &&\exp(-l_3 \,p_T^{l_{4}} |x_F|) \;\;p_T^{l_5} \exp\left[ -l_6 \;  p_T^{l_7} \right] \;, 
\end{eqnarray}
where $l_i$ are the fit parameters. The energy dependent normalization $A_{\Lambda}(s)$ is taken to be:
\begin{eqnarray}
  A_{\Lambda} (s) &=&  A_{\Lambda,0} \left(1-\sqrt{\frac{l_8}{s}}^{l_{9}-l_{10}} \right) \sqrt{s}^{l_{10}} 
  \label{eq:As_lambda}
\end{eqnarray}
where the $A_{\Lambda,0}$ is determined by the condition $A_{\Lambda}(\sqrt{s_0}=17.3 {\rm GeV})=1$ and the best-fit parameters $l_8$, $l_9$ and $l_{10}$ are determined by a second fit to the multiplicities at different $\sqrt{s}$. For this, we extract the collection of data on the multiplicity reported in Fig.~16 of \cite{NA61SHINE:2015haq} and fit it by the multiplicity obtained from our parametrization in Eq.~\eqref{eq:main_equation_lambda}.
We obtain a good result for both fits. The $\chi^2$/d.o.f. of the best fits converges to 27/49 and 53/23 for the fit to NA61/SHINE cross section and the multiplicities, respectively. All values of the best-fit parameters are reported in Tab.~\ref{tab::Fit_results_hyperons}. 
Fig.~\ref{fig:lambda} shows that our parametrization provides a good description of the data. In the top panel, we compare the NA61/SHINE data with the result of Eq.~\eqref{eq:main_equation_lambda}, while in the bottom panel we show the multiplicity as a function of $\sqrt{s}$ together with the best-fit of our parametrization in Eq.~\eqref{eq:main_equation_lambda}.

For comparison, we checked the predictions of the multiplicity using the Pythia event generators. 
Pythia produces predictions for the multiplicity which are close to the data with a shape slightly different with respect to the best fit obtained with Eq.~\eqref{eq:main_equation_lambda}.

\begin{table*}[t!]
  \caption{
            Summary of the fit quality in the $p+p$ channel. The first row corresponds to the fit of NA49 data (as detailed in Sec.~\ref{sec:piplus_fitNA49} for $\pi^+$). Then, the second row states the $\chi^2$ of the fit to other center of mass energies (see Sec.~\ref{sec:piplus_differentCME}) with the individual contributions from rows three to six. In the last row we give the total $\chi^2$ and the d.o.f..
          }
  \label{tab::chi2_pp}
  \begin{tabular}{ l c c c c}
     \hline \hline
                                                       &   $\pi^+$        &   $\pi^-$   &   $K^+$      &  $K^-$    \\ \hline
     $\chi^2_\mathrm{NA49}$/d.o.f.                        &   338/263        &   287/290.  &   146/151    &  197/151  \\
     $\chi^2_n$/d.o.f.                                    &   189/129        &   169/96    &   160/102    &  135/100  \\
     $\chi^2_\mathrm{ALICE}$                           &    77 (33)       &   -         &    42 (27)   &  36 (27)  \\
     $\chi^2_\mathrm{CMS}$                             &   100 (88)       &   154 (88)  &    77 (68)   &  54 (68)  \\
     $\chi^2_\mathrm{NA61,Antinucci}$                       &    10 (12)       &    15 (12)  &    39 (11)   &   44 (9)  \\
     $\chi^2_\mathrm{tot}$/d.o.f.                         &    527/392       &   456/386   &   306/253    &  332/251  \\
  \hline \hline
  \end{tabular}
\end{table*}

\subsection{Subdominant channels}
\label{sec:eplus_otherchannels}

Other channels contribute with a subdominant amount to the $e^{+}$ and $e^{-}$ yield.
The $\bar{\Lambda}$, the charged $\Sigma$ and $\Xi$ hyperons have typical decay times of the order of $10^{-10}$ s and their pion contributions are usually removed with the feed-down correction. We thus have to add it to our calculations. The multiplicities of $\Omega$ baryons in $p+p$ collisions are a factor of about 3-4 orders of magnitude smaller than the one of $\Lambda$ particles, so we neglect them.

Unfortunately, no data are available at the energies of interest for the secondary source term. We decide thus to estimate the contribution of the $\bar{\Lambda}$, $\Sigma$ and $\Xi$ baryons using the Pythia code \cite{Sjostrand:2014zea}.
In particular, we run simulations of $p+p$ collisions for $\sqrt{s}$ ranging from a few GeV to a few TeV, i.e.~$E_p=[20,10^6]$ GeV.
We calculate the multiplicities of these particles, $n_{i}$, where $i$ runs over $\Sigma^+$, $\Sigma^-$, $\Xi^0$, $\Xi^-$ and their antiparticles as well as $\bar{\Lambda}$. Then, we calculate the ratio $n_i/n_\Lambda$, both derived with Pythia for consistency. We decide to proceed in this way because for $\Lambda$ we have a model for the invariant cross section (see Sec.~\ref{sec:eminus_lambda}) and its mass is similar or equal to the $\bar{\Lambda}$, $\Sigma$ and $\Xi$, so we expect the dependence of the cross section with the kinematic parameters to be similar. Then, we use the ratio $n_{i}/n_{\Lambda}$ to add these subdominant channels to the total yield of $e^{\pm}$ by rescaling the $\Lambda$ cross sections into $e^{\pm}$ as follows:
\begin{equation}
  \frac{d\sigma}{dT_e}(T_p,T_e) = \frac{d\sigma}{dT_e}(T_p,T_e)_{\Lambda} \times \sum_i \mathcal{F}^i(T_p),
 \label{eq:sub}
\end{equation}
where $\mathcal{F}^i(T_p)$ represents the correction factor that we use to rescale the cross section for the production of $e^{-}$ or $e^{+}$ for the $i$-th hyperon from the one of $\Lambda$ particles. 
For example, charged $\Sigma$ particles can decay into protons or neutrons and pions, so $\mathcal{F}^{\Sigma}(T_p)$ can be written as:
\begin{equation}
  \mathcal{F}^{\Sigma}(T_p) =  \frac{n_{\Sigma}(T_p)\cdot {B_r}^{\pi}_{\Sigma}} {n_{\Lambda}(T_p)\cdot {B_r}^{\pi}_{\Lambda}}
 \label{eq:subfactor}
\end{equation}
where ${B_r}^{\pi}_\Sigma$ is the branching ratio for the decay of the hyperons into charged pions.
In contrast, the $\Xi$ particles decay into pions and $\Lambda$ particles so $\mathcal{F}^\Xi(T_p)$ takes a different form that we will report below.
Finally, since $\bar{\Lambda}$ is the antiparticle of the $\Lambda$ the correction factor for this particle is simply the ratio $n_{\bar{\Lambda}}/n_{\Lambda}$ between the multiplicity for the production of $\bar{\Lambda}$ with respect to $\Lambda$.

We list below the particles we consider in this section reporting the branching ratios into pions.
\begin{itemize}

    \item The $\bar{\Lambda}$ hyperon decays mainly into $\bar{p}\pi^+$ with $B_r= 63.9\%$ and $\bar{n}\pi^0$ with $B_r=35.8\%$. The former contributes only to the $e^+$ through the decay of the $\pi^+$, while the latter would contribute negligibly through the $\pi^0$ decay. Instead, the part related to the antineutron decay would contribute only at energies below 100 MeV \cite{Kamae:2006bf}. However, in this case the branching ratio exactly cancels with the one of $\Lambda$ such that we get $\mathcal{F}^{\Lambda}(T_p)  = n_{\bar{\Lambda}}/n_{\Lambda}$.

    \item The $\Sigma^{+}$ baryon decays with $B_r=51.6\%$ into $p\pi^0$ and $48.3\%$ into $n\pi^+$. The former contributes less than the per-mille level to the total source term through the decay of the $\pi^0$ (see Sec.~\ref{sec:eplus_pi0}). Instead, the latter is relevant for the $e^+$ production. For this particle thus ${B_r}^{\pi^+}_{\Sigma^{+}}=0.48$ and the correction factor $\mathcal{F}$ is given by Eq.~\eqref{eq:subfactor}. The antiparticle of $\Sigma^{+}$ is $\bar{\Sigma}^-$ and contributes to the electron yield.
    
    \item $\Sigma^{-}$ decays with almost $B_r=100\%$ into $\bar{n}\pi^{-}$ and contributes to the electron yield. For this source thus we have $Br^{\pi^-}_{\Sigma^{-}}=1$. Its antiparticle is $\bar{\Sigma}^{+}$ and has to be included for the  positron production. 
    
    \item The $\Xi^0$ decays with almost 100$\%$ into $\pi^0 \Lambda$ thus producing $e^-$ through the $\Lambda$ decay. We use for the correction factor in Eq.~\eqref{eq:subfactor} $\mathcal{F}_{\Xi^0} = (Br^{\pi^-}_{\Lambda}\cdot n_{\Xi^0})/(Br^{\pi^-}_{\Lambda} \cdot n_{\Lambda}) = n_{\Xi^0}/n_{\Lambda}$. The antiparticle of $\Xi^0$ is $\bar{\Xi}^0$. Since $\bar{\Xi}^0$ decays into $\pi^0 \bar{\Lambda}$ we rescale by $\mathcal{F}_{\bar{\Xi}^0} = (Br^{\pi^+}_{\bar{\Lambda}}\cdot n_{\bar{\Xi}^0})/(Br^{\pi^-}_{\Lambda} \cdot n_{\Lambda}) = n_{\bar{\Xi}^0}/n_{\Lambda}$.
    
    \item The $\Xi^-$ decays with almost 100$\%$ into $\pi^-\Lambda$. We use for this particle $\mathcal{F}_{\Xi^-} = ((1+Br^{\pi^-}_{\Lambda})\cdot n_{\Xi^-})/(Br^{\pi^-}_{\Lambda} \cdot n_{\Lambda})$. The antiparticle is $\bar{\Xi}^+$ for which we take $\mathcal{F}_{\bar{\Xi}^+} = ((1+Br^{\pi^+}_{\bar{\Lambda}})\cdot n_{\bar{\Xi}^+})/(Br^{\pi^-}_{\Lambda} \cdot n_{\Lambda})$.
\end{itemize}

\begin{figure}[t]
    \includegraphics[width=0.49\textwidth]{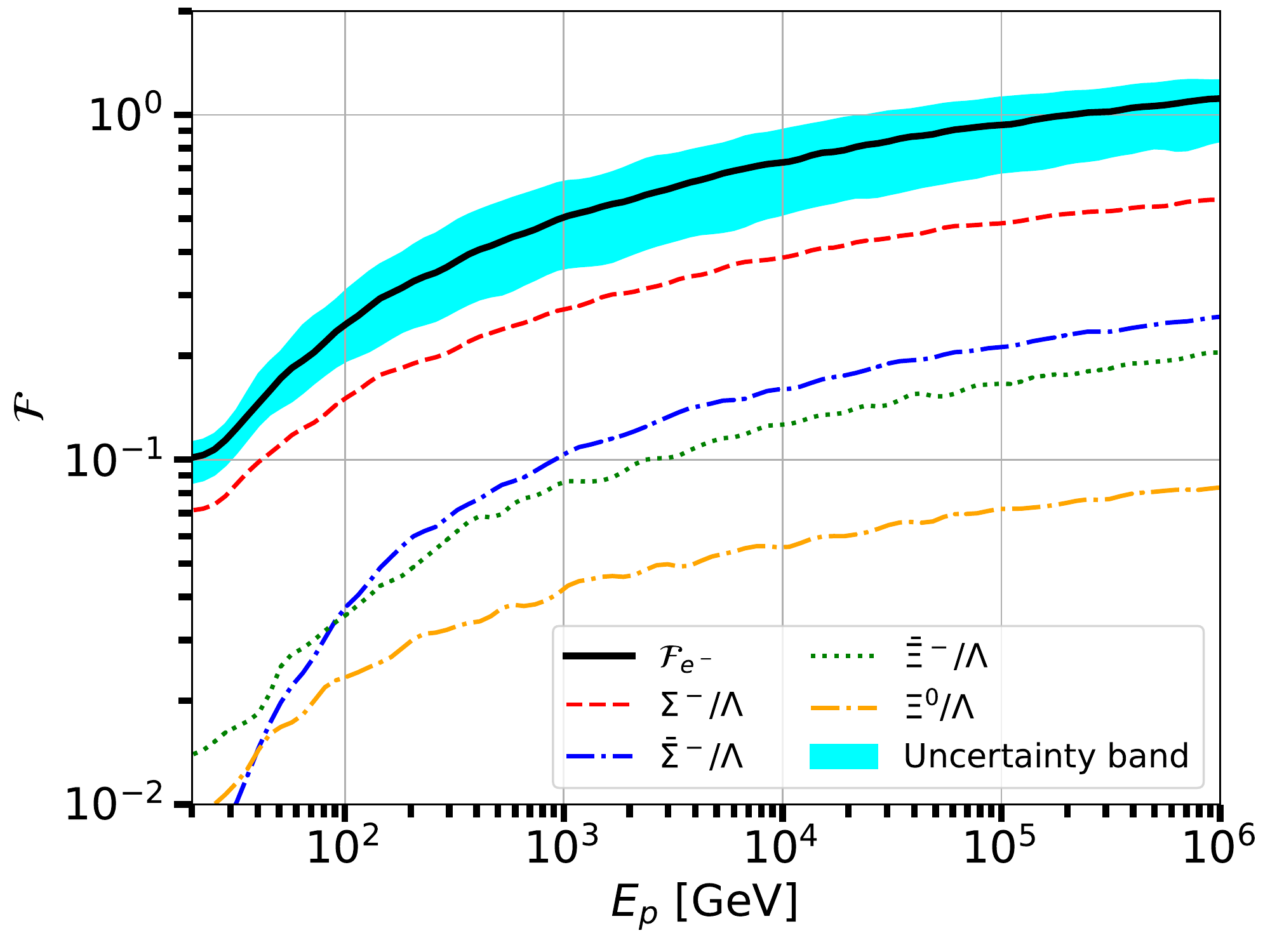}
    \includegraphics[width=0.49\textwidth]{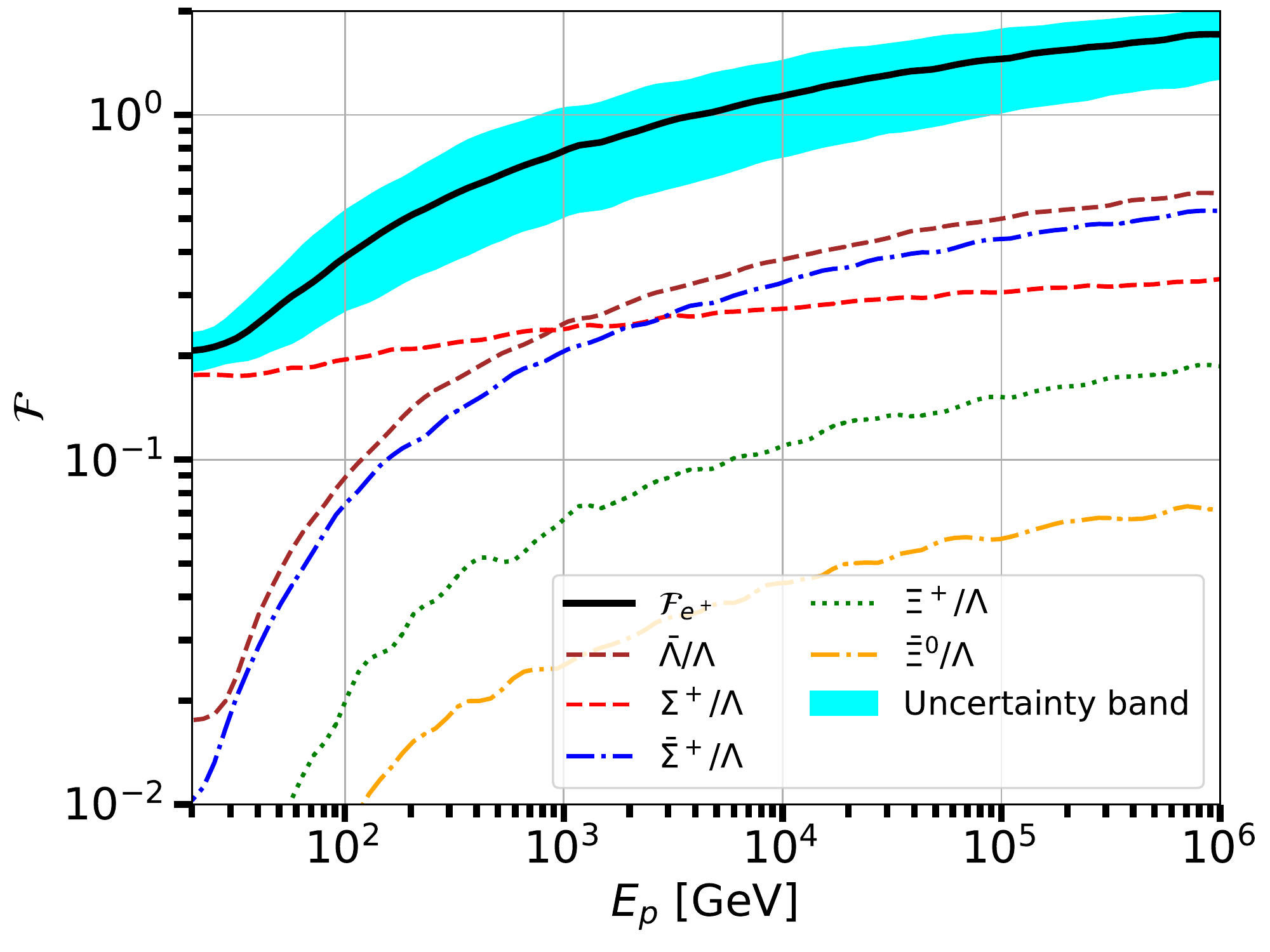}
    \caption{Correction factor $\mathcal{F}$ for the contribution of $\bar{\Lambda}$, $\Sigma$ and $\Xi$ from $p+p$ collisions at different proton energies in the LAB frame $E_p$. We show the result obtained for each individual contribution and total one obtained with Eq.~\eqref{eq:subfactor}. We also display the uncertainty band found by running Pythia using different setup parameters and tunings. The top (bottom) panel is for the correction factor applied to secondary $e^-$ ($e^+$).} 
    \label{Fig:subdominantchannels}
\end{figure}

In Fig.~\ref{Fig:subdominantchannels}, we show the correction factor $\mathcal{F}$ for the subdominant channels that contribute to $e^+$ and $e^-$.
In particular, we see that the $\Sigma^+$  and $\Sigma^-$ are the hyperons that contribute the most to the $e^+$ and $e^-$ production, respectively, with about 10--30$\%$ of the $\Lambda$ particles.
Instead, the $\Xi$ baryon contribution is well below the $10\%$ of the $\Lambda$.
At the $\sqrt{s}$ of NA49 the results we find for $\Sigma^+$, $\Sigma^-$ and $\bar{\Lambda}$ are consistent with the multiplicities calculated from the $dn/dx_F$ shown in Fig.~22 of \cite{NA49_2007}\footnote{Fig.~22 of \cite{NA49_2007} reports the result of a Monte Carlo simulation for the $dn/dx_F$ of hyperons that the NA49 collaboration used to correct the data for the feed-down.}.
At low energy, $\mathcal{F}$ is between $10\%$ and $50\%$ for $e^+$ and $e^-$,  while at high energy it reaches 1 for $e^-$ and 2 for $e^+$.
We also show in the same figure the variation to $\mathcal{F}$ obtained from different Pythia setups (uncertainty band). We explain the details of this in App.~\ref{sec:pythia}. The correction factor can change by $40\%$ depending on the setup of the Monte Carlo. Therefore, we decide to associate a systematic uncertainty of $40\%$ to these channels at all energies.
\subsubsection{Contribution from $\pi^0$}
\label{sec:eplus_pi0}

\begin{figure}[t]
    \includegraphics[width=0.49\textwidth]{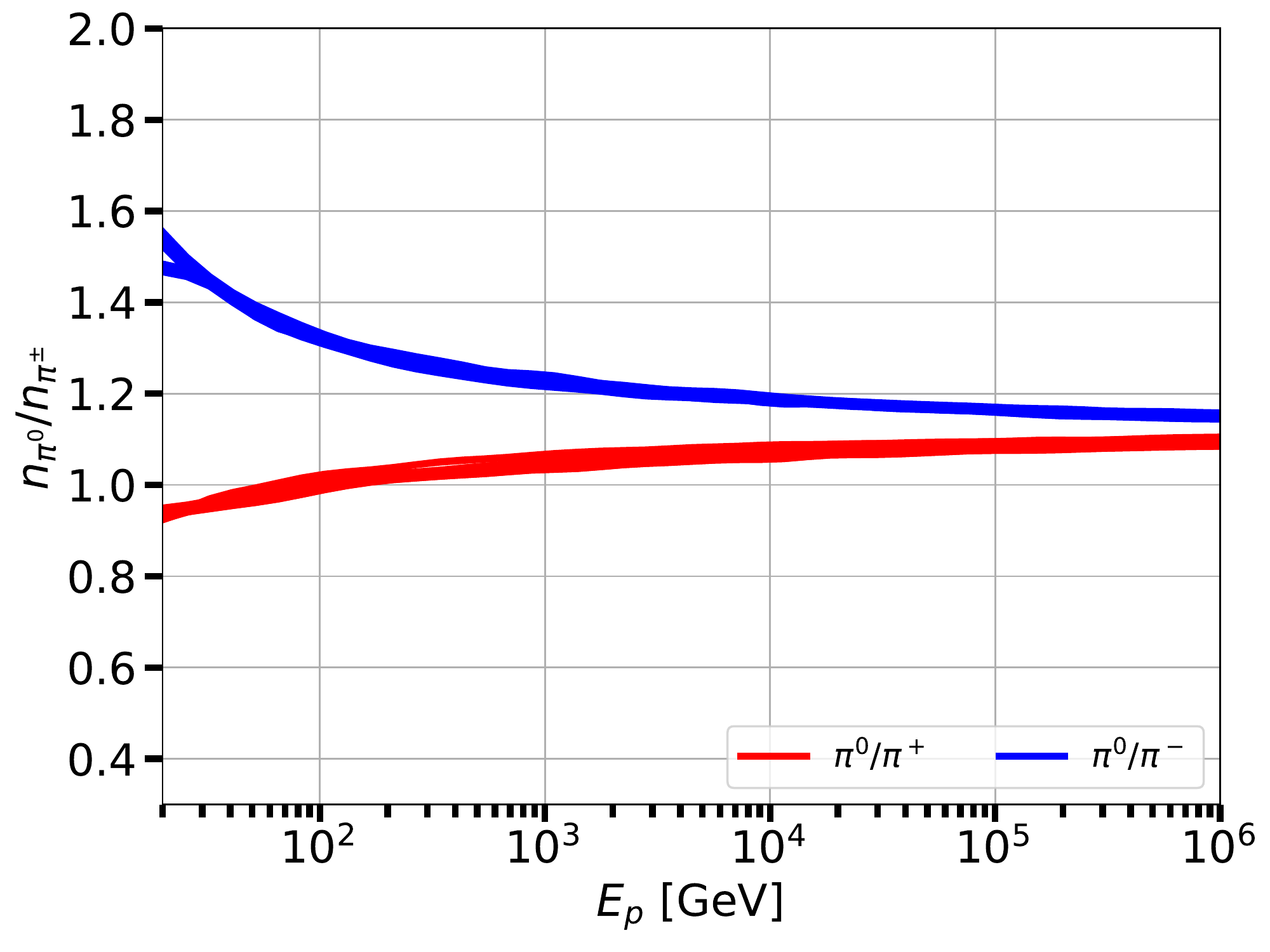}
    \caption{
    Ratio of the multiplicity for $\pi^0$ and $\pi^+$ (red band) and $\pi^-$ (blue band) from $p+p$ collisions. The bands represent the envelop of the results by changing the setup of Pythia as explained in details in App.~\ref{sec:pythia}. 
    } 
    \label{Fig:pi0}
\end{figure}

Neutral pions  are expected to be produced in $p+p$ collisions with a similar rate as charged pions.
However, $\pi^0$s decay with a branching ratio of $98.82\%$ into two photons and only with $1.17\%$ into $e^+ e^- \gamma$.
Therefore, the contribution of the $\pi^0$ to the $e^{\pm}$ production is expected to be at the $1\%$ level. 
Since no data are available for the $e^{\pm}$ from $\pi^0$ at the energy of interest, we use the Pythia event generator to derive the $p_T$ and $x_F$ dependence of $e^{\pm}$ produced from $\pi^{\pm}$ and $\pi^0$.
We find that the $dn/dx_F$ and $dn/dp_T$ are very similar in shape for the production of $e^{\pm}$ from $\pi^0$ and from $\pi^{\pm}$. The difference is just a normalization factor that depends on the different multiplicity of $\pi^0$ ($n_{\pi^0}$) and $\pi^{\pm}$ ($n_{\pi^{\pm}}$) for the production of pions from $p+p$ collisions.
We show in Fig.~\ref{Fig:pi0} the result obtained for $n_{\pi^0}/n_{\pi^+}$ and $n_{\pi^0}/n_{\pi^-}$ as a function of the incoming proton energy and for different Pythia setups.
As expected, $n_{\pi^0}/n_{\pi^+}$ is smaller than $n_{\pi^0}/n_{\pi^-}$ and they both tend to 1 for very high energies.
The variations in the ratio of the multiplicities with respect to the average are between $1.0\%$ and $1.5\%$. 
We decide to add the contribution from $\pi^0$ to the $e^{\pm}$ yield by multiplying the charged pions cross sections by a factor $(1+n_{\pi^0}\cdot B^{\pi^0}_r/n_{\pi^{\pm}})$, where $B^{\pi^0}_r=0.017$.
We associated to this contribution an uncertainty of $1\%$.

\section{Contribution from nuclei collisions}
\label{sec:nuclei}
In the Galaxy, nuclei interactions ($p+A$, $A+p$, and $A+A$) give a significant contribution to the production of secondary particles. 
Many former analyses relied on a simple, overall rescaling of the $p+p$ cross section by a geometric factor or mass number~\cite{Delahaye_2009,1986ApJ...307...47D,1976ApJ...206..312O}. Here we go beyond this approximations by using the data of NA49 for the production of $\pi^+$ in $p+$C collisions at $p_p = 158$ GeV \cite{NA49_2007}. While pion production in $p+p$ collisions is by definition symmetric under a reflection along the beam axis in the CM frame, this is not necessarily the case in $p+A$ collisions (in the nucleon-nucleon CM frame). Actually, the NA49 $p$+C data reveals an asymmetry in the cross section between forward and backward production~\cite{NA49_2007_discussion}, which is plausible, because the carbon target contains not only protons but also neutrons and the binding of the nucleons could play a role. The asymmetry makes a description of the cross section in terms of $x_R$, an intrinsically symmetric variable, inconsistent. Thus, we will use $x_F$ instead of $x_R$ to parametrize $p+A$ collisions.

In principle, it would be useful to determine a standalone parametrization for the pion production of each $p+A$ initial state, especially for $p+$He, which is most relevant in the context of CRs. However, the currently available data on $\pi^+$ production measurements in $p+A$ collisions are not sufficient to obtain independent descriptions.
Especially for $p+$He collisions the available data is very scarce. A few measurements of pion production in $p+$He collisions were taken in the 1980s \cite{Baldin:1982my}, however, with the goal to study the nuclear quark structure and in a kinematic regime where the production is forbidden in single-nucleon collisions. This kinematic regime is highly suppressed in the Galaxy.
So, we will rely on an $x_F$ and $A$-dependent rescaling. Inspired by the treatment for antiprotons in \cite{Korsmeier_2018} we exploit a rescaling of $p+p$ cross section in terms of overlap functions. The idea is to split the $\pi^+$ production into two components produced by either the projectile or the target, where the $\pi^+$ from each component are mainly produced in forward direction. Adjusting the normalization of the overlap functions separately allows accommodating an asymmetry.

We model the inclusive Lorentz invariant cross section of the $A_1 + A_2 \rightarrow \pi^+ + X$ scattering by:
\begin{eqnarray}
  \label{eqn::param_pA}
  &&\sigmaInv^{A_1 A_2} (\sS, x_F, \pT)  = \\ \nonumber
  && \qquad f^{A_1 A_2}(A_1, A_2, x_F, D_1, D_2,D_3) \,\,  \sigmaInv^{pp}(\sS, \xR, \pT), \quad
\end{eqnarray}
where $A_1$ and $A_2$ are the mass numbers of the projectile and target nucleus, respectively, and $D_1$, $D_2$, and $D_3$ are three fit parameters. Explicitly, the factor $f^{A_1 A_2}$ is defined by:
\begin{equation}
  \label{eqn::param_pA_2}
  f^{A_1 A_2} (x_F)=  A_1^{D_1} A_2^{D_1} \left[ A_1^{D_2} F_\mathrm{pro}(x_F) +   A_2^{D_2} F_\mathrm{tar}(x_F) \right],
\end{equation}
with $F_\mathrm{pro}(x_F)$ and $F_\mathrm{tar}(x_F)$ given by
\begin{equation}
    F_\mathrm{pro/tar}(x_F)=\frac{1\pm\tanh(D_3 x_F)}{2}.
    \label{eq:proj}
\end{equation}
In the above equations, the kinetic variables $x_F$ and $\sS$ refer to the nucleon-nucleon CM frame. We  do not claim that $F_\mathrm{pro}(x_F)$ and $F_\mathrm{tar}(x_F)$ are the actual projectile and target overlap functions. They are rather an effective treatment that we have introduced to describe the NA49 data.
\begin{figure}[t]
    \includegraphics[width=0.49\textwidth]{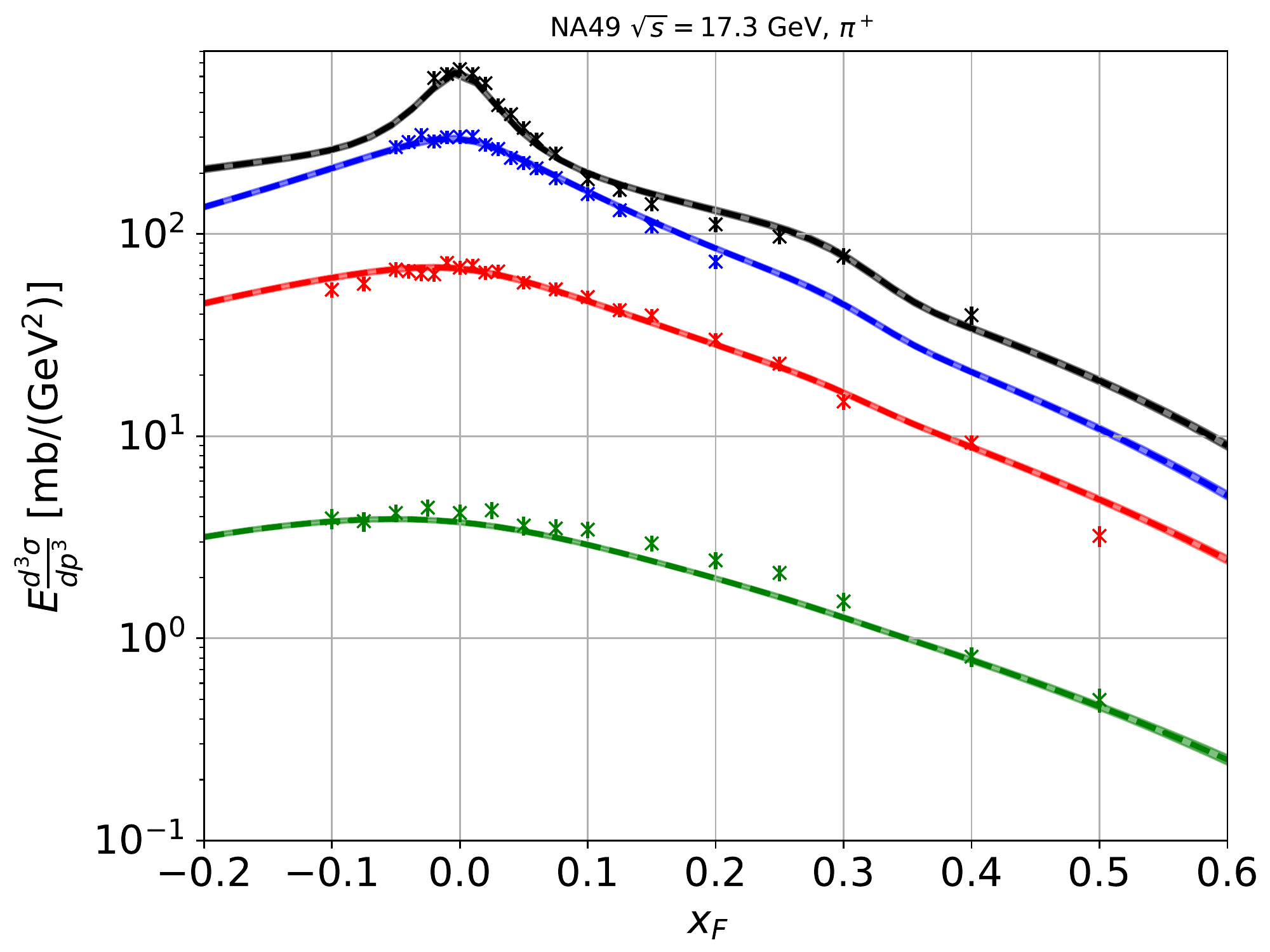}
    \caption{
             Results of the fit on the NA49 data \cite{NA49_2007} invariant cross section for the inclusive $\pi^+$ production in $p+C$ collisions. We show the NA49 data together with our fit results as a function of $x_F$ for some representative values of $p_T$. Shaded bands show the $1\sigma$ uncertainty band.
    } 
    \label{Fig:pi-plus-carbon-NA49}
\end{figure}
To determine $\sigmaInv^{pA}$, we fit the $x_F$-dependent rescaling factor $f^{A_1 A_2} (x_F)$ of Eq.~\eqref{eqn::param_pA_2}, while $\sigmaInv^{pp}(\sS, x_R, p_T)$ is fixed to the best-fit values of Sec.~\ref{sec:piplus}. In other words, we fix the three free parameters that are $D_1$ to $D_3$ performing a $\chi^2$ fit using the NA49 data on $\sigmaInv$ for the inclusive $\pi^+$ production in $p+C$ collisions at $\sS = 17.3$ GeV \cite{NA49_2007}. We obtain a good fit with a $\chi^2$/d.o.f. of 400/265. The best-fit parameters are reported in Tab.~\ref{tab::Fit_results_pC}. 
The result of the fitted parametrization is compared to the NA49 data of $\sigmaInv$ in Fig.~\ref{Fig:pi-plus-carbon-NA49}. The cross section is plotted as a function of $x_F$ for a few representative values of $p_T$. We observe a good agreement of the data with the parametrization, especially at low values of $p_T$, which are the most important for positron production in the Galaxy. The uncertainties on the model turn out to be about 5\%, which mostly comes from the uncertainty in the $p+p$ collisions.
Finally, we also check, a posteriori (as for $p+p$ collisions), that our parametrization is qualitatively in good agreement with NA61/SHINE data \cite{NA61_2016} of $p+C$ scattering provided at $\sS = 7.7$ GeV. 
Using the rescaling relation of Eq.~\eqref{eqn::param_pA} we obtain the cross sections for $p+$He and all other nuclei collisions. 

\begin{figure*}[t]
    \includegraphics[width=0.49\textwidth]{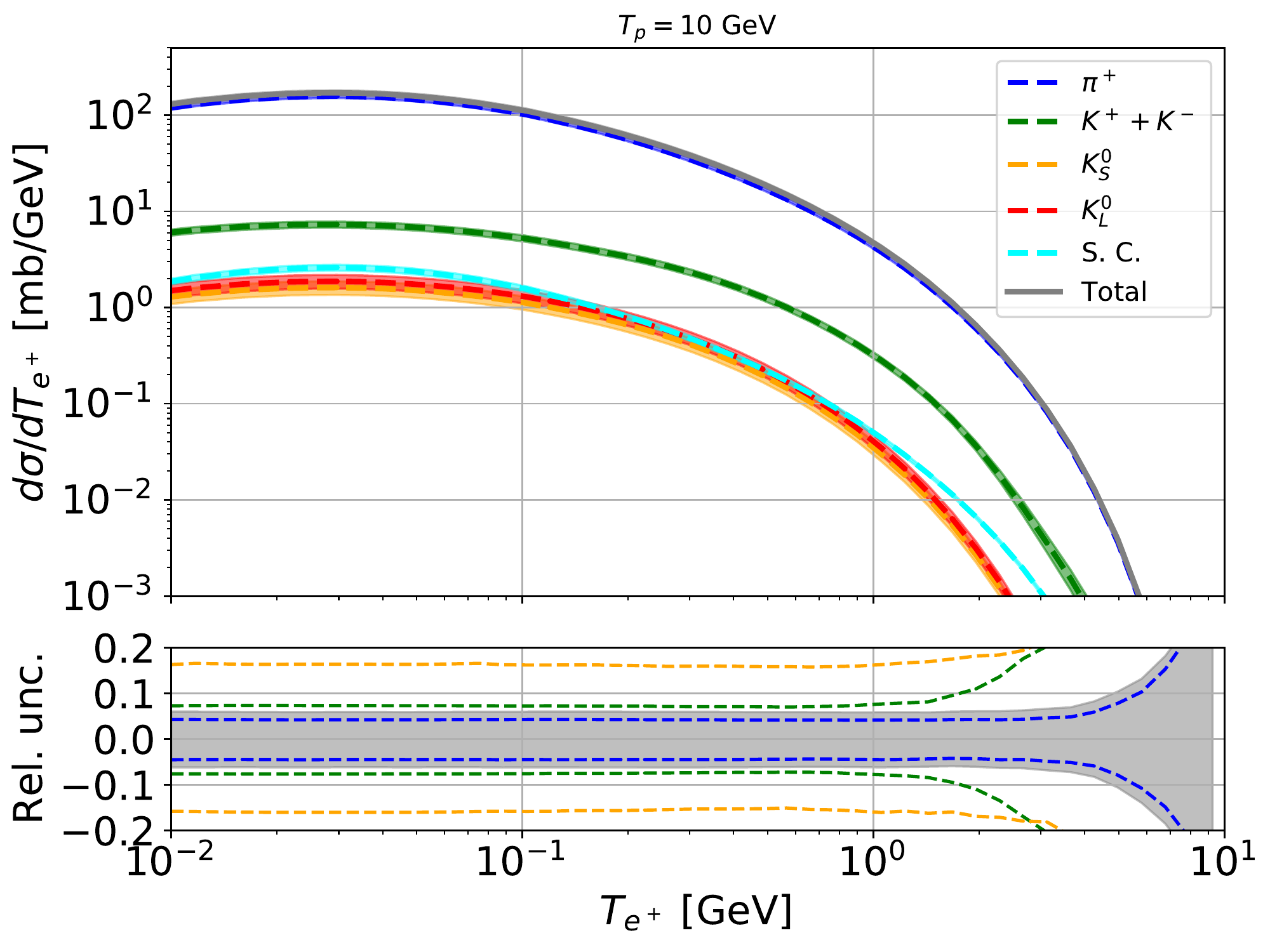}
    \includegraphics[width=0.49\textwidth]{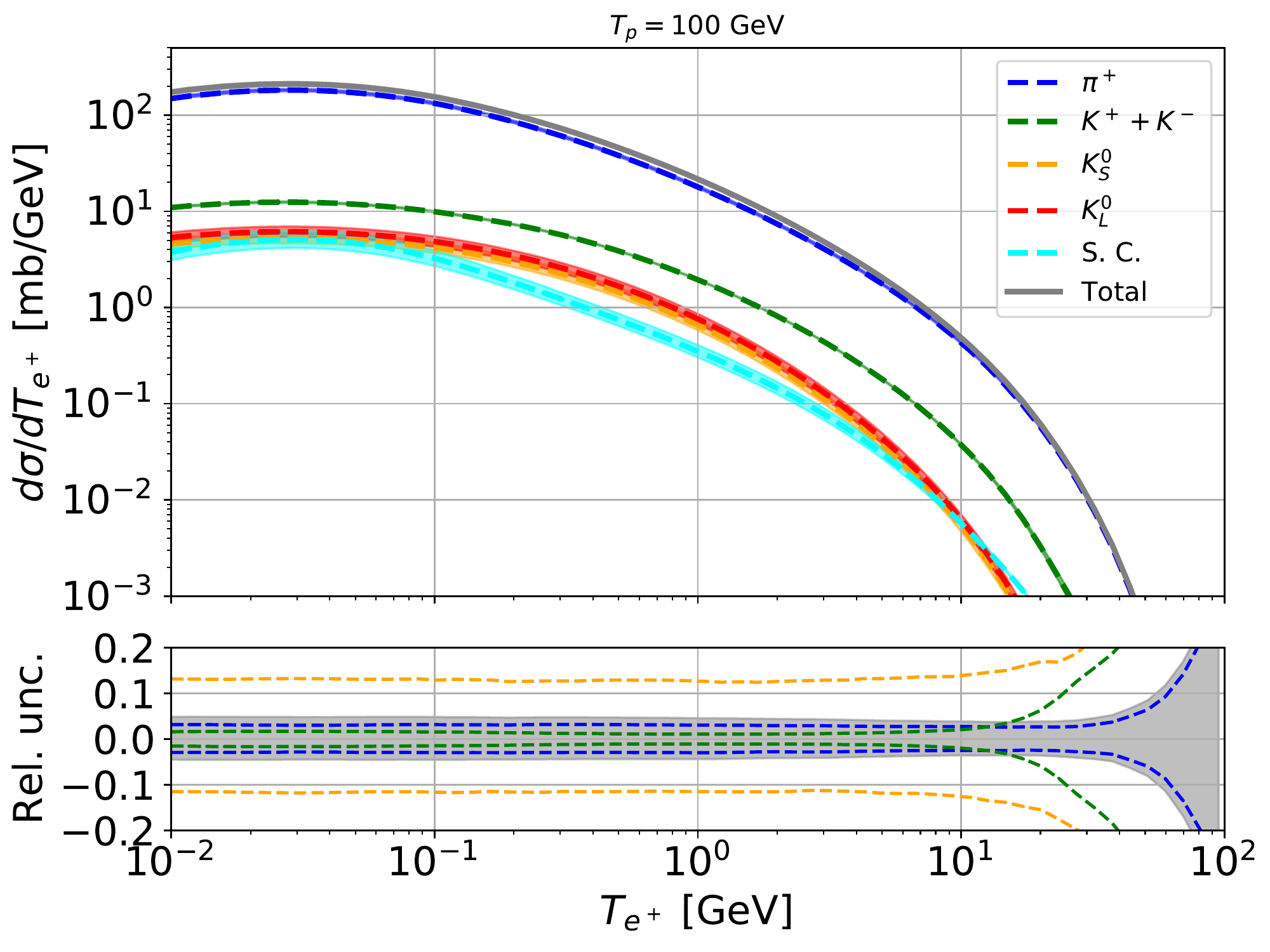}
    \includegraphics[width=0.49\textwidth]{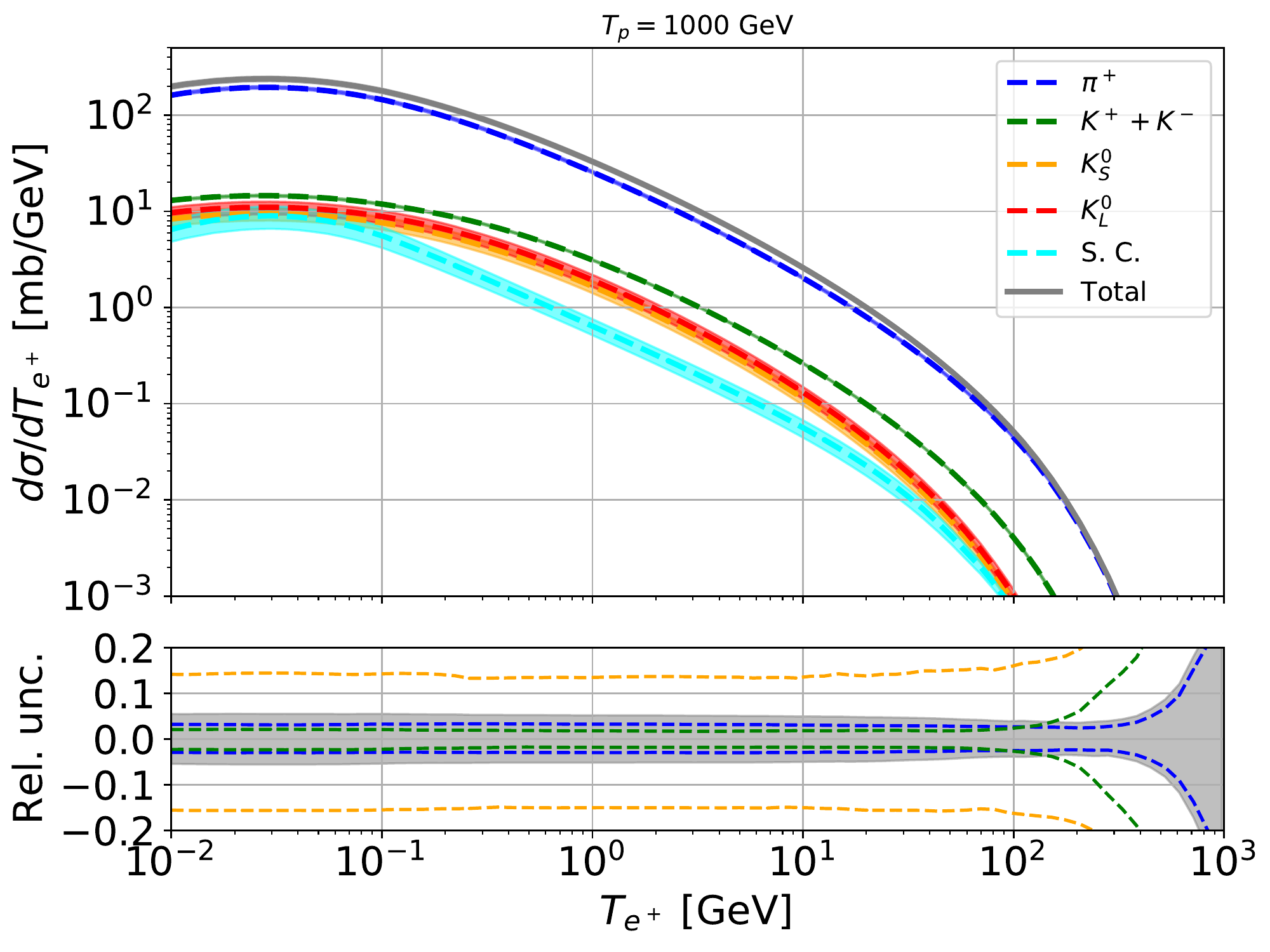}
    \includegraphics[width=0.49\textwidth]{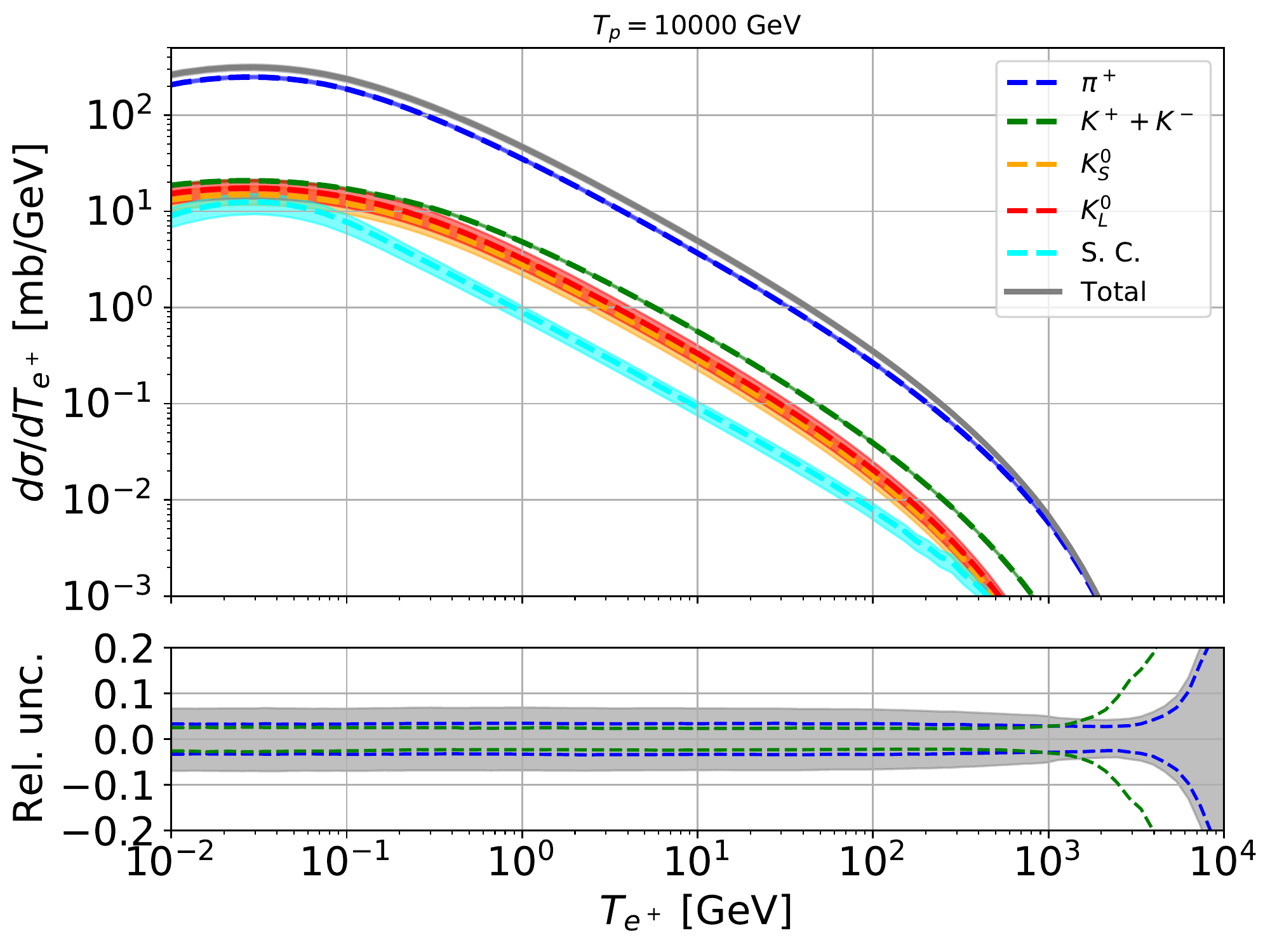}
    \caption{Differential cross section for the inclusive production of $e^+$ in $p+p$ collisions, derived from fits to the data as described in Sec.~\ref{sec:cross_section_eplus} and \ref{sec:other_channels}. We plot separate production of $\pi^+$, $K^+$ and $K^-$, $K^S_0$, $K^L_0$ and subdominant channels (S.C.), and their sum. Each plot is computed for incident proton energies $T_p$, of 10, 100, 1000, and 10000 GeV. The curves are displayed along with their  $1\sigma$ error band. At the bottom of each panel the $1\sigma$ uncertainty band is displayed around the best fit individually for each contribution. 
    } 
    \label{Fig:final_cross_positron}
\end{figure*}
While we are improving the state of the art \cite{Delahaye_2009,1986ApJ...307...47D,1986A&A...157..223D}, which is based on a rescaling of the normalization of $p+p$ cross section by a simple geometrical factor, our result points to the need of collecting data of the $p+{\rm He} \rightarrow \pi^+ + X$ cross section. This might allow disentangling $p+p$ and $p+A$ fits in the future by performing separate fits of the parametrizations for each $p+$He and $p+A$ that avoid rescaling from $p+p$. 
Actually, one reason for the small uncertainty bands in Fig.~\ref{Fig:pi-plus-carbon-NA49} can be related to the fact that the kinematic shape of our parametrization for $p+A$ is already partly fixed by $p+p$, see Eq.~\eqref{eqn::param_pA}. In this sense, more data in the $p+$He (and more general $p+A$) collisions might allow a more correct estimation of uncertainties. 
We also note the absence of data for $x_F<0.1$ in Fig.~\ref{Fig:pi-plus-carbon-NA49}, a kinematic regime which is important for the production of pions in $A+p$ collision in CRs. Here we rely on an extrapolation of our parametrization.
We also tried a fit to $p+C$ data, but considering  all the parameters in Eqs.~\eqref{eq:main_equation}, \eqref{eq:function_prompt}, \eqref{eq:fr} and \eqref{eqn::param_pA}. In this case uncertainties rise to 7-8 \%.

For the $K^+$ production channel, we refer to NA61/SHINE \cite{NA61_2016} data at  $\sS = 7.7$ GeV. We found that a simple rescaling from the $p+p$ case ($f^{pA} = A^{D_1}$) is sufficient. The best fit converges to a $\chi^2/{\rm d.o.f.}$ of 151/93 and the best-fit value of $D_1$ is reported in Tab.~\ref{tab::Fit_results_pC}. For the remaining subdominant production channels discussed in Sec.~\ref{sec:other_channels} we adopt the same rescaling as for $K^+$.

\begin{table}[b]
\caption{Best fit result and 1$\sigma$ error for the parameters in Eqs.~\eqref{eqn::param_pA_2} and \eqref{eq:proj}. 
}
\label{tab::Fit_results_pC}
\begin{tabular}{ l c c c c}
 \hline \hline
              &  $\pi^+$           &   $\pi^-$       & $K^+$              & $K^-$             \\  \hline
$D_1$         &   $0.73\pm 0.01$   & $0.72\pm 0.02$  &  $0.835\pm 0.004$  & $0.829\pm 0.007$  \\
$D_2$         &   $0.30\pm 0.02$   & $0.35\pm 0.03$  &  $0.0$             & $0.0$             \\
$D_3$         &   $3.93\pm 0.43$   & $4.21\pm 0.50$  &  -                 & -                 \\

 \hline \hline
\end{tabular}
\end{table}

\section{Results on the $e^+$ production cross section and source spectrum }
\label{sec:source_term_eplus}

We now have all the elements to compute the total differential cross section
$d\sigma/dT_{e^+}$ for the inclusive production of $e^+$ in $p+p$ inelastic collisions. 
The result is obtained by summing all the contributions of $\pi^+$, $K^+$ and $K^-$, $K^S_0$, $K^L_0$, and subdominant channels (S. C.) fitted on the data as discussed in Sec.~\ref{sec:cross_section_eplus} and \ref{sec:other_channels}. 
This is the main result of our paper and it is displayed in Fig.~\ref{Fig:final_cross_positron}. We plot $d\sigma/dT_{e^+}$ for the separate production channels, and their sum, along with the relevant $1\sigma$ uncertainty band. At the bottom of each panel we display the $1\sigma$ uncertainty band around the best fit for the total $d\sigma/dT_{e^+}$. The four plots are for incident proton energies $T_p$ of 10, 100, 1000, and 10000 GeV. 
The $\pi^+$ channel dominates the total cross section, being about 10 times higher than the $K^+$ (and $K^-$contributing few \% of $K^+$) channels. Positron production from $K^S_0$, $K^L_0$, and S. C. are of the order, all contributing at a few \% level, slightly depending on $T_{e^+}$ and $T_p$. 
The main comment to these results is the smallness of the uncertainty with which we determine $d\sigma/dT_{e^+}$. At $1\sigma$ the uncertainty band around the best fit is 4\% to 7\% at all $T_p$ energies. For $T_e^+$ values close to $T_p$, the error band spreads up since data for this limit (which corresponds to $x_R=1$) are not available.

\begin{figure*}[t]
    \includegraphics[width=0.49\textwidth]{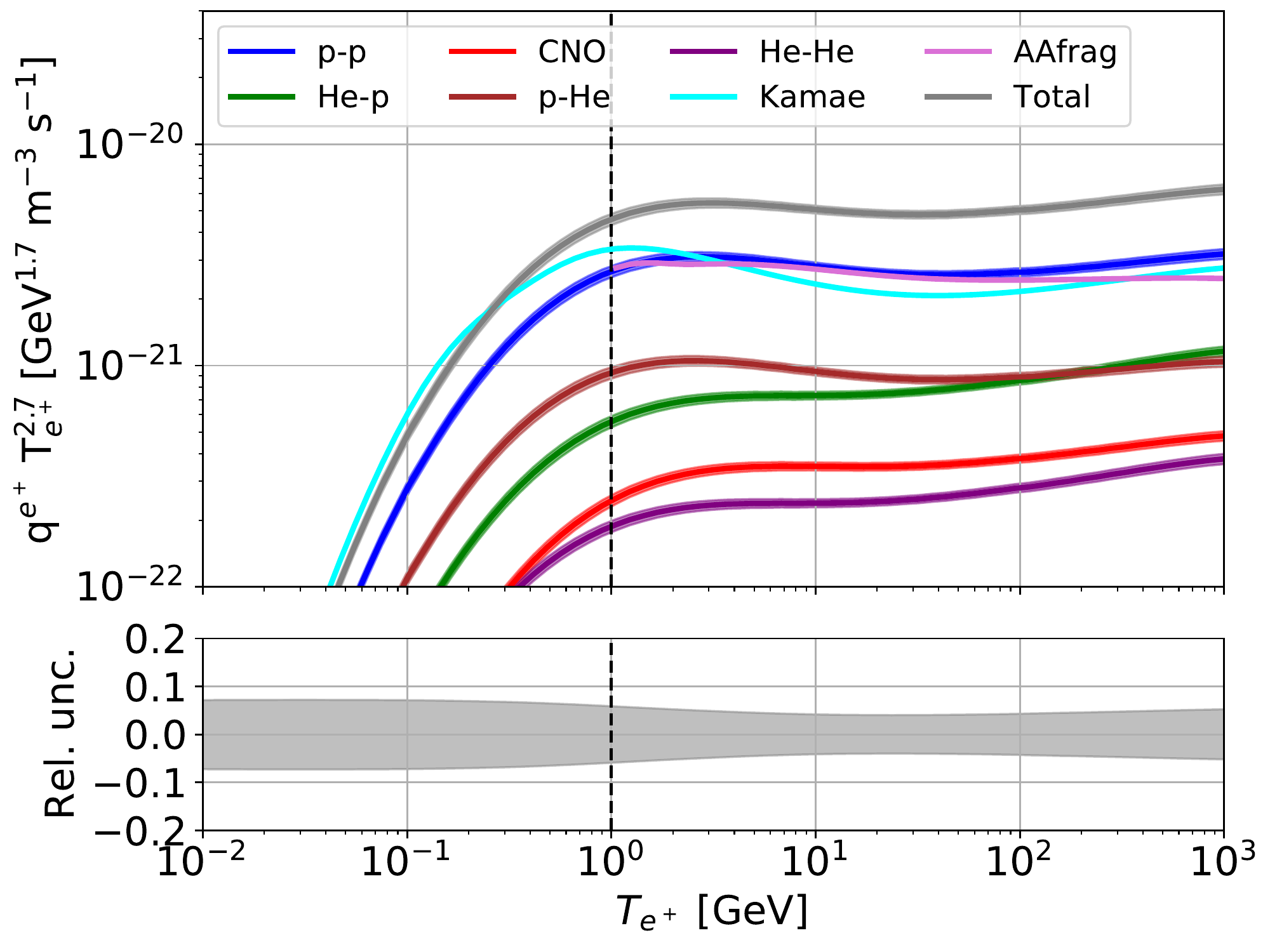}
    \includegraphics[width=0.49\textwidth]{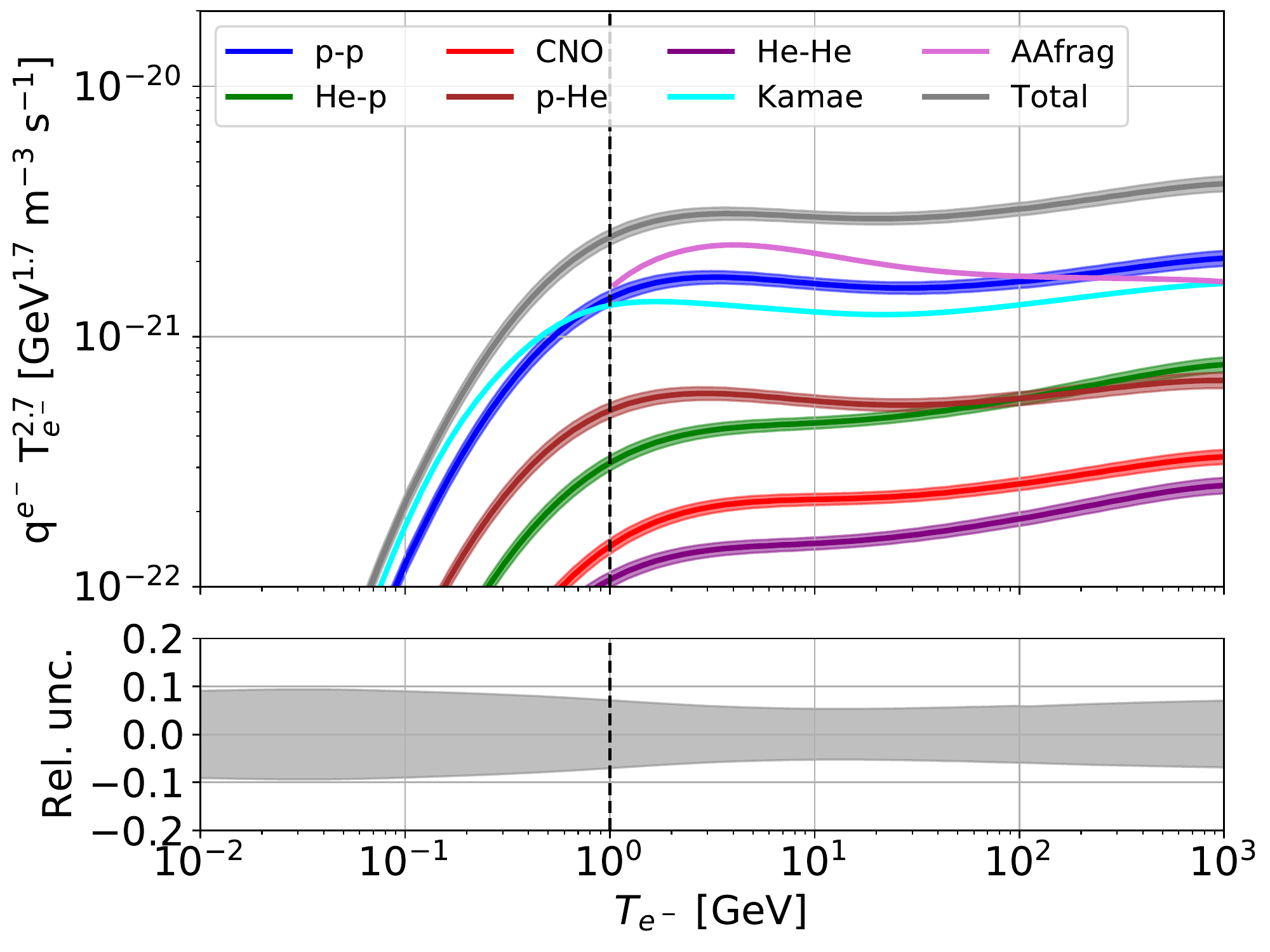}
    \caption{
              Source terms of CR $e^+$ (left panel) and $e^-$ (right panel). Next to the total source term we show the separate CR-ISM contributions. In the bottom panels, we display the relative uncertainty of the total source term. 
              We note, however, that for $T_{e^+}\lesssim 1$~GeV (black dashed line) the source term is not constrained by cross section data but rather an extrapolation of our parametrization which could possibly be affected by systematics.
            } 
    \label{Fig:final_source_positron}
\end{figure*}

We conclude that the $e^+$ production cross section from $p+p$ collisions is determined with very high precision. This result is mainly due to the precision of the data at our disposal, and also to the appropriate empirical description provided by our algebraic model.

In Fig.~\ref{Fig:final_source_positron}, we present the computation of the source spectrum of $e^+$ in the Galaxy as a function of $T_{e^+}$, implementing Eq.~\eqref{eq:source_term}. We fix $n_{\rm H}=0.9 \,{\rm cm^{-3}}$ and $n_{\rm He}=0.1 \, {\rm cm^{-3}}$. The CR fluxes $\phi_i$ for a nucleus $i$ are taken from \cite{Korsmeier:2021bkw}. We plot separate results for the collision of $p+p$, $p+$He, He$+p$, He-He and C, N and O CR scattering off H, with their uncertainty due the production cross sections computed in this paper. The $q(E)$ is predicted with a remarkably small uncertainty, ranging from 5\% to 8\% depending on the energy. 
We nevertheless remind that the different estimations and parametrizations used in the literature pointed out differences by a factor of two. Our results definitively exclude that $e^+$ cross sections can gauge the source spectrum, and consequently the flux at Earth, by more than a factor of few \%. 
We compare our results for the $p+p$ channel with \cite{Kamae:2006bf} (labeled Kamae) and \cite{Koldobskiy:2021nld} (labeled AAfrag). The Kamae cross section predicts an about 20\% smaller source term above 5 GeV, while it predicts a significantly larger source term below 1 GeV. In contrast, for AAfrag, we only report results for $T_{e^+}$ above 1 GeV, since they report cross sections only for $T_p \geq 3.1$~GeV meaning that the source term cannot be predicted accurately at lower energies. The differences are within 10\% for most of the energy range between 1 and 100 GeV, while our prediction becomes about 20\% higher at 1 TeV. We also checked the predictions for $p+$He, He$+p$, and He$+$He, finding differences at a similar level.

Finally, we note that the available cross section data (especially for pion production) constrain the positron source term down to about 1 GeV. Below this energy, the prediction of the source term relies on an extrapolation from our parametrization and could be affected by larger systematics.

\section{Results on the $e-$ production cross section and source spectrum }
\label{sec:electrons}
Secondary $e^-$ are produced in the Galaxy from the same $p+p$ collisions as $e^+$. In this paper, we also provide new results for the $e^-$ production cross section. We mirror the same analysis performed for $e^+$ and described at length in the previous sections. In particular, for the $\sigmaInv$ for $\pi^-$ production we adopt the parametrizations reported in Eqs.~\eqref{eq:main_equation}, \eqref{eq:function_prompt}, \eqref{eq:fr} and \eqref{eq:As}. The data employed in the fits are taken from NA49 \cite{2005_NA49}, NA61/SHINE \cite{Aduszkiewicz:2017sei}, Antinucci \cite{osti_4593576}, ALICE \cite{2011_ALICE} and CMS \cite{2012_CMS}, \cite{2017_CMS}, as reported in Tab.~\ref{tab::pp_data}.

\begin{figure*}[t]
\setlength{\unitlength}{1\textwidth}
\begin{picture}(1,0.4)
  \put(-0.01 , 0.0){\includegraphics[width=0.505\textwidth]{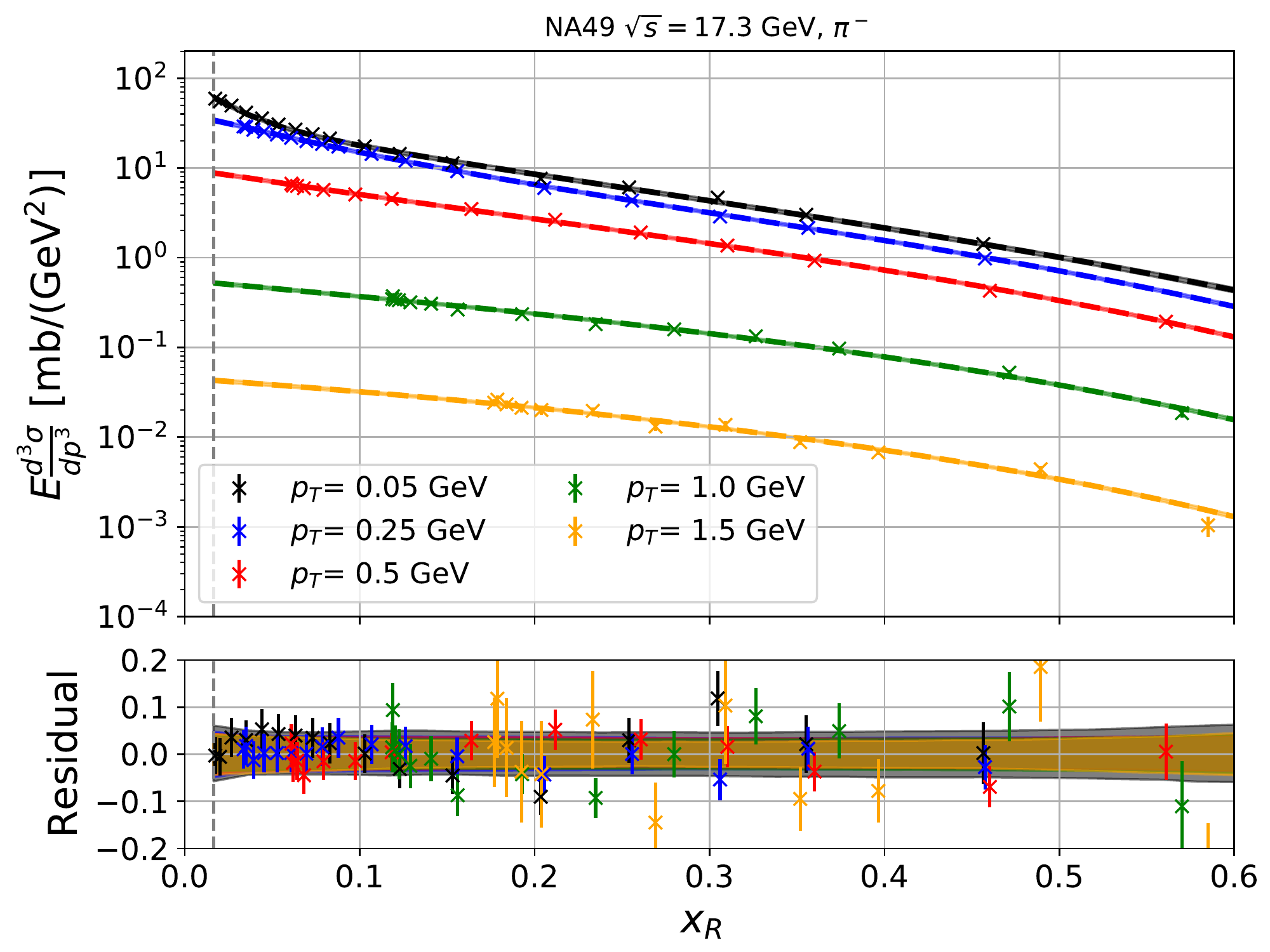}}
  \put( 0.50,-0.004){\includegraphics[width=0.25\textwidth  ]{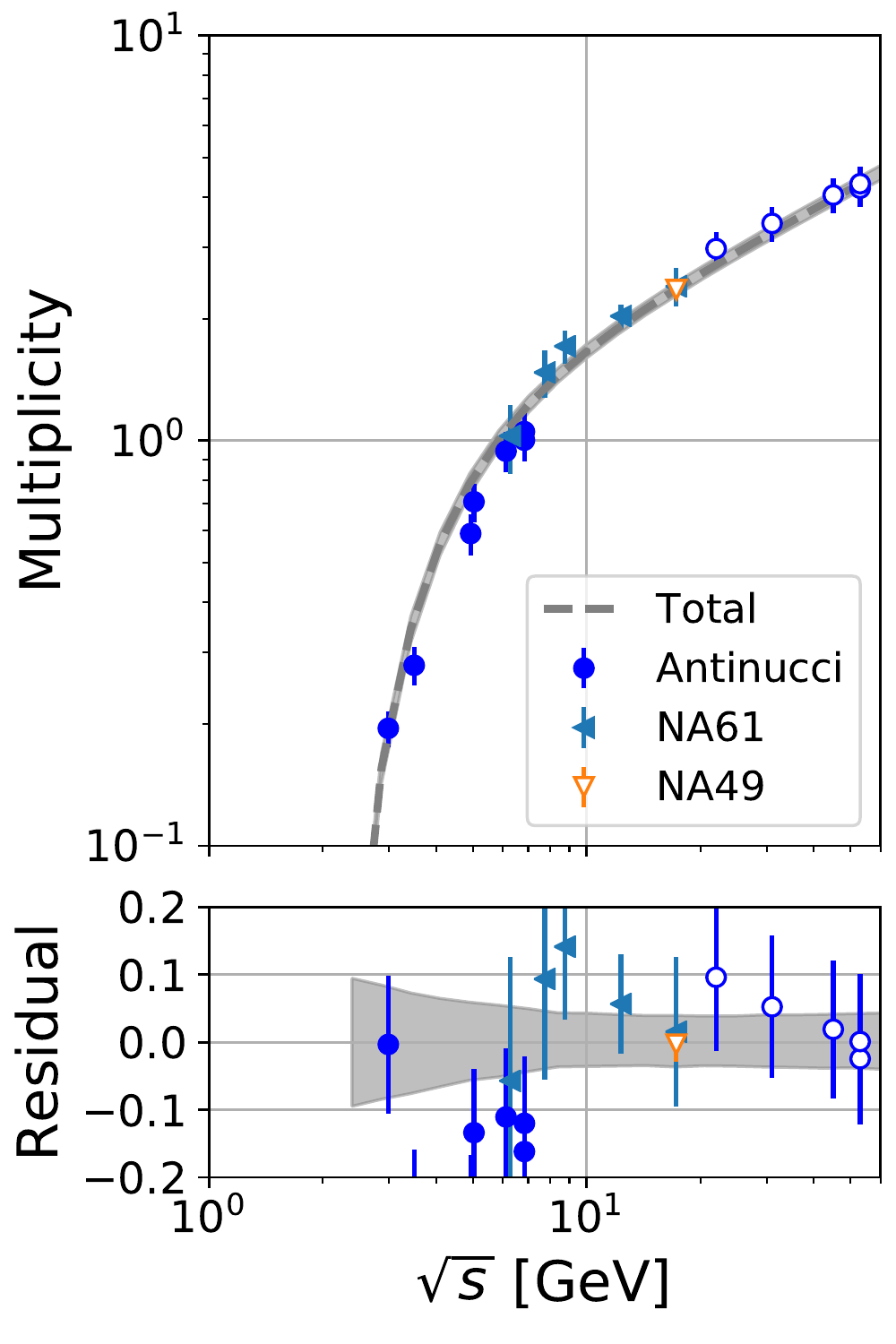}}
  \put( 0.75,-0.004){\includegraphics[width=0.25\textwidth  ]{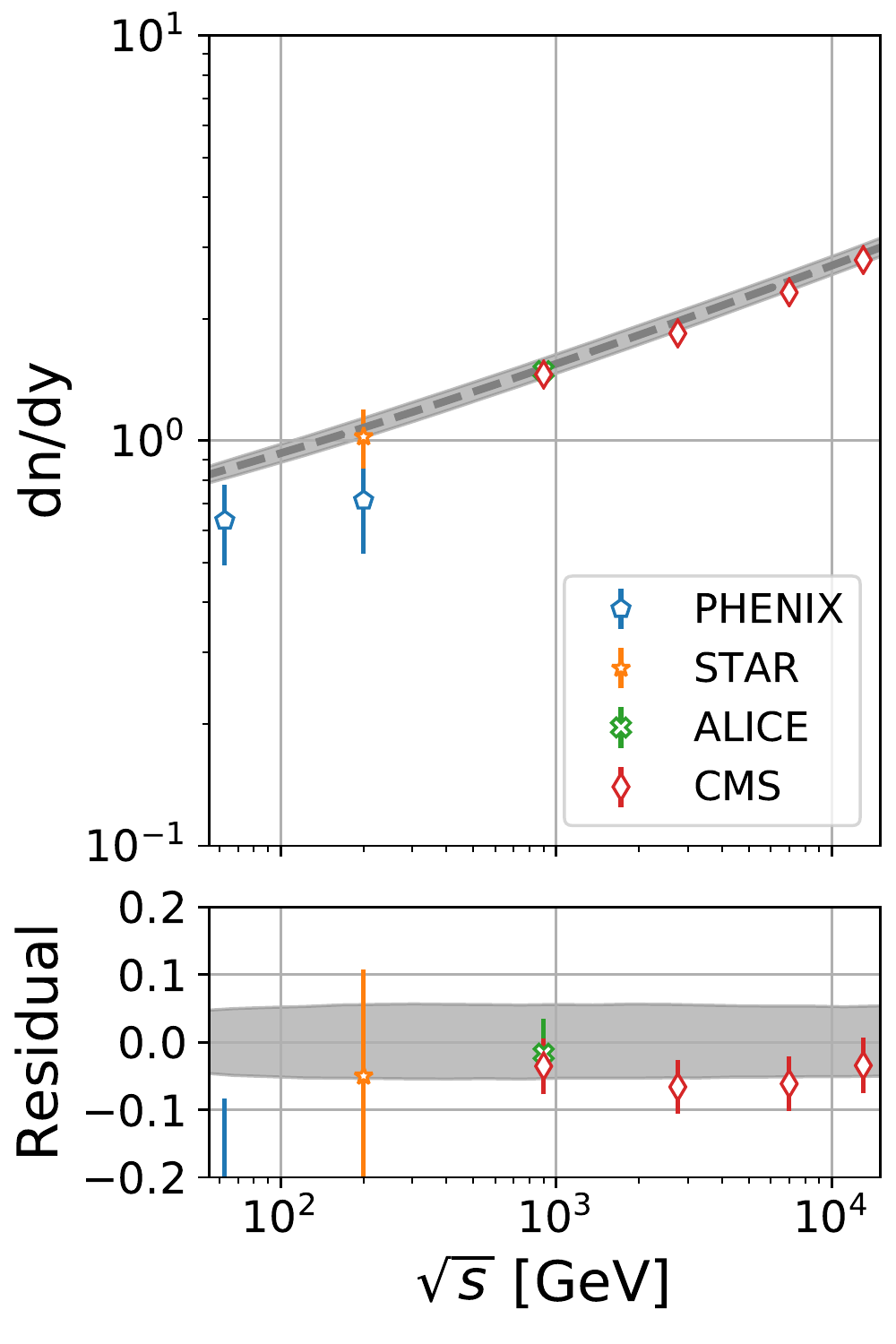}}
\end{picture}
  \caption{
        Left panel: same as Fig.~\ref{Fig:pi-plus-NA49} (left panel) but for $\pi^-$ production in $p+p$ collisions. Right panel: same as Fig.~\ref{fig:sqrt_s_sigma} (right panel) but for $\pi^-$ production in $p+p$ collisions at various $\sqrt{s}$, as described  by Eqs.~\eqref{eqn::chiSquare}, \eqref{eqn::chiSquare_stat}, and \eqref{eqn::chiSquare_scale} (see text for details).
  }
  \label{fig:sqrt_s_pi_minus_mult}
\end{figure*}

\begin{figure*}[t]
    \includegraphics[width=0.49\textwidth]{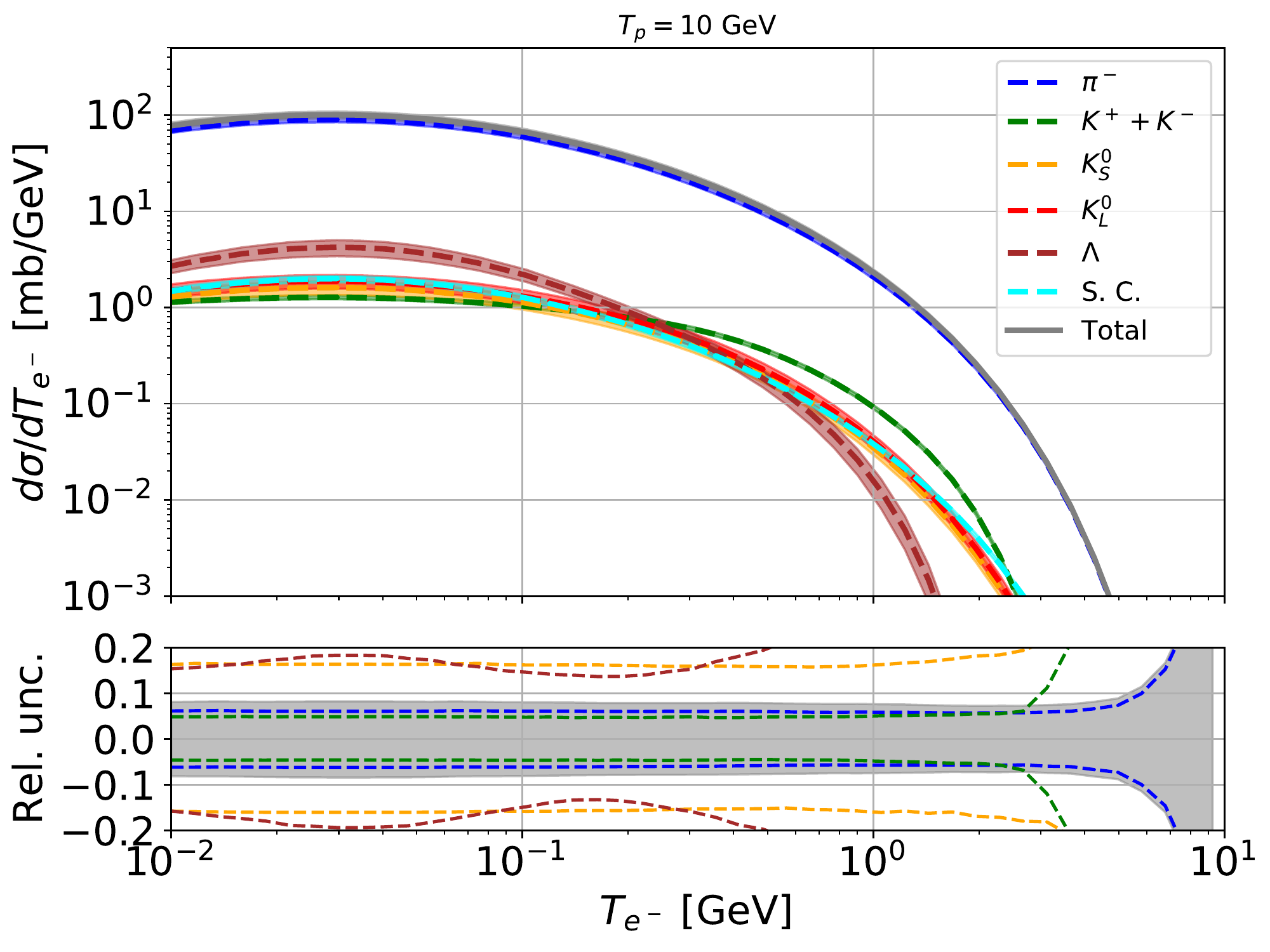}
    \includegraphics[width=0.49\textwidth]{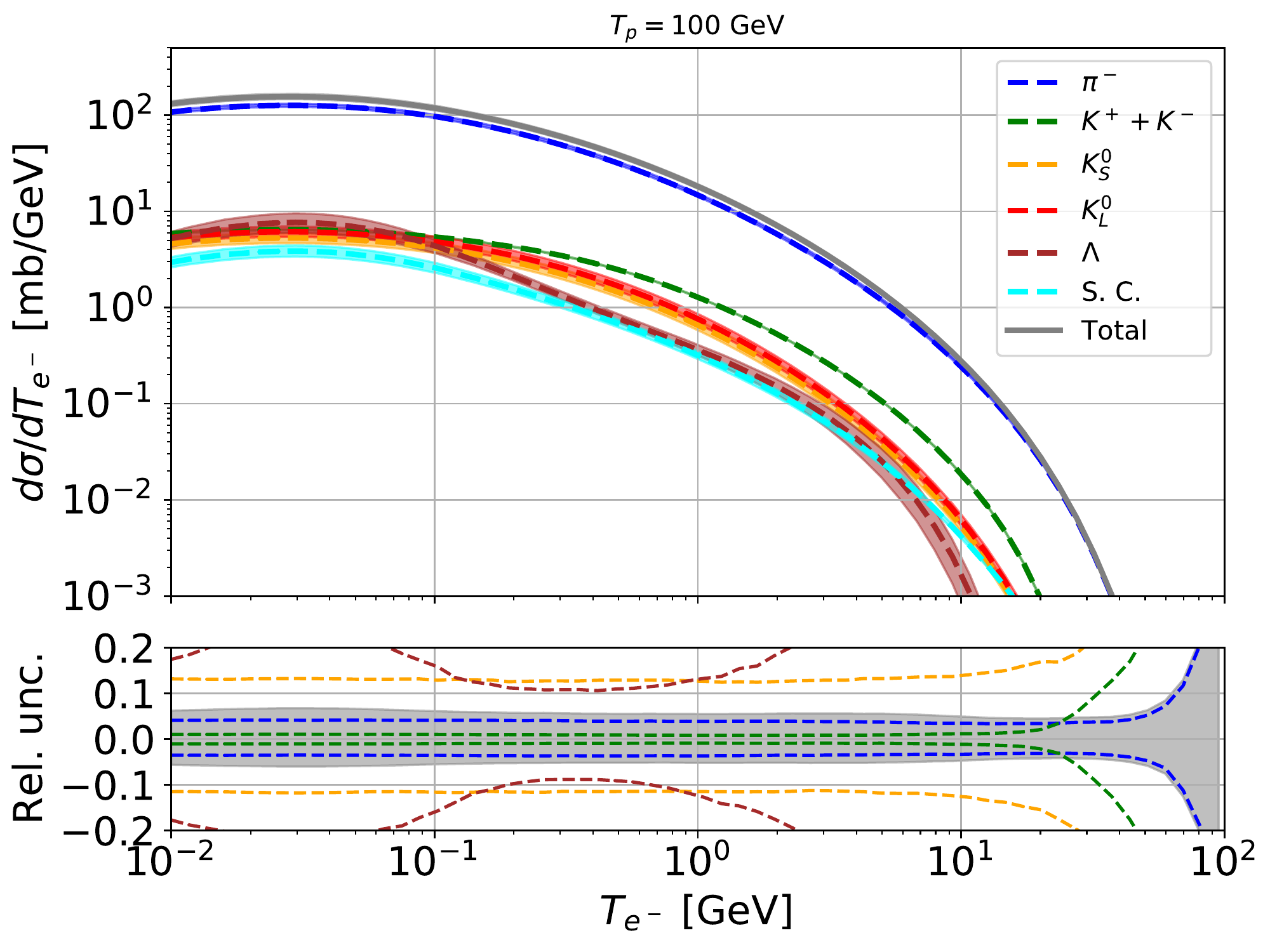}
    \caption{Differential cross section for the inclusive production of $e^-$ in $p+p$ collisions, derived from fits to the data as described in Sec.~\ref{sec:cross_section_eplus} and \ref{sec:other_channels}. We plot separately the contribution from $\pi^-$, $K^+$ and $K^-$, $K^S_0$,
    $K^L_0$, $\Lambda$, subdominat channels(S.C.), and their sum. We provide the result for incident proton energy $T_p$ of  10 and 100 GeV. The curves are displayed along with their  $1\sigma$ uncertainty band. At the bottom of each panel it is displayed the $1\sigma$ uncertainty band around the best fit for the total 
    $d\sigma/dT_{e^-}$. 
    } 
    \label{Fig:final_cross_electrons}
\end{figure*}

The results of the fit to the NA49 production cross section $\pi^-$ data are displayed in Fig.~\ref{fig:sqrt_s_pi_minus_mult} (left panel), as a function of $x_R$ and for a few representative values of $p_T$. The fit is globally very good, and the resulting uncertainties are about 5-6\%, as shown in the bottom panel and similar to what we obtained for the $\pi^+$ fit. 
The energy dependence of the cross section has been fixed as for $\pi^+$, see Sec.~\ref{sec:piplus_differentCME}. The only difference is that the ALICE data points have not been considered in the final fit, because they are inconsistent with CMS data at $\sqrt{s}= 0.9$ TeV. 
The results on the multiplicity are shown in Fig.~\ref{fig:sqrt_s_pi_minus_mult} (right panel). 
Again, the fit is pretty good, and the uncertainty is below 10\%.
The results on goodness of the fit are summarized in Tab.~\ref{tab::chi2_pp} and the best-fit parameters are reported in Tab.~\ref{tab::Fit_results_pp_pion}.

The contribution from $K^-$ is computed following the same procedure as for the $e^+$ from $K^+$, see Sec.~\ref{sec:eplus_kplus}. In particular, we fitted data from the same experiments (and same references) to Eqs.~\eqref{eq:main_equation_K}, \eqref{eq:function_prompt_K} and \eqref{eq:As_K}. The fit to the data is very good, see Tab.~\ref{tab::chi2_pp}, and the uncertainty band is similar to the one found for the $K^+$ channel. The contributions from $K^0_S$, $K^0_L$ and $\pi^0$ decays are symmetric for both $e^+$ and $e^-$, and have been 
discussed in Secs.~\ref{sec:eplus_K0s}, \ref{sec:eplus_K0l} and \ref{sec:eplus_pi0}. In addition, we consider also the contribution from the 
$\Lambda$ baryon as explained in Sec.~\ref{sec:eminus_lambda}.

In Fig.~\ref{Fig:final_source_positron} we present the computation of the source spectrum of $e^-$ in the Galaxy as a function of $T_{e^-}$,
as discussed for $e^+$ in Sec.~\ref{sec:source_term_eplus}. 
It is predicted with a remarkably small uncertainty, ranging from 6\% to 10\% depending on the energy. 
With respect to \cite{Kamae:2006bf}, we obtain for the $p+p$ channel a higher prediction between 20-30$\%$ between 1 GeV to 1 TeV. Instead, at lower energies our cross sections are lower. However, at such low energies our results, in particular below 1 GeV, as well as the ones from \cite{Kamae:2006bf}, are driven by extrapolation.
In contrast, the AAfrag cross sections predict a 30-40\% larger source term compared to our cross section between 1 GeV and 100 GeV. The large difference for $e^-$ between the AAfrag model and Kamae  was already observed in \cite{Koldobskiy:2021nld}.

\section{Discussion and conclusions}
\label{sec:conclusions}

The secondary production of $e^\pm$ in our Galaxy presents a significant contribution to the $e^\pm$  fluxes measured at Earth. In particular, the $e^+$ flux is dominated by secondaries below 10 GeV. At higher energies, several primary contributions are discussed in the literature, the most popular being pulsars and dark matter annihilation or decay. The correct interpretation of those primary contributions depends on the accurate description of the secondary production.

Most of the secondary $e^\pm$ are produced in $p+p$ collisions, nonetheless, the contributions from collisions involving helium, both as a target and as a projectile, are relevant. The main production channels of the secondary $e^\pm$ involve the intermediate production and decay of $\pi^\pm$ and $K^\pm$, while some additional channels can contribute to the source term at the percent level each. 

In the last years, new experimental data have become available covering large portions of the kinematic phase space. In this paper, we determine an analytical description of the Lorentz invariant cross section for the production of $\pi^\pm$ and $K^\pm$, especially focusing on $p+p$ collisions. Then, we also evaluate, either by exploiting further data or by referring to Monte Carlo generators, the inclusive cross section into $K_0^S$, $K_0^L$, ${\Lambda}$, $\bar{\Lambda}$, $\pi^0$, $\Sigma$ and $\Xi$. For all these particles, we implement the relevant 2 and 3 body decay channels, which finally contribute to $e^\pm$. The most important decay of polarized $\mu^\pm$ is computed including NLO corrections. 

The most relevant data are provided by the NA49 experiment which measured $\pi^\pm$ and $K^\pm$ production in $p+p$ fixed-target collisions at proton momenta of $158$ GeV. These data are intrinsically precise at a level of a few percent (maximum 10\%). Our analytical expressions for the invariant cross section fit this data very well. For the important $\pi^\pm$ channels the invariant cross section is determined with an uncertainty of about 5\% in the relevant kinematic parameter space. Further data at lower and higher $\sS$ are also described well by our parametrizations. 
The differential cross section $d\sigma/dT_{e^\pm}(p+p\rightarrow \pi^\pm+X)$, which enters in the computation of the $e^\pm$ source term, is determined with about 5\% precision. Including all the production and decay channels, the total $d\sigma/dT_{e^\pm}(p+p\rightarrow e^\pm+X)$ is predicted from 10 MeV up to tens of TeV of $e^\pm$ energy,  with an uncertainty of about 5-7\%. 

The cross section for scattering of nuclei heavier than protons is obtained by fitting the NA49 data for the production of $\pi^\pm$ on $p+$C collisions. 
The statistical uncertainties are very small, however, we cannot exclude systematic effects, for example, due to the rescaling from the $p+$C. Future measurements of pion production in the $p$+He could help to remove this ambiguity. 

Finally, we provide a prediction for the Galactic $e^\pm$ source spectrum, which is obtained from a convolution of the differential production cross section with the incident CR flux and the ISM density. We include CR nuclei up to O and $p$ and He ISM targets. Our major result resides in the precision with which this source term is predicted, which ranges between 5\% and 8\% for $e^+$ and 7\% and 10\% for $e^-$. The uncertainty in the secondary $e^+$ and $e^-$ production is therefore dramatically decreased with respect to the state of the art, where different descriptions of the cross section vary by a factor of about two, posing a large systematic uncertainty due to spallation reactions. 
We note, however, that for $T_{e^+}\lesssim 1$~GeV the source term is not constrained by cross section data but rather an extrapolation of our parametrization which could possibly be affected by systematics.
Our results, especially in the $e^+$ sector, finally open the door to interpretations of CR data, especially from the AMS-02 experiment, in which the second component is no longer a limiting factor in pinpointing primary components. 

We provide numerical tables for the energy-differential cross sections $d\sigma/dT_{e^{\pm}}$ as a function of the $e^{\pm}$ and proton energies and a script to read them. The material is available at \url{https://github.com/lucaorusa/positron_electron_cross_section}.

\section*{Acknowledgements}
We warmly thank Roberto Mussa, Francesca Bellini, Hans Gerhard Fischer and Torbj\"orn Sj\"ostrand for useful discussions. 
MDM research is supported by Fellini - Fellowship for Innovation at INFN, funded by the European Union’s Horizon 2020 research program under the Marie Skłodowska-Curie Cofund Action, grant agreement no.~754496.
The work of FD and LO is supported by the {\sc Departments of Excellence} grant awarded by the Italian Ministry of Education,
University and Research ({\sc Miur}). FD and LO acknowledge the support the Research grant {\sc TAsP} (Theoretical Astroparticle Physics) funded by Istituto Nazionale di Fisica Nucleare.
MK is partially supported by the Swedish National Space Agency under contract 117/19 and the European Research Council under grant 742104.

\bibliography{paper}

\begin{thebibliography}{95}%
\makeatletter
\providecommand \@ifxundefined [1]{%
 \@ifx{#1\undefined}
}%
\providecommand \@ifnum [1]{%
 \ifnum #1\expandafter \@firstoftwo
 \else \expandafter \@secondoftwo
 \fi
}%
\providecommand \@ifx [1]{%
 \ifx #1\expandafter \@firstoftwo
 \else \expandafter \@secondoftwo
 \fi
}%
\providecommand \natexlab [1]{#1}%
\providecommand \enquote  [1]{``#1''}%
\providecommand \bibnamefont  [1]{#1}%
\providecommand \bibfnamefont [1]{#1}%
\providecommand \citenamefont [1]{#1}%
\providecommand \href@noop [0]{\@secondoftwo}%
\providecommand \href [0]{\begingroup \@sanitize@url \@href}%
\providecommand \@href[1]{\@@startlink{#1}\@@href}%
\providecommand \@@href[1]{\endgroup#1\@@endlink}%
\providecommand \@sanitize@url [0]{\catcode `\\12\catcode `\$12\catcode
  `\&12\catcode `\#12\catcode `\^12\catcode `\_12\catcode `\%12\relax}%
\providecommand \@@startlink[1]{}%
\providecommand \@@endlink[0]{}%
\providecommand \url  [0]{\begingroup\@sanitize@url \@url }%
\providecommand \@url [1]{\endgroup\@href {#1}{\urlprefix }}%
\providecommand \urlprefix  [0]{URL }%
\providecommand \Eprint [0]{\href }%
\providecommand \doibase [0]{http://dx.doi.org/}%
\providecommand \selectlanguage [0]{\@gobble}%
\providecommand \bibinfo  [0]{\@secondoftwo}%
\providecommand \bibfield  [0]{\@secondoftwo}%
\providecommand \translation [1]{[#1]}%
\providecommand \BibitemOpen [0]{}%
\providecommand \bibitemStop [0]{}%
\providecommand \bibitemNoStop [0]{.\EOS\space}%
\providecommand \EOS [0]{\spacefactor3000\relax}%
\providecommand \BibitemShut  [1]{\csname bibitem#1\endcsname}%
\let\auto@bib@innerbib\@empty
\bibitem [{\citenamefont {Adriani}\ \emph
  {et~al.}(2011{\natexlab{a}})\citenamefont {Adriani}, \citenamefont
  {Barbarino}, \citenamefont {Bazilevskaya}, \citenamefont {Bellotti},
  \citenamefont {Boezio}, \citenamefont {Bogomolov}, \citenamefont {Bonechi},
  \citenamefont {Bongi}, \citenamefont {Bonvicini}, \citenamefont {Borisov},\
  and\ \citenamefont {et~al.}}]{Adriani_2011}%
  \BibitemOpen
  \bibfield  {author} {\bibinfo {author} {\bibfnamefont {O.}~\bibnamefont
  {Adriani}}, \bibinfo {author} {\bibfnamefont {G.~C.}\ \bibnamefont
  {Barbarino}}, \bibinfo {author} {\bibfnamefont {G.~A.}\ \bibnamefont
  {Bazilevskaya}}, \bibinfo {author} {\bibfnamefont {R.}~\bibnamefont
  {Bellotti}}, \bibinfo {author} {\bibfnamefont {M.}~\bibnamefont {Boezio}},
  \bibinfo {author} {\bibfnamefont {E.~A.}\ \bibnamefont {Bogomolov}}, \bibinfo
  {author} {\bibfnamefont {L.}~\bibnamefont {Bonechi}}, \bibinfo {author}
  {\bibfnamefont {M.}~\bibnamefont {Bongi}}, \bibinfo {author} {\bibfnamefont
  {V.}~\bibnamefont {Bonvicini}}, \bibinfo {author} {\bibfnamefont
  {S.}~\bibnamefont {Borisov}}, \ and\ \bibinfo {author} {\bibnamefont
  {et~al.}},\ }\href {\doibase 10.1126/science.1199172} {\bibfield  {journal}
  {\bibinfo  {journal} {Science}\ }\textbf {\bibinfo {volume} {332}},\ \bibinfo
  {pages} {69–72} (\bibinfo {year} {2011}{\natexlab{a}})}\BibitemShut
  {NoStop}%
\bibitem [{\citenamefont {Adriani}\ \emph {et~al.}(2014)\citenamefont
  {Adriani}, \citenamefont {Barbarino}, \citenamefont {Bazilevskaya},
  \citenamefont {Bellotti}, \citenamefont {Boezio}, \citenamefont {Bogomolov},
  \citenamefont {Bongi}, \citenamefont {Bonvicini}, \citenamefont {Bottai},
  \citenamefont {Bruno},\ and\ \citenamefont {et~al.}}]{Adriani_2014}%
  \BibitemOpen
  \bibfield  {author} {\bibinfo {author} {\bibfnamefont {O.}~\bibnamefont
  {Adriani}}, \bibinfo {author} {\bibfnamefont {G.~C.}\ \bibnamefont
  {Barbarino}}, \bibinfo {author} {\bibfnamefont {G.~A.}\ \bibnamefont
  {Bazilevskaya}}, \bibinfo {author} {\bibfnamefont {R.}~\bibnamefont
  {Bellotti}}, \bibinfo {author} {\bibfnamefont {M.}~\bibnamefont {Boezio}},
  \bibinfo {author} {\bibfnamefont {E.~A.}\ \bibnamefont {Bogomolov}}, \bibinfo
  {author} {\bibfnamefont {M.}~\bibnamefont {Bongi}}, \bibinfo {author}
  {\bibfnamefont {V.}~\bibnamefont {Bonvicini}}, \bibinfo {author}
  {\bibfnamefont {S.}~\bibnamefont {Bottai}}, \bibinfo {author} {\bibfnamefont
  {A.}~\bibnamefont {Bruno}}, \ and\ \bibinfo {author} {\bibnamefont
  {et~al.}},\ }\href {\doibase 10.1088/0004-637x/791/2/93} {\bibfield
  {journal} {\bibinfo  {journal} {The Astrophysical Journal}\ }\textbf
  {\bibinfo {volume} {791}},\ \bibinfo {pages} {93} (\bibinfo {year}
  {2014})}\BibitemShut {NoStop}%
\bibitem [{\citenamefont {Aguilar}\ \emph
  {et~al.}(2015{\natexlab{a}})\citenamefont {Aguilar} \emph
  {et~al.}}]{Aguilar:2015ooa}%
  \BibitemOpen
  \bibfield  {author} {\bibinfo {author} {\bibfnamefont {M.}~\bibnamefont
  {Aguilar}} \emph {et~al.} (\bibinfo {collaboration} {AMS}),\ }\href {\doibase
  10.1103/PhysRevLett.114.171103} {\bibfield  {journal} {\bibinfo  {journal}
  {Phys. Rev. Lett.}\ }\textbf {\bibinfo {volume} {114}},\ \bibinfo {pages}
  {171103} (\bibinfo {year} {2015}{\natexlab{a}})}\BibitemShut {NoStop}%
\bibitem [{\citenamefont {Aguilar}\ \emph
  {et~al.}(2015{\natexlab{b}})\citenamefont {Aguilar}, \citenamefont {Aisa},
  \citenamefont {Alpat}, \citenamefont {Alvino}, \citenamefont {Ambrosi},
  \citenamefont {Andeen}, \citenamefont {Arruda}, \citenamefont {Attig},
  \citenamefont {Azzarello}, \citenamefont {Bachlechner}, \citenamefont
  {Barao}, \citenamefont {Barrau}, \citenamefont {Barrin}, \citenamefont
  {Bartoloni}, \citenamefont {Basara}, \citenamefont {Battarbee}, \citenamefont
  {Battiston}, \citenamefont {Bazo}, \citenamefont {Becker},\ and\
  \citenamefont {Zuccon}}]{article}%
  \BibitemOpen
  \bibfield  {author} {\bibinfo {author} {\bibfnamefont {M.}~\bibnamefont
  {Aguilar}}, \bibinfo {author} {\bibfnamefont {D.}~\bibnamefont {Aisa}},
  \bibinfo {author} {\bibfnamefont {B.}~\bibnamefont {Alpat}}, \bibinfo
  {author} {\bibfnamefont {A.}~\bibnamefont {Alvino}}, \bibinfo {author}
  {\bibfnamefont {G.}~\bibnamefont {Ambrosi}}, \bibinfo {author} {\bibfnamefont
  {K.}~\bibnamefont {Andeen}}, \bibinfo {author} {\bibfnamefont
  {L.}~\bibnamefont {Arruda}}, \bibinfo {author} {\bibfnamefont
  {N.}~\bibnamefont {Attig}}, \bibinfo {author} {\bibfnamefont
  {P.}~\bibnamefont {Azzarello}}, \bibinfo {author} {\bibfnamefont
  {A.}~\bibnamefont {Bachlechner}}, \bibinfo {author} {\bibfnamefont
  {F.}~\bibnamefont {Barao}}, \bibinfo {author} {\bibfnamefont
  {A.}~\bibnamefont {Barrau}}, \bibinfo {author} {\bibfnamefont
  {L.}~\bibnamefont {Barrin}}, \bibinfo {author} {\bibfnamefont
  {A.}~\bibnamefont {Bartoloni}}, \bibinfo {author} {\bibfnamefont
  {L.}~\bibnamefont {Basara}}, \bibinfo {author} {\bibfnamefont
  {M.}~\bibnamefont {Battarbee}}, \bibinfo {author} {\bibfnamefont
  {R.}~\bibnamefont {Battiston}}, \bibinfo {author} {\bibfnamefont
  {A.}~\bibnamefont {Bazo}}, \bibinfo {author} {\bibfnamefont {B.}~\bibnamefont
  {Becker}}, \ and\ \bibinfo {author} {\bibfnamefont {P.}~\bibnamefont
  {Zuccon}},\ }\href {\doibase 10.1103/PhysRevLett.115.211101} {\bibfield
  {journal} {\bibinfo  {journal} {Physical Review Letters}\ }\textbf {\bibinfo
  {volume} {115}},\ \bibinfo {pages} {211101} (\bibinfo {year}
  {2015}{\natexlab{b}})}\BibitemShut {NoStop}%
\bibitem [{\citenamefont {Aguilar}\ \emph
  {et~al.}(2016{\natexlab{a}})\citenamefont {Aguilar} \emph
  {et~al.}}]{Aguilar:2016vqr}%
  \BibitemOpen
  \bibfield  {author} {\bibinfo {author} {\bibfnamefont {M.}~\bibnamefont
  {Aguilar}} \emph {et~al.} (\bibinfo {collaboration} {AMS}),\ }\href {\doibase
  10.1103/PhysRevLett.117.231102} {\bibfield  {journal} {\bibinfo  {journal}
  {Phys. Rev. Lett.}\ }\textbf {\bibinfo {volume} {117}},\ \bibinfo {pages}
  {231102} (\bibinfo {year} {2016}{\natexlab{a}})}\BibitemShut {NoStop}%
\bibitem [{\citenamefont {An}\ \emph {et~al.}(2019)\citenamefont {An},
  \citenamefont {Asfandiyarov}, \citenamefont {Azzarello}, \citenamefont
  {Bernardini}, \citenamefont {Bi}, \citenamefont {Cai}, \citenamefont {Chang},
  \citenamefont {Chen}, \citenamefont {Chen},\ and\ \citenamefont
  {et~al.}}]{2019}%
  \BibitemOpen
  \bibfield  {author} {\bibinfo {author} {\bibfnamefont {Q.}~\bibnamefont
  {An}}, \bibinfo {author} {\bibfnamefont {R.}~\bibnamefont {Asfandiyarov}},
  \bibinfo {author} {\bibfnamefont {P.}~\bibnamefont {Azzarello}}, \bibinfo
  {author} {\bibfnamefont {P.}~\bibnamefont {Bernardini}}, \bibinfo {author}
  {\bibfnamefont {X.~J.}\ \bibnamefont {Bi}}, \bibinfo {author} {\bibfnamefont
  {M.~S.}\ \bibnamefont {Cai}}, \bibinfo {author} {\bibfnamefont
  {J.}~\bibnamefont {Chang}}, \bibinfo {author} {\bibfnamefont {D.~Y.}\
  \bibnamefont {Chen}}, \bibinfo {author} {\bibfnamefont {H.~F.}\ \bibnamefont
  {Chen}}, \ and\ \bibinfo {author} {\bibnamefont {et~al.}},\ }\href {\doibase
  10.1126/sciadv.aax3793} {\bibfield  {journal} {\bibinfo  {journal} {Science
  Advances}\ }\textbf {\bibinfo {volume} {5}},\ \bibinfo {pages} {eaax3793}
  (\bibinfo {year} {2019})}\BibitemShut {NoStop}%
\bibitem [{\citenamefont {Adriani}\ \emph {et~al.}(2019)\citenamefont
  {Adriani}, \citenamefont {Akaike}, \citenamefont {Asano}, \citenamefont
  {Asaoka}, \citenamefont {Bagliesi}, \citenamefont {Berti}, \citenamefont
  {Bigongiari}, \citenamefont {Binns}, \citenamefont {Bonechi}, \citenamefont
  {Bongi},\ and\ \citenamefont {et~al.}}]{Adriani_2019}%
  \BibitemOpen
  \bibfield  {author} {\bibinfo {author} {\bibfnamefont {O.}~\bibnamefont
  {Adriani}}, \bibinfo {author} {\bibfnamefont {Y.}~\bibnamefont {Akaike}},
  \bibinfo {author} {\bibfnamefont {K.}~\bibnamefont {Asano}}, \bibinfo
  {author} {\bibfnamefont {Y.}~\bibnamefont {Asaoka}}, \bibinfo {author}
  {\bibfnamefont {M.}~\bibnamefont {Bagliesi}}, \bibinfo {author}
  {\bibfnamefont {E.}~\bibnamefont {Berti}}, \bibinfo {author} {\bibfnamefont
  {G.}~\bibnamefont {Bigongiari}}, \bibinfo {author} {\bibfnamefont
  {W.}~\bibnamefont {Binns}}, \bibinfo {author} {\bibfnamefont
  {S.}~\bibnamefont {Bonechi}}, \bibinfo {author} {\bibfnamefont
  {M.}~\bibnamefont {Bongi}}, \ and\ \bibinfo {author} {\bibnamefont
  {et~al.}},\ }\href {\doibase 10.1103/physrevlett.122.181102} {\bibfield
  {journal} {\bibinfo  {journal} {Physical Review Letters}\ }\textbf {\bibinfo
  {volume} {122}} (\bibinfo {year} {2019}),\
  10.1103/physrevlett.122.181102}\BibitemShut {NoStop}%
\bibitem [{\citenamefont {Adriani}\ \emph
  {et~al.}(2011{\natexlab{b}})\citenamefont {Adriani}, \citenamefont
  {Barbarino}, \citenamefont {Bazilevskaya}, \citenamefont {Bellotti},
  \citenamefont {Boezio}, \citenamefont {Bogomolov}, \citenamefont {Bongi},
  \citenamefont {Bonvicini}, \citenamefont {Borisov}, \citenamefont {Bottai},\
  and\ \citenamefont {et~al.}}]{Adriani2011}%
  \BibitemOpen
  \bibfield  {author} {\bibinfo {author} {\bibfnamefont {O.}~\bibnamefont
  {Adriani}}, \bibinfo {author} {\bibfnamefont {G.~C.}\ \bibnamefont
  {Barbarino}}, \bibinfo {author} {\bibfnamefont {G.~A.}\ \bibnamefont
  {Bazilevskaya}}, \bibinfo {author} {\bibfnamefont {R.}~\bibnamefont
  {Bellotti}}, \bibinfo {author} {\bibfnamefont {M.}~\bibnamefont {Boezio}},
  \bibinfo {author} {\bibfnamefont {E.~A.}\ \bibnamefont {Bogomolov}}, \bibinfo
  {author} {\bibfnamefont {M.}~\bibnamefont {Bongi}}, \bibinfo {author}
  {\bibfnamefont {V.}~\bibnamefont {Bonvicini}}, \bibinfo {author}
  {\bibfnamefont {S.}~\bibnamefont {Borisov}}, \bibinfo {author} {\bibfnamefont
  {S.}~\bibnamefont {Bottai}}, \ and\ \bibinfo {author} {\bibnamefont
  {et~al.}},\ }\href {\doibase 10.1103/physrevlett.106.201101} {\bibfield
  {journal} {\bibinfo  {journal} {Physical Review Letters}\ }\textbf {\bibinfo
  {volume} {106}} (\bibinfo {year} {2011}{\natexlab{b}}),\
  10.1103/physrevlett.106.201101}\BibitemShut {NoStop}%
\bibitem [{\citenamefont {Aguilar}\ \emph
  {et~al.}(2019{\natexlab{a}})\citenamefont {Aguilar} \emph
  {et~al.}}]{Aguilar:2019ksn}%
  \BibitemOpen
  \bibfield  {author} {\bibinfo {author} {\bibfnamefont {M.}~\bibnamefont
  {Aguilar}} \emph {et~al.} (\bibinfo {collaboration} {AMS}),\ }\href {\doibase
  10.1103/PhysRevLett.122.101101} {\bibfield  {journal} {\bibinfo  {journal}
  {Phys. Rev. Lett.}\ }\textbf {\bibinfo {volume} {122}},\ \bibinfo {pages}
  {101101} (\bibinfo {year} {2019}{\natexlab{a}})}\BibitemShut {NoStop}%
\bibitem [{\citenamefont {{DAMPE Collaboration}}\ \emph
  {et~al.}(2017)\citenamefont {{DAMPE Collaboration}}, \citenamefont
  {{Ambrosi}} \emph {et~al.}}]{2017Natur.552...63D}%
  \BibitemOpen
  \bibfield  {author} {\bibinfo {author} {\bibnamefont {{DAMPE
  Collaboration}}}, \bibinfo {author} {\bibfnamefont {G.}~\bibnamefont
  {{Ambrosi}}},  \emph {et~al.},\ }\href {\doibase 10.1038/nature24475}
  {\bibfield  {journal} {\bibinfo  {journal} {\nat}\ }\textbf {\bibinfo
  {volume} {552}},\ \bibinfo {pages} {63} (\bibinfo {year} {2017})},\ \Eprint
  {http://arxiv.org/abs/1711.10981} {arXiv:1711.10981 [astro-ph.HE]}
  \BibitemShut {NoStop}%
\bibitem [{\citenamefont {Adriani}\ \emph {et~al.}(2018)\citenamefont
  {Adriani}, \citenamefont {Akaike}, \citenamefont {Asano}, \citenamefont
  {Asaoka}, \citenamefont {Bagliesi}, \citenamefont {Berti}, \citenamefont
  {Bigongiari}, \citenamefont {Binns}, \citenamefont {Bonechi}, \citenamefont
  {Bongi},\ and\ \citenamefont {et~al.}}]{Adriani_2018}%
  \BibitemOpen
  \bibfield  {author} {\bibinfo {author} {\bibfnamefont {O.}~\bibnamefont
  {Adriani}}, \bibinfo {author} {\bibfnamefont {Y.}~\bibnamefont {Akaike}},
  \bibinfo {author} {\bibfnamefont {K.}~\bibnamefont {Asano}}, \bibinfo
  {author} {\bibfnamefont {Y.}~\bibnamefont {Asaoka}}, \bibinfo {author}
  {\bibfnamefont {M.}~\bibnamefont {Bagliesi}}, \bibinfo {author}
  {\bibfnamefont {E.}~\bibnamefont {Berti}}, \bibinfo {author} {\bibfnamefont
  {G.}~\bibnamefont {Bigongiari}}, \bibinfo {author} {\bibfnamefont
  {W.}~\bibnamefont {Binns}}, \bibinfo {author} {\bibfnamefont
  {S.}~\bibnamefont {Bonechi}}, \bibinfo {author} {\bibfnamefont
  {M.}~\bibnamefont {Bongi}}, \ and\ \bibinfo {author} {\bibnamefont
  {et~al.}},\ }\href {\doibase 10.1103/physrevlett.120.261102} {\bibfield
  {journal} {\bibinfo  {journal} {Physical Review Letters}\ }\textbf {\bibinfo
  {volume} {120}} (\bibinfo {year} {2018}),\
  10.1103/physrevlett.120.261102}\BibitemShut {NoStop}%
\bibitem [{\citenamefont {Aguilar}\ \emph
  {et~al.}(2019{\natexlab{b}})\citenamefont {Aguilar}, \citenamefont
  {Ali~Cavasonza}, \citenamefont {Ambrosi} \emph
  {et~al.}}]{PhysRevLett.122.041102}%
  \BibitemOpen
  \bibfield  {author} {\bibinfo {author} {\bibfnamefont {M.}~\bibnamefont
  {Aguilar}}, \bibinfo {author} {\bibfnamefont {L.}~\bibnamefont
  {Ali~Cavasonza}}, \bibinfo {author} {\bibfnamefont {G.}~\bibnamefont
  {Ambrosi}},  \emph {et~al.} (\bibinfo {collaboration} {AMS Collaboration}),\
  }\href {\doibase 10.1103/PhysRevLett.122.041102} {\bibfield  {journal}
  {\bibinfo  {journal} {Phys. Rev. Lett.}\ }\textbf {\bibinfo {volume} {122}},\
  \bibinfo {pages} {041102} (\bibinfo {year} {2019}{\natexlab{b}})}\BibitemShut
  {NoStop}%
\bibitem [{\citenamefont {Aguilar}\ \emph
  {et~al.}(2016{\natexlab{b}})\citenamefont {Aguilar} \emph
  {et~al.}}]{Aguilar:2016kjl}%
  \BibitemOpen
  \bibfield  {author} {\bibinfo {author} {\bibfnamefont {M.}~\bibnamefont
  {Aguilar}} \emph {et~al.} (\bibinfo {collaboration} {AMS}),\ }\href {\doibase
  10.1103/PhysRevLett.117.091103} {\bibfield  {journal} {\bibinfo  {journal}
  {Phys. Rev. Lett.}\ }\textbf {\bibinfo {volume} {117}},\ \bibinfo {pages}
  {091103} (\bibinfo {year} {2016}{\natexlab{b}})}\BibitemShut {NoStop}%
\bibitem [{\citenamefont {Adriani}\ \emph {et~al.}(2010)\citenamefont
  {Adriani}, \citenamefont {Barbarino}, \citenamefont {Bazilevskaya},
  \citenamefont {Bellotti}, \citenamefont {Boezio}, \citenamefont {Bogomolov},
  \citenamefont {Bonechi}, \citenamefont {Bongi}, \citenamefont {Bonvicini},
  \citenamefont {Borisov},\ and\ \citenamefont {et~al.}}]{Adriani_2010}%
  \BibitemOpen
  \bibfield  {author} {\bibinfo {author} {\bibfnamefont {O.}~\bibnamefont
  {Adriani}}, \bibinfo {author} {\bibfnamefont {G.~C.}\ \bibnamefont
  {Barbarino}}, \bibinfo {author} {\bibfnamefont {G.~A.}\ \bibnamefont
  {Bazilevskaya}}, \bibinfo {author} {\bibfnamefont {R.}~\bibnamefont
  {Bellotti}}, \bibinfo {author} {\bibfnamefont {M.}~\bibnamefont {Boezio}},
  \bibinfo {author} {\bibfnamefont {E.~A.}\ \bibnamefont {Bogomolov}}, \bibinfo
  {author} {\bibfnamefont {L.}~\bibnamefont {Bonechi}}, \bibinfo {author}
  {\bibfnamefont {M.}~\bibnamefont {Bongi}}, \bibinfo {author} {\bibfnamefont
  {V.}~\bibnamefont {Bonvicini}}, \bibinfo {author} {\bibfnamefont
  {S.}~\bibnamefont {Borisov}}, \ and\ \bibinfo {author} {\bibnamefont
  {et~al.}},\ }\href {\doibase 10.1103/physrevlett.105.121101} {\bibfield
  {journal} {\bibinfo  {journal} {Physical Review Letters}\ }\textbf {\bibinfo
  {volume} {105}} (\bibinfo {year} {2010}),\
  10.1103/physrevlett.105.121101}\BibitemShut {NoStop}%
\bibitem [{\citenamefont {Korsmeier}\ and\ \citenamefont
  {Cuoco}(2016)}]{Korsmeier:2016kha}%
  \BibitemOpen
  \bibfield  {author} {\bibinfo {author} {\bibfnamefont {M.}~\bibnamefont
  {Korsmeier}}\ and\ \bibinfo {author} {\bibfnamefont {A.}~\bibnamefont
  {Cuoco}},\ }\href {\doibase 10.1103/PhysRevD.94.123019} {\bibfield  {journal}
  {\bibinfo  {journal} {Phys. Rev. D}\ }\textbf {\bibinfo {volume} {94}},\
  \bibinfo {pages} {123019} (\bibinfo {year} {2016})},\ \Eprint
  {http://arxiv.org/abs/1607.06093} {arXiv:1607.06093 [astro-ph.HE]}
  \BibitemShut {NoStop}%
\bibitem [{\citenamefont {Tomassetti}(2017)}]{Tomassetti:2017hbe}%
  \BibitemOpen
  \bibfield  {author} {\bibinfo {author} {\bibfnamefont {N.}~\bibnamefont
  {Tomassetti}},\ }\href {\doibase 10.1103/PhysRevD.96.103005} {\bibfield
  {journal} {\bibinfo  {journal} {Phys. Rev. D}\ }\textbf {\bibinfo {volume}
  {96}},\ \bibinfo {pages} {103005} (\bibinfo {year} {2017})},\ \Eprint
  {http://arxiv.org/abs/1707.06917} {arXiv:1707.06917 [astro-ph.HE]}
  \BibitemShut {NoStop}%
\bibitem [{\citenamefont {Liu}\ \emph {et~al.}(2018)\citenamefont {Liu},
  \citenamefont {Yao},\ and\ \citenamefont {Guo}}]{Liu:2018ujp}%
  \BibitemOpen
  \bibfield  {author} {\bibinfo {author} {\bibfnamefont {W.}~\bibnamefont
  {Liu}}, \bibinfo {author} {\bibfnamefont {Y.-h.}\ \bibnamefont {Yao}}, \ and\
  \bibinfo {author} {\bibfnamefont {Y.-Q.}\ \bibnamefont {Guo}},\ }\href
  {\doibase 10.3847/1538-4357/aaef39} {\bibfield  {journal} {\bibinfo
  {journal} {Astrophys. J.}\ }\textbf {\bibinfo {volume} {869}},\ \bibinfo
  {pages} {176} (\bibinfo {year} {2018})},\ \Eprint
  {http://arxiv.org/abs/1802.03602} {arXiv:1802.03602 [astro-ph.HE]}
  \BibitemShut {NoStop}%
\bibitem [{\citenamefont {G\'enolini}\ \emph {et~al.}(2019)\citenamefont
  {G\'enolini} \emph {et~al.}}]{Genolini:2019ewc}%
  \BibitemOpen
  \bibfield  {author} {\bibinfo {author} {\bibfnamefont {Y.}~\bibnamefont
  {G\'enolini}} \emph {et~al.},\ }\href {\doibase 10.1103/PhysRevD.99.123028}
  {\bibfield  {journal} {\bibinfo  {journal} {Phys. Rev. D}\ }\textbf {\bibinfo
  {volume} {99}},\ \bibinfo {pages} {123028} (\bibinfo {year} {2019})},\
  \Eprint {http://arxiv.org/abs/1904.08917} {arXiv:1904.08917 [astro-ph.HE]}
  \BibitemShut {NoStop}%
\bibitem [{\citenamefont {Weinrich}\ \emph
  {et~al.}(2020{\natexlab{a}})\citenamefont {Weinrich}, \citenamefont
  {G\'enolini}, \citenamefont {Boudaud}, \citenamefont {Derome},\ and\
  \citenamefont {Maurin}}]{Weinrich:2020cmw}%
  \BibitemOpen
  \bibfield  {author} {\bibinfo {author} {\bibfnamefont {N.}~\bibnamefont
  {Weinrich}}, \bibinfo {author} {\bibfnamefont {Y.}~\bibnamefont
  {G\'enolini}}, \bibinfo {author} {\bibfnamefont {M.}~\bibnamefont {Boudaud}},
  \bibinfo {author} {\bibfnamefont {L.}~\bibnamefont {Derome}}, \ and\ \bibinfo
  {author} {\bibfnamefont {D.}~\bibnamefont {Maurin}},\ }\href {\doibase
  10.1051/0004-6361/202037875} {\bibfield  {journal} {\bibinfo  {journal}
  {Astron. Astrophys.}\ }\textbf {\bibinfo {volume} {639}},\ \bibinfo {pages}
  {A131} (\bibinfo {year} {2020}{\natexlab{a}})},\ \Eprint
  {http://arxiv.org/abs/2002.11406} {arXiv:2002.11406 [astro-ph.HE]}
  \BibitemShut {NoStop}%
\bibitem [{\citenamefont {Weinrich}\ \emph
  {et~al.}(2020{\natexlab{b}})\citenamefont {Weinrich}, \citenamefont
  {Boudaud}, \citenamefont {Derome}, \citenamefont {Genolini}, \citenamefont
  {Lavalle}, \citenamefont {Maurin}, \citenamefont {Salati}, \citenamefont
  {Serpico},\ and\ \citenamefont {Weymann-Despres}}]{Weinrich:2020ftb}%
  \BibitemOpen
  \bibfield  {author} {\bibinfo {author} {\bibfnamefont {N.}~\bibnamefont
  {Weinrich}}, \bibinfo {author} {\bibfnamefont {M.}~\bibnamefont {Boudaud}},
  \bibinfo {author} {\bibfnamefont {L.}~\bibnamefont {Derome}}, \bibinfo
  {author} {\bibfnamefont {Y.}~\bibnamefont {Genolini}}, \bibinfo {author}
  {\bibfnamefont {J.}~\bibnamefont {Lavalle}}, \bibinfo {author} {\bibfnamefont
  {D.}~\bibnamefont {Maurin}}, \bibinfo {author} {\bibfnamefont
  {P.}~\bibnamefont {Salati}}, \bibinfo {author} {\bibfnamefont
  {P.}~\bibnamefont {Serpico}}, \ and\ \bibinfo {author} {\bibfnamefont
  {G.}~\bibnamefont {Weymann-Despres}},\ }\href {\doibase
  10.1051/0004-6361/202038064} {\bibfield  {journal} {\bibinfo  {journal}
  {Astron. Astrophys.}\ }\textbf {\bibinfo {volume} {639}},\ \bibinfo {pages}
  {A74} (\bibinfo {year} {2020}{\natexlab{b}})},\ \Eprint
  {http://arxiv.org/abs/2004.00441} {arXiv:2004.00441 [astro-ph.HE]}
  \BibitemShut {NoStop}%
\bibitem [{\citenamefont {Evoli}\ \emph {et~al.}(2019)\citenamefont {Evoli},
  \citenamefont {Aloisio},\ and\ \citenamefont {Blasi}}]{Evoli:2019wwu}%
  \BibitemOpen
  \bibfield  {author} {\bibinfo {author} {\bibfnamefont {C.}~\bibnamefont
  {Evoli}}, \bibinfo {author} {\bibfnamefont {R.}~\bibnamefont {Aloisio}}, \
  and\ \bibinfo {author} {\bibfnamefont {P.}~\bibnamefont {Blasi}},\ }\href
  {\doibase 10.1103/PhysRevD.99.103023} {\bibfield  {journal} {\bibinfo
  {journal} {Phys. Rev. D}\ }\textbf {\bibinfo {volume} {99}},\ \bibinfo
  {pages} {103023} (\bibinfo {year} {2019})},\ \Eprint
  {http://arxiv.org/abs/1904.10220} {arXiv:1904.10220 [astro-ph.HE]}
  \BibitemShut {NoStop}%
\bibitem [{\citenamefont {Evoli}\ \emph {et~al.}(2020)\citenamefont {Evoli},
  \citenamefont {Morlino}, \citenamefont {Blasi},\ and\ \citenamefont
  {Aloisio}}]{Evoli:2019iih}%
  \BibitemOpen
  \bibfield  {author} {\bibinfo {author} {\bibfnamefont {C.}~\bibnamefont
  {Evoli}}, \bibinfo {author} {\bibfnamefont {G.}~\bibnamefont {Morlino}},
  \bibinfo {author} {\bibfnamefont {P.}~\bibnamefont {Blasi}}, \ and\ \bibinfo
  {author} {\bibfnamefont {R.}~\bibnamefont {Aloisio}},\ }\href {\doibase
  10.1103/PhysRevD.101.023013} {\bibfield  {journal} {\bibinfo  {journal}
  {Phys. Rev. D}\ }\textbf {\bibinfo {volume} {101}},\ \bibinfo {pages}
  {023013} (\bibinfo {year} {2020})},\ \Eprint
  {http://arxiv.org/abs/1910.04113} {arXiv:1910.04113 [astro-ph.HE]}
  \BibitemShut {NoStop}%
\bibitem [{\citenamefont {Boschini}\ \emph {et~al.}(2018)\citenamefont
  {Boschini} \emph {et~al.}}]{Boschini:2018baj}%
  \BibitemOpen
  \bibfield  {author} {\bibinfo {author} {\bibfnamefont {M.~J.}\ \bibnamefont
  {Boschini}} \emph {et~al.},\ }\href {\doibase 10.3847/1538-4357/aabc54}
  {\bibfield  {journal} {\bibinfo  {journal} {Astrophys. J.}\ }\textbf
  {\bibinfo {volume} {858}},\ \bibinfo {pages} {61} (\bibinfo {year} {2018})},\
  \Eprint {http://arxiv.org/abs/1804.06956} {arXiv:1804.06956 [astro-ph.HE]}
  \BibitemShut {NoStop}%
\bibitem [{\citenamefont {Boschini}\ \emph
  {et~al.}(2020{\natexlab{a}})\citenamefont {Boschini} \emph
  {et~al.}}]{Boschini:2019gow}%
  \BibitemOpen
  \bibfield  {author} {\bibinfo {author} {\bibfnamefont {M.~J.}\ \bibnamefont
  {Boschini}} \emph {et~al.},\ }\href {\doibase 10.3847/1538-4357/ab64f1}
  {\bibfield  {journal} {\bibinfo  {journal} {Astrophys. J.}\ }\textbf
  {\bibinfo {volume} {889}},\ \bibinfo {pages} {167} (\bibinfo {year}
  {2020}{\natexlab{a}})},\ \Eprint {http://arxiv.org/abs/1911.03108}
  {arXiv:1911.03108 [astro-ph.HE]} \BibitemShut {NoStop}%
\bibitem [{\citenamefont {Boschini}\ \emph
  {et~al.}(2020{\natexlab{b}})\citenamefont {Boschini} \emph
  {et~al.}}]{Boschini:2020jty}%
  \BibitemOpen
  \bibfield  {author} {\bibinfo {author} {\bibfnamefont {M.~J.}\ \bibnamefont
  {Boschini}} \emph {et~al.},\ }\href {\doibase 10.3847/1538-4365/aba901}
  {\bibfield  {journal} {\bibinfo  {journal} {Astrophys. J. Suppl.}\ }\textbf
  {\bibinfo {volume} {250}},\ \bibinfo {pages} {27} (\bibinfo {year}
  {2020}{\natexlab{b}})},\ \Eprint {http://arxiv.org/abs/2006.01337}
  {arXiv:2006.01337 [astro-ph.HE]} \BibitemShut {NoStop}%
\bibitem [{\citenamefont {Di~Mauro}\ and\ \citenamefont
  {Winkler}(2021)}]{Di_Mauro_2021}%
  \BibitemOpen
  \bibfield  {author} {\bibinfo {author} {\bibfnamefont {M.}~\bibnamefont
  {Di~Mauro}}\ and\ \bibinfo {author} {\bibfnamefont {M.~W.}\ \bibnamefont
  {Winkler}},\ }\href {\doibase 10.1103/physrevd.103.123005} {\bibfield
  {journal} {\bibinfo  {journal} {Physical Review D}\ }\textbf {\bibinfo
  {volume} {103}} (\bibinfo {year} {2021}),\
  10.1103/physrevd.103.123005}\BibitemShut {NoStop}%
\bibitem [{\citenamefont {Luque}\ \emph {et~al.}(2021)\citenamefont {Luque},
  \citenamefont {Mazziotta}, \citenamefont {Loparco}, \citenamefont {Gargano},\
  and\ \citenamefont {Serini}}]{Luque:2021nxb}%
  \BibitemOpen
  \bibfield  {author} {\bibinfo {author} {\bibfnamefont {P.~D. L.~T.}\
  \bibnamefont {Luque}}, \bibinfo {author} {\bibfnamefont {M.}~\bibnamefont
  {Mazziotta}}, \bibinfo {author} {\bibfnamefont {F.}~\bibnamefont {Loparco}},
  \bibinfo {author} {\bibfnamefont {F.}~\bibnamefont {Gargano}}, \ and\
  \bibinfo {author} {\bibfnamefont {D.}~\bibnamefont {Serini}},\ }\href
  {\doibase 10.1088/1475-7516/2021/07/010} {\bibfield  {journal} {\bibinfo
  {journal} {Journal of Cosmology and Astroparticle Physics}\ }\textbf
  {\bibinfo {volume} {2021}},\ \bibinfo {pages} {010} (\bibinfo {year}
  {2021})}\BibitemShut {NoStop}%
\bibitem [{\citenamefont {De~La Torre~Luque}\ \emph {et~al.}(2021)\citenamefont
  {De~La Torre~Luque}, \citenamefont {Mazziotta}, \citenamefont {Loparco},
  \citenamefont {Gargano},\ and\ \citenamefont
  {Serini}}]{DeLaTorreLuque:2021yfq}%
  \BibitemOpen
  \bibfield  {author} {\bibinfo {author} {\bibfnamefont {P.}~\bibnamefont
  {De~La Torre~Luque}}, \bibinfo {author} {\bibfnamefont {M.~N.}\ \bibnamefont
  {Mazziotta}}, \bibinfo {author} {\bibfnamefont {F.}~\bibnamefont {Loparco}},
  \bibinfo {author} {\bibfnamefont {F.}~\bibnamefont {Gargano}}, \ and\
  \bibinfo {author} {\bibfnamefont {D.}~\bibnamefont {Serini}},\ }\href
  {\doibase 10.1088/1475-7516/2021/03/099} {\bibfield  {journal} {\bibinfo
  {journal} {JCAP}\ }\textbf {\bibinfo {volume} {03}},\ \bibinfo {pages} {099}
  (\bibinfo {year} {2021})},\ \Eprint {http://arxiv.org/abs/2101.01547}
  {arXiv:2101.01547 [astro-ph.HE]} \BibitemShut {NoStop}%
\bibitem [{\citenamefont {Schroer}\ \emph {et~al.}(2021)\citenamefont
  {Schroer}, \citenamefont {Evoli},\ and\ \citenamefont
  {Blasi}}]{Schroer:2021ojh}%
  \BibitemOpen
  \bibfield  {author} {\bibinfo {author} {\bibfnamefont {B.}~\bibnamefont
  {Schroer}}, \bibinfo {author} {\bibfnamefont {C.}~\bibnamefont {Evoli}}, \
  and\ \bibinfo {author} {\bibfnamefont {P.}~\bibnamefont {Blasi}},\ }\href
  {\doibase 10.1103/PhysRevD.103.123010} {\bibfield  {journal} {\bibinfo
  {journal} {Phys. Rev. D}\ }\textbf {\bibinfo {volume} {103}},\ \bibinfo
  {pages} {123010} (\bibinfo {year} {2021})},\ \Eprint
  {http://arxiv.org/abs/2102.12576} {arXiv:2102.12576 [astro-ph.HE]}
  \BibitemShut {NoStop}%
\bibitem [{\citenamefont {Korsmeier}\ and\ \citenamefont
  {Cuoco}(2021{\natexlab{a}})}]{Korsmeier:2021brc}%
  \BibitemOpen
  \bibfield  {author} {\bibinfo {author} {\bibfnamefont {M.}~\bibnamefont
  {Korsmeier}}\ and\ \bibinfo {author} {\bibfnamefont {A.}~\bibnamefont
  {Cuoco}},\ }\href {\doibase 10.1103/PhysRevD.103.103016} {\bibfield
  {journal} {\bibinfo  {journal} {Phys. Rev. D}\ }\textbf {\bibinfo {volume}
  {103}},\ \bibinfo {pages} {103016} (\bibinfo {year} {2021}{\natexlab{a}})},\
  \Eprint {http://arxiv.org/abs/2103.09824} {arXiv:2103.09824 [astro-ph.HE]}
  \BibitemShut {NoStop}%
\bibitem [{\citenamefont {Korsmeier}\ and\ \citenamefont
  {Cuoco}(2021{\natexlab{b}})}]{Korsmeier:2021bkw}%
  \BibitemOpen
  \bibfield  {author} {\bibinfo {author} {\bibfnamefont {M.}~\bibnamefont
  {Korsmeier}}\ and\ \bibinfo {author} {\bibfnamefont {A.}~\bibnamefont
  {Cuoco}},\ }\href@noop {} {\  (\bibinfo {year} {2021}{\natexlab{b}})},\
  \Eprint {http://arxiv.org/abs/2112.08381} {arXiv:2112.08381 [astro-ph.HE]}
  \BibitemShut {NoStop}%
\bibitem [{\citenamefont {Hooper}\ \emph {et~al.}(2009)\citenamefont {Hooper},
  \citenamefont {Blasi},\ and\ \citenamefont {Serpico}}]{Hooper:2008kg}%
  \BibitemOpen
  \bibfield  {author} {\bibinfo {author} {\bibfnamefont {D.}~\bibnamefont
  {Hooper}}, \bibinfo {author} {\bibfnamefont {P.}~\bibnamefont {Blasi}}, \
  and\ \bibinfo {author} {\bibfnamefont {P.~D.}\ \bibnamefont {Serpico}},\
  }\href {\doibase 10.1088/1475-7516/2009/01/025} {\bibfield  {journal}
  {\bibinfo  {journal} {JCAP}\ }\textbf {\bibinfo {volume} {01}},\ \bibinfo
  {pages} {025} (\bibinfo {year} {2009})},\ \Eprint
  {http://arxiv.org/abs/0810.1527} {arXiv:0810.1527 [astro-ph]} \BibitemShut
  {NoStop}%
\bibitem [{\citenamefont {Ahlers}\ \emph {et~al.}(2009)\citenamefont {Ahlers},
  \citenamefont {Mertsch},\ and\ \citenamefont {Sarkar}}]{Ahlers:2009ae}%
  \BibitemOpen
  \bibfield  {author} {\bibinfo {author} {\bibfnamefont {M.}~\bibnamefont
  {Ahlers}}, \bibinfo {author} {\bibfnamefont {P.}~\bibnamefont {Mertsch}}, \
  and\ \bibinfo {author} {\bibfnamefont {S.}~\bibnamefont {Sarkar}},\ }\href
  {\doibase 10.1103/PhysRevD.80.123017} {\bibfield  {journal} {\bibinfo
  {journal} {Phys. Rev. D}\ }\textbf {\bibinfo {volume} {80}},\ \bibinfo
  {pages} {123017} (\bibinfo {year} {2009})},\ \Eprint
  {http://arxiv.org/abs/0909.4060} {arXiv:0909.4060 [astro-ph.HE]} \BibitemShut
  {NoStop}%
\bibitem [{\citenamefont {Boudaud}\ \emph {et~al.}(2015)\citenamefont {Boudaud}
  \emph {et~al.}}]{Boudaud:2014dta}%
  \BibitemOpen
  \bibfield  {author} {\bibinfo {author} {\bibfnamefont {M.}~\bibnamefont
  {Boudaud}} \emph {et~al.},\ }\href {\doibase 10.1051/0004-6361/201425197}
  {\bibfield  {journal} {\bibinfo  {journal} {Astron. Astrophys.}\ }\textbf
  {\bibinfo {volume} {575}},\ \bibinfo {pages} {A67} (\bibinfo {year}
  {2015})},\ \Eprint {http://arxiv.org/abs/1410.3799} {arXiv:1410.3799
  [astro-ph.HE]} \BibitemShut {NoStop}%
\bibitem [{\citenamefont {Boudaud}\ \emph {et~al.}(2017)\citenamefont
  {Boudaud}, \citenamefont {Bueno}, \citenamefont {Caroff}, \citenamefont
  {Genolini}, \citenamefont {Poulin}, \citenamefont {Poireau}, \citenamefont
  {Putze}, \citenamefont {Rosier}, \citenamefont {Salati},\ and\ \citenamefont
  {Vecchi}}]{Boudaud:2016jvj}%
  \BibitemOpen
  \bibfield  {author} {\bibinfo {author} {\bibfnamefont {M.}~\bibnamefont
  {Boudaud}}, \bibinfo {author} {\bibfnamefont {E.~F.}\ \bibnamefont {Bueno}},
  \bibinfo {author} {\bibfnamefont {S.}~\bibnamefont {Caroff}}, \bibinfo
  {author} {\bibfnamefont {Y.}~\bibnamefont {Genolini}}, \bibinfo {author}
  {\bibfnamefont {V.}~\bibnamefont {Poulin}}, \bibinfo {author} {\bibfnamefont
  {V.}~\bibnamefont {Poireau}}, \bibinfo {author} {\bibfnamefont
  {A.}~\bibnamefont {Putze}}, \bibinfo {author} {\bibfnamefont
  {S.}~\bibnamefont {Rosier}}, \bibinfo {author} {\bibfnamefont
  {P.}~\bibnamefont {Salati}}, \ and\ \bibinfo {author} {\bibfnamefont
  {M.}~\bibnamefont {Vecchi}},\ }\href {\doibase 10.1051/0004-6361/201630321}
  {\bibfield  {journal} {\bibinfo  {journal} {Astron. Astrophys.}\ }\textbf
  {\bibinfo {volume} {605}},\ \bibinfo {pages} {A17} (\bibinfo {year}
  {2017})},\ \Eprint {http://arxiv.org/abs/1612.03924} {arXiv:1612.03924
  [astro-ph.HE]} \BibitemShut {NoStop}%
\bibitem [{\citenamefont {Manconi}\ \emph {et~al.}(2017)\citenamefont
  {Manconi}, \citenamefont {Mauro},\ and\ \citenamefont
  {Donato}}]{Manconi_2017}%
  \BibitemOpen
  \bibfield  {author} {\bibinfo {author} {\bibfnamefont {S.}~\bibnamefont
  {Manconi}}, \bibinfo {author} {\bibfnamefont {M.~D.}\ \bibnamefont {Mauro}},
  \ and\ \bibinfo {author} {\bibfnamefont {F.}~\bibnamefont {Donato}},\ }\href
  {\doibase 10.1088/1475-7516/2017/01/006} {\bibfield  {journal} {\bibinfo
  {journal} {Journal of Cosmology and Astroparticle Physics}\ }\textbf
  {\bibinfo {volume} {2017}},\ \bibinfo {pages} {006–006} (\bibinfo {year}
  {2017})}\BibitemShut {NoStop}%
\bibitem [{\citenamefont {Manconi}\ \emph {et~al.}(2019)\citenamefont
  {Manconi}, \citenamefont {Di~Mauro},\ and\ \citenamefont
  {Donato}}]{Manconi:2018azw}%
  \BibitemOpen
  \bibfield  {author} {\bibinfo {author} {\bibfnamefont {S.}~\bibnamefont
  {Manconi}}, \bibinfo {author} {\bibfnamefont {M.}~\bibnamefont {Di~Mauro}}, \
  and\ \bibinfo {author} {\bibfnamefont {F.}~\bibnamefont {Donato}},\ }\href
  {\doibase 10.1088/1475-7516/2019/04/024} {\bibfield  {journal} {\bibinfo
  {journal} {JCAP}\ }\textbf {\bibinfo {volume} {04}},\ \bibinfo {pages} {024}
  (\bibinfo {year} {2019})},\ \Eprint {http://arxiv.org/abs/1803.01009}
  {arXiv:1803.01009 [astro-ph.HE]} \BibitemShut {NoStop}%
\bibitem [{\citenamefont {Fornieri}\ \emph {et~al.}(2020)\citenamefont
  {Fornieri}, \citenamefont {Gaggero},\ and\ \citenamefont
  {Grasso}}]{Fornieri_2020}%
  \BibitemOpen
  \bibfield  {author} {\bibinfo {author} {\bibfnamefont {O.}~\bibnamefont
  {Fornieri}}, \bibinfo {author} {\bibfnamefont {D.}~\bibnamefont {Gaggero}}, \
  and\ \bibinfo {author} {\bibfnamefont {D.}~\bibnamefont {Grasso}},\ }\href
  {\doibase 10.1088/1475-7516/2020/02/009} {\bibfield  {journal} {\bibinfo
  {journal} {Journal of Cosmology and Astroparticle Physics}\ }\textbf
  {\bibinfo {volume} {2020}},\ \bibinfo {pages} {009} (\bibinfo {year}
  {2020})}\BibitemShut {NoStop}%
\bibitem [{\citenamefont {Manconi}\ \emph {et~al.}(2020)\citenamefont
  {Manconi}, \citenamefont {Di~Mauro},\ and\ \citenamefont
  {Donato}}]{Manconi:2020ipm}%
  \BibitemOpen
  \bibfield  {author} {\bibinfo {author} {\bibfnamefont {S.}~\bibnamefont
  {Manconi}}, \bibinfo {author} {\bibfnamefont {M.}~\bibnamefont {Di~Mauro}}, \
  and\ \bibinfo {author} {\bibfnamefont {F.}~\bibnamefont {Donato}},\ }\href
  {\doibase 10.1103/PhysRevD.102.023015} {\bibfield  {journal} {\bibinfo
  {journal} {Phys. Rev. D}\ }\textbf {\bibinfo {volume} {102}},\ \bibinfo
  {pages} {023015} (\bibinfo {year} {2020})},\ \Eprint
  {http://arxiv.org/abs/2001.09985} {arXiv:2001.09985 [astro-ph.HE]}
  \BibitemShut {NoStop}%
\bibitem [{\citenamefont {Di~Mauro}\ \emph {et~al.}(2019)\citenamefont
  {Di~Mauro}, \citenamefont {Manconi},\ and\ \citenamefont
  {Donato}}]{DiMauro:2019yvh}%
  \BibitemOpen
  \bibfield  {author} {\bibinfo {author} {\bibfnamefont {M.}~\bibnamefont
  {Di~Mauro}}, \bibinfo {author} {\bibfnamefont {S.}~\bibnamefont {Manconi}}, \
  and\ \bibinfo {author} {\bibfnamefont {F.}~\bibnamefont {Donato}},\ }\href
  {\doibase 10.1103/PhysRevD.100.123015} {\bibfield  {journal} {\bibinfo
  {journal} {Phys. Rev. D}\ }\textbf {\bibinfo {volume} {100}},\ \bibinfo
  {pages} {123015} (\bibinfo {year} {2019})},\ \Eprint
  {http://arxiv.org/abs/1903.05647} {arXiv:1903.05647 [astro-ph.HE]}
  \BibitemShut {NoStop}%
\bibitem [{\citenamefont {Orusa}\ \emph {et~al.}(2021)\citenamefont {Orusa},
  \citenamefont {Manconi}, \citenamefont {Donato},\ and\ \citenamefont
  {Di~Mauro}}]{Orusa_2021}%
  \BibitemOpen
  \bibfield  {author} {\bibinfo {author} {\bibfnamefont {L.}~\bibnamefont
  {Orusa}}, \bibinfo {author} {\bibfnamefont {S.}~\bibnamefont {Manconi}},
  \bibinfo {author} {\bibfnamefont {F.}~\bibnamefont {Donato}}, \ and\ \bibinfo
  {author} {\bibfnamefont {M.}~\bibnamefont {Di~Mauro}},\ }\href {\doibase
  10.1088/1475-7516/2021/12/014} {\bibfield  {journal} {\bibinfo  {journal}
  {Journal of Cosmology and Astroparticle Physics}\ }\textbf {\bibinfo {volume}
  {2021}},\ \bibinfo {pages} {014} (\bibinfo {year} {2021})}\BibitemShut
  {NoStop}%
\bibitem [{\citenamefont {Evoli}\ \emph {et~al.}(2021)\citenamefont {Evoli},
  \citenamefont {Amato}, \citenamefont {Blasi},\ and\ \citenamefont
  {Aloisio}}]{Evoli_2021}%
  \BibitemOpen
  \bibfield  {author} {\bibinfo {author} {\bibfnamefont {C.}~\bibnamefont
  {Evoli}}, \bibinfo {author} {\bibfnamefont {E.}~\bibnamefont {Amato}},
  \bibinfo {author} {\bibfnamefont {P.}~\bibnamefont {Blasi}}, \ and\ \bibinfo
  {author} {\bibfnamefont {R.}~\bibnamefont {Aloisio}},\ }\href {\doibase
  10.1103/physrevd.103.083010} {\bibfield  {journal} {\bibinfo  {journal}
  {Physical Review D}\ }\textbf {\bibinfo {volume} {103}} (\bibinfo {year}
  {2021}),\ 10.1103/physrevd.103.083010}\BibitemShut {NoStop}%
\bibitem [{\citenamefont {Diesing}\ and\ \citenamefont
  {Caprioli}(2020)}]{Diesing:2020jtm}%
  \BibitemOpen
  \bibfield  {author} {\bibinfo {author} {\bibfnamefont {R.}~\bibnamefont
  {Diesing}}\ and\ \bibinfo {author} {\bibfnamefont {D.}~\bibnamefont
  {Caprioli}},\ }\href {\doibase 10.1103/PhysRevD.101.103030} {\bibfield
  {journal} {\bibinfo  {journal} {Phys. Rev. D}\ }\textbf {\bibinfo {volume}
  {101}},\ \bibinfo {pages} {103030} (\bibinfo {year} {2020})},\ \Eprint
  {http://arxiv.org/abs/2001.02240} {arXiv:2001.02240 [astro-ph.HE]}
  \BibitemShut {NoStop}%
\bibitem [{\citenamefont {Cholis}\ \emph {et~al.}(2018)\citenamefont {Cholis},
  \citenamefont {Karwal},\ and\ \citenamefont {Kamionkowski}}]{Cholis_2018}%
  \BibitemOpen
  \bibfield  {author} {\bibinfo {author} {\bibfnamefont {I.}~\bibnamefont
  {Cholis}}, \bibinfo {author} {\bibfnamefont {T.}~\bibnamefont {Karwal}}, \
  and\ \bibinfo {author} {\bibfnamefont {M.}~\bibnamefont {Kamionkowski}},\
  }\href {\doibase 10.1103/physrevd.98.063008} {\bibfield  {journal} {\bibinfo
  {journal} {Physical Review D}\ }\textbf {\bibinfo {volume} {98}} (\bibinfo
  {year} {2018}),\ 10.1103/physrevd.98.063008}\BibitemShut {NoStop}%
\bibitem [{\citenamefont {Cholis}\ and\ \citenamefont
  {Krommydas}(2022)}]{Cholis:2021kqk}%
  \BibitemOpen
  \bibfield  {author} {\bibinfo {author} {\bibfnamefont {I.}~\bibnamefont
  {Cholis}}\ and\ \bibinfo {author} {\bibfnamefont {I.}~\bibnamefont
  {Krommydas}},\ }\href {\doibase 10.1103/PhysRevD.105.023015} {\bibfield
  {journal} {\bibinfo  {journal} {Phys. Rev. D}\ }\textbf {\bibinfo {volume}
  {105}},\ \bibinfo {pages} {023015} (\bibinfo {year} {2022})},\ \Eprint
  {http://arxiv.org/abs/2111.05864} {arXiv:2111.05864 [astro-ph.HE]}
  \BibitemShut {NoStop}%
\bibitem [{\citenamefont {Cirelli}\ \emph {et~al.}(2008)\citenamefont
  {Cirelli}, \citenamefont {Franceschini},\ and\ \citenamefont
  {Strumia}}]{Cirelli:2008id}%
  \BibitemOpen
  \bibfield  {author} {\bibinfo {author} {\bibfnamefont {M.}~\bibnamefont
  {Cirelli}}, \bibinfo {author} {\bibfnamefont {R.}~\bibnamefont
  {Franceschini}}, \ and\ \bibinfo {author} {\bibfnamefont {A.}~\bibnamefont
  {Strumia}},\ }\href {\doibase 10.1016/j.nuclphysb.2008.03.013} {\bibfield
  {journal} {\bibinfo  {journal} {Nucl. Phys. B}\ }\textbf {\bibinfo {volume}
  {800}},\ \bibinfo {pages} {204} (\bibinfo {year} {2008})},\ \Eprint
  {http://arxiv.org/abs/0802.3378} {arXiv:0802.3378 [hep-ph]} \BibitemShut
  {NoStop}%
\bibitem [{\citenamefont {Bergstrom}\ \emph {et~al.}(2013)\citenamefont
  {Bergstrom}, \citenamefont {Bringmann}, \citenamefont {Cholis}, \citenamefont
  {Hooper},\ and\ \citenamefont {Weniger}}]{Bergstrom:2013jra}%
  \BibitemOpen
  \bibfield  {author} {\bibinfo {author} {\bibfnamefont {L.}~\bibnamefont
  {Bergstrom}}, \bibinfo {author} {\bibfnamefont {T.}~\bibnamefont
  {Bringmann}}, \bibinfo {author} {\bibfnamefont {I.}~\bibnamefont {Cholis}},
  \bibinfo {author} {\bibfnamefont {D.}~\bibnamefont {Hooper}}, \ and\ \bibinfo
  {author} {\bibfnamefont {C.}~\bibnamefont {Weniger}},\ }\href {\doibase
  10.1103/PhysRevLett.111.171101} {\bibfield  {journal} {\bibinfo  {journal}
  {Phys. Rev. Lett.}\ }\textbf {\bibinfo {volume} {111}},\ \bibinfo {pages}
  {171101} (\bibinfo {year} {2013})},\ \Eprint {http://arxiv.org/abs/1306.3983}
  {arXiv:1306.3983 [astro-ph.HE]} \BibitemShut {NoStop}%
\bibitem [{\citenamefont {Di~Mauro}\ \emph {et~al.}(2016)\citenamefont
  {Di~Mauro}, \citenamefont {Donato}, \citenamefont {Fornengo},\ and\
  \citenamefont {Vittino}}]{DiMauro:2015jxa}%
  \BibitemOpen
  \bibfield  {author} {\bibinfo {author} {\bibfnamefont {M.}~\bibnamefont
  {Di~Mauro}}, \bibinfo {author} {\bibfnamefont {F.}~\bibnamefont {Donato}},
  \bibinfo {author} {\bibfnamefont {N.}~\bibnamefont {Fornengo}}, \ and\
  \bibinfo {author} {\bibfnamefont {A.}~\bibnamefont {Vittino}},\ }\href
  {\doibase 10.1088/1475-7516/2016/05/031} {\bibfield  {journal} {\bibinfo
  {journal} {JCAP}\ }\textbf {\bibinfo {volume} {05}},\ \bibinfo {pages} {031}
  (\bibinfo {year} {2016})},\ \Eprint {http://arxiv.org/abs/1507.07001}
  {arXiv:1507.07001 [astro-ph.HE]} \BibitemShut {NoStop}%
\bibitem [{\citenamefont {Delahaye}\ \emph {et~al.}(2009)\citenamefont
  {Delahaye}, \citenamefont {Lineros}, \citenamefont {Donato}, \citenamefont
  {Fornengo}, \citenamefont {Lavalle}, \citenamefont {Salati},\ and\
  \citenamefont {Taillet}}]{Delahaye_2009}%
  \BibitemOpen
  \bibfield  {author} {\bibinfo {author} {\bibfnamefont {T.}~\bibnamefont
  {Delahaye}}, \bibinfo {author} {\bibfnamefont {R.}~\bibnamefont {Lineros}},
  \bibinfo {author} {\bibfnamefont {F.}~\bibnamefont {Donato}}, \bibinfo
  {author} {\bibfnamefont {N.}~\bibnamefont {Fornengo}}, \bibinfo {author}
  {\bibfnamefont {J.}~\bibnamefont {Lavalle}}, \bibinfo {author} {\bibfnamefont
  {P.}~\bibnamefont {Salati}}, \ and\ \bibinfo {author} {\bibfnamefont
  {R.}~\bibnamefont {Taillet}},\ }\href {\doibase 10.1051/0004-6361/200811130}
  {\bibfield  {journal} {\bibinfo  {journal} {Astronomy \& Astrophysics}\
  }\textbf {\bibinfo {volume} {501}},\ \bibinfo {pages} {821–833} (\bibinfo
  {year} {2009})}\BibitemShut {NoStop}%
\bibitem [{\citenamefont {Korsmeier}\ \emph {et~al.}(2018)\citenamefont
  {Korsmeier}, \citenamefont {Donato},\ and\ \citenamefont
  {Di~Mauro}}]{Korsmeier_2018}%
  \BibitemOpen
  \bibfield  {author} {\bibinfo {author} {\bibfnamefont {M.}~\bibnamefont
  {Korsmeier}}, \bibinfo {author} {\bibfnamefont {F.}~\bibnamefont {Donato}}, \
  and\ \bibinfo {author} {\bibfnamefont {M.}~\bibnamefont {Di~Mauro}},\ }\href
  {\doibase 10.1103/physrevd.97.103019} {\bibfield  {journal} {\bibinfo
  {journal} {Physical Review D}\ }\textbf {\bibinfo {volume} {97}} (\bibinfo
  {year} {2018}),\ 10.1103/physrevd.97.103019}\BibitemShut {NoStop}%
\bibitem [{\citenamefont {Tan}\ and\ \citenamefont {Ng}(1983)}]{Tan:1984ha}%
  \BibitemOpen
  \bibfield  {author} {\bibinfo {author} {\bibfnamefont {L.~C.}\ \bibnamefont
  {Tan}}\ and\ \bibinfo {author} {\bibfnamefont {L.~K.}\ \bibnamefont {Ng}},\
  }\href {\doibase 10.1088/0305-4616/9/10/015} {\bibfield  {journal} {\bibinfo
  {journal} {J. Phys. G}\ }\textbf {\bibinfo {volume} {9}},\ \bibinfo {pages}
  {1289} (\bibinfo {year} {1983})}\BibitemShut {NoStop}%
\bibitem [{\citenamefont {Blattnig}\ \emph {et~al.}(2000)\citenamefont
  {Blattnig}, \citenamefont {Swaminathan}, \citenamefont {Kruger},
  \citenamefont {Ngom},\ and\ \citenamefont {Norbury}}]{Blattnig:2000zf}%
  \BibitemOpen
  \bibfield  {author} {\bibinfo {author} {\bibfnamefont {S.~R.}\ \bibnamefont
  {Blattnig}}, \bibinfo {author} {\bibfnamefont {S.~R.}\ \bibnamefont
  {Swaminathan}}, \bibinfo {author} {\bibfnamefont {A.~T.}\ \bibnamefont
  {Kruger}}, \bibinfo {author} {\bibfnamefont {M.}~\bibnamefont {Ngom}}, \ and\
  \bibinfo {author} {\bibfnamefont {J.~W.}\ \bibnamefont {Norbury}},\ }\href
  {\doibase 10.1103/PhysRevD.62.094030} {\bibfield  {journal} {\bibinfo
  {journal} {Phys. Rev. D}\ }\textbf {\bibinfo {volume} {62}},\ \bibinfo
  {pages} {094030} (\bibinfo {year} {2000})},\ \Eprint
  {http://arxiv.org/abs/hep-ph/0010170} {arXiv:hep-ph/0010170} \BibitemShut
  {NoStop}%
\bibitem [{\citenamefont {Sj\"ostrand}\ \emph {et~al.}(2015)\citenamefont
  {Sj\"ostrand}, \citenamefont {Ask}, \citenamefont {Christiansen},
  \citenamefont {Corke}, \citenamefont {Desai}, \citenamefont {Ilten},
  \citenamefont {Mrenna}, \citenamefont {Prestel}, \citenamefont {Rasmussen},\
  and\ \citenamefont {Skands}}]{Sjostrand:2014zea}%
  \BibitemOpen
  \bibfield  {author} {\bibinfo {author} {\bibfnamefont {T.}~\bibnamefont
  {Sj\"ostrand}}, \bibinfo {author} {\bibfnamefont {S.}~\bibnamefont {Ask}},
  \bibinfo {author} {\bibfnamefont {J.~R.}\ \bibnamefont {Christiansen}},
  \bibinfo {author} {\bibfnamefont {R.}~\bibnamefont {Corke}}, \bibinfo
  {author} {\bibfnamefont {N.}~\bibnamefont {Desai}}, \bibinfo {author}
  {\bibfnamefont {P.}~\bibnamefont {Ilten}}, \bibinfo {author} {\bibfnamefont
  {S.}~\bibnamefont {Mrenna}}, \bibinfo {author} {\bibfnamefont
  {S.}~\bibnamefont {Prestel}}, \bibinfo {author} {\bibfnamefont {C.~O.}\
  \bibnamefont {Rasmussen}}, \ and\ \bibinfo {author} {\bibfnamefont {P.~Z.}\
  \bibnamefont {Skands}},\ }\href {\doibase 10.1016/j.cpc.2015.01.024}
  {\bibfield  {journal} {\bibinfo  {journal} {Comput. Phys. Commun.}\ }\textbf
  {\bibinfo {volume} {191}},\ \bibinfo {pages} {159} (\bibinfo {year}
  {2015})},\ \Eprint {http://arxiv.org/abs/1410.3012} {arXiv:1410.3012
  [hep-ph]} \BibitemShut {NoStop}%
\bibitem [{\citenamefont {Kelner}\ \emph {et~al.}(2006)\citenamefont {Kelner},
  \citenamefont {Aharonian},\ and\ \citenamefont {Bugayov}}]{Kelner:2006tc}%
  \BibitemOpen
  \bibfield  {author} {\bibinfo {author} {\bibfnamefont {S.~R.}\ \bibnamefont
  {Kelner}}, \bibinfo {author} {\bibfnamefont {F.~A.}\ \bibnamefont
  {Aharonian}}, \ and\ \bibinfo {author} {\bibfnamefont {V.~V.}\ \bibnamefont
  {Bugayov}},\ }\href {\doibase 10.1103/PhysRevD.74.034018} {\bibfield
  {journal} {\bibinfo  {journal} {Phys. Rev. D}\ }\textbf {\bibinfo {volume}
  {74}},\ \bibinfo {pages} {034018} (\bibinfo {year} {2006})},\ \bibinfo {note}
  {[Erratum: Phys.Rev.D 79, 039901 (2009)]},\ \Eprint
  {http://arxiv.org/abs/astro-ph/0606058} {arXiv:astro-ph/0606058} \BibitemShut
  {NoStop}%
\bibitem [{\citenamefont {Koldobskiy}\ \emph {et~al.}(2021)\citenamefont
  {Koldobskiy}, \citenamefont {Kachelrie\ss{}}, \citenamefont {Lskavyan},
  \citenamefont {Neronov}, \citenamefont {Ostapchenko},\ and\ \citenamefont
  {Semikoz}}]{Koldobskiy:2021nld}%
  \BibitemOpen
  \bibfield  {author} {\bibinfo {author} {\bibfnamefont {S.}~\bibnamefont
  {Koldobskiy}}, \bibinfo {author} {\bibfnamefont {M.}~\bibnamefont
  {Kachelrie\ss{}}}, \bibinfo {author} {\bibfnamefont {A.}~\bibnamefont
  {Lskavyan}}, \bibinfo {author} {\bibfnamefont {A.}~\bibnamefont {Neronov}},
  \bibinfo {author} {\bibfnamefont {S.}~\bibnamefont {Ostapchenko}}, \ and\
  \bibinfo {author} {\bibfnamefont {D.~V.}\ \bibnamefont {Semikoz}},\ }\href
  {\doibase 10.1103/PhysRevD.104.123027} {\bibfield  {journal} {\bibinfo
  {journal} {Phys. Rev. D}\ }\textbf {\bibinfo {volume} {104}},\ \bibinfo
  {pages} {123027} (\bibinfo {year} {2021})},\ \Eprint
  {http://arxiv.org/abs/2110.00496} {arXiv:2110.00496 [astro-ph.HE]}
  \BibitemShut {NoStop}%
\bibitem [{\citenamefont {Kamae}\ \emph {et~al.}(2006)\citenamefont {Kamae},
  \citenamefont {Karlsson}, \citenamefont {Mizuno}, \citenamefont {Abe},\ and\
  \citenamefont {Koi}}]{Kamae:2006bf}%
  \BibitemOpen
  \bibfield  {author} {\bibinfo {author} {\bibfnamefont {T.}~\bibnamefont
  {Kamae}}, \bibinfo {author} {\bibfnamefont {N.}~\bibnamefont {Karlsson}},
  \bibinfo {author} {\bibfnamefont {T.}~\bibnamefont {Mizuno}}, \bibinfo
  {author} {\bibfnamefont {T.}~\bibnamefont {Abe}}, \ and\ \bibinfo {author}
  {\bibfnamefont {T.}~\bibnamefont {Koi}},\ }\href {\doibase 10.1086/513602}
  {\bibfield  {journal} {\bibinfo  {journal} {Astrophys. J.}\ }\textbf
  {\bibinfo {volume} {647}},\ \bibinfo {pages} {692} (\bibinfo {year}
  {2006})},\ \bibinfo {note} {[Erratum: Astrophys.J. 662, 779 (2007)]},\
  \Eprint {http://arxiv.org/abs/astro-ph/0605581} {arXiv:astro-ph/0605581}
  \BibitemShut {NoStop}%
\bibitem [{\citenamefont {Kachelriess}\ \emph {et~al.}(2015)\citenamefont
  {Kachelriess}, \citenamefont {Moskalenko},\ and\ \citenamefont
  {Ostapchenko}}]{Kachelriess:2015wpa}%
  \BibitemOpen
  \bibfield  {author} {\bibinfo {author} {\bibfnamefont {M.}~\bibnamefont
  {Kachelriess}}, \bibinfo {author} {\bibfnamefont {I.~V.}\ \bibnamefont
  {Moskalenko}}, \ and\ \bibinfo {author} {\bibfnamefont {S.~S.}\ \bibnamefont
  {Ostapchenko}},\ }\href {\doibase 10.1088/0004-637X/803/2/54} {\bibfield
  {journal} {\bibinfo  {journal} {Astrophys. J.}\ }\textbf {\bibinfo {volume}
  {803}},\ \bibinfo {pages} {54} (\bibinfo {year} {2015})},\ \Eprint
  {http://arxiv.org/abs/1502.04158} {arXiv:1502.04158 [astro-ph.HE]}
  \BibitemShut {NoStop}%
\bibitem [{\citenamefont {Kachelrie\ss{}}\ \emph {et~al.}(2019)\citenamefont
  {Kachelrie\ss{}}, \citenamefont {Moskalenko},\ and\ \citenamefont
  {Ostapchenko}}]{Kachelriess:2019ifk}%
  \BibitemOpen
  \bibfield  {author} {\bibinfo {author} {\bibfnamefont {M.}~\bibnamefont
  {Kachelrie\ss{}}}, \bibinfo {author} {\bibfnamefont {I.~V.}\ \bibnamefont
  {Moskalenko}}, \ and\ \bibinfo {author} {\bibfnamefont {S.}~\bibnamefont
  {Ostapchenko}},\ }\href {\doibase 10.1016/j.cpc.2019.08.001} {\bibfield
  {journal} {\bibinfo  {journal} {Comput. Phys. Commun.}\ }\textbf {\bibinfo
  {volume} {245}},\ \bibinfo {pages} {106846} (\bibinfo {year} {2019})},\
  \Eprint {http://arxiv.org/abs/1904.05129} {arXiv:1904.05129 [hep-ph]}
  \BibitemShut {NoStop}%
\bibitem [{\citenamefont {Badhwar}\ \emph {et~al.}(1977)\citenamefont
  {Badhwar}, \citenamefont {Stephens},\ and\ \citenamefont
  {Golden}}]{PhysRevD.15.820}%
  \BibitemOpen
  \bibfield  {author} {\bibinfo {author} {\bibfnamefont {G.~D.}\ \bibnamefont
  {Badhwar}}, \bibinfo {author} {\bibfnamefont {S.~A.}\ \bibnamefont
  {Stephens}}, \ and\ \bibinfo {author} {\bibfnamefont {R.~L.}\ \bibnamefont
  {Golden}},\ }\href {\doibase 10.1103/PhysRevD.15.820} {\bibfield  {journal}
  {\bibinfo  {journal} {Phys. Rev. D}\ }\textbf {\bibinfo {volume} {15}},\
  \bibinfo {pages} {820} (\bibinfo {year} {1977})}\BibitemShut {NoStop}%
\bibitem [{\citenamefont {Strong}\ \emph {et~al.}(2009)\citenamefont {Strong},
  \citenamefont {Moskalenko}, \citenamefont {Porter}, \citenamefont
  {Jóhannesson}, \citenamefont {Orlando},\ and\ \citenamefont
  {Digel}}]{strong2009galprop}%
  \BibitemOpen
  \bibfield  {author} {\bibinfo {author} {\bibfnamefont {A.~W.}\ \bibnamefont
  {Strong}}, \bibinfo {author} {\bibfnamefont {I.~V.}\ \bibnamefont
  {Moskalenko}}, \bibinfo {author} {\bibfnamefont {T.~A.}\ \bibnamefont
  {Porter}}, \bibinfo {author} {\bibfnamefont {G.}~\bibnamefont
  {Jóhannesson}}, \bibinfo {author} {\bibfnamefont {E.}~\bibnamefont
  {Orlando}}, \ and\ \bibinfo {author} {\bibfnamefont {S.~W.}\ \bibnamefont
  {Digel}},\ }\href@noop {} {\enquote {\bibinfo {title} {The galprop cosmic-ray
  propagation code},}\ } (\bibinfo {year} {2009}),\ \Eprint
  {http://arxiv.org/abs/0907.0559} {arXiv:0907.0559 [astro-ph.HE]} \BibitemShut
  {NoStop}%
\bibitem [{\citenamefont {{Dermer}}(1986{\natexlab{a}})}]{1986ApJ...307...47D}%
  \BibitemOpen
  \bibfield  {author} {\bibinfo {author} {\bibfnamefont {C.~D.}\ \bibnamefont
  {{Dermer}}},\ }\href {\doibase 10.1086/164391} {\bibfield  {journal}
  {\bibinfo  {journal} {\apj}\ }\textbf {\bibinfo {volume} {307}},\ \bibinfo
  {pages} {47} (\bibinfo {year} {1986}{\natexlab{a}})}\BibitemShut {NoStop}%
\bibitem [{\citenamefont {{Dermer}}(1986{\natexlab{b}})}]{1986A&A...157..223D}%
  \BibitemOpen
  \bibfield  {author} {\bibinfo {author} {\bibfnamefont {C.~D.}\ \bibnamefont
  {{Dermer}}},\ }\href@noop {} {\bibfield  {journal} {\bibinfo  {journal}
  {\aap}\ }\textbf {\bibinfo {volume} {157}},\ \bibinfo {pages} {223} (\bibinfo
  {year} {1986}{\natexlab{b}})}\BibitemShut {NoStop}%
\bibitem [{\citenamefont {Evoli}\ \emph {et~al.}(2017)\citenamefont {Evoli},
  \citenamefont {Gaggero}, \citenamefont {Vittino}, \citenamefont
  {Di~Bernardo}, \citenamefont {Di~Mauro}, \citenamefont {Ligorini},
  \citenamefont {Ullio},\ and\ \citenamefont {Grasso}}]{Evoli:2016xgn}%
  \BibitemOpen
  \bibfield  {author} {\bibinfo {author} {\bibfnamefont {C.}~\bibnamefont
  {Evoli}}, \bibinfo {author} {\bibfnamefont {D.}~\bibnamefont {Gaggero}},
  \bibinfo {author} {\bibfnamefont {A.}~\bibnamefont {Vittino}}, \bibinfo
  {author} {\bibfnamefont {G.}~\bibnamefont {Di~Bernardo}}, \bibinfo {author}
  {\bibfnamefont {M.}~\bibnamefont {Di~Mauro}}, \bibinfo {author}
  {\bibfnamefont {A.}~\bibnamefont {Ligorini}}, \bibinfo {author}
  {\bibfnamefont {P.}~\bibnamefont {Ullio}}, \ and\ \bibinfo {author}
  {\bibfnamefont {D.}~\bibnamefont {Grasso}},\ }\href {\doibase
  10.1088/1475-7516/2017/02/015} {\bibfield  {journal} {\bibinfo  {journal}
  {JCAP}\ }\textbf {\bibinfo {volume} {02}},\ \bibinfo {pages} {015} (\bibinfo
  {year} {2017})},\ \Eprint {http://arxiv.org/abs/1607.07886} {arXiv:1607.07886
  [astro-ph.HE]} \BibitemShut {NoStop}%
\bibitem [{\citenamefont {Evoli}\ \emph {et~al.}(2018)\citenamefont {Evoli},
  \citenamefont {Gaggero}, \citenamefont {Vittino}, \citenamefont {Di~Mauro},
  \citenamefont {Grasso},\ and\ \citenamefont {Mazziotta}}]{Evoli:2017vim}%
  \BibitemOpen
  \bibfield  {author} {\bibinfo {author} {\bibfnamefont {C.}~\bibnamefont
  {Evoli}}, \bibinfo {author} {\bibfnamefont {D.}~\bibnamefont {Gaggero}},
  \bibinfo {author} {\bibfnamefont {A.}~\bibnamefont {Vittino}}, \bibinfo
  {author} {\bibfnamefont {M.}~\bibnamefont {Di~Mauro}}, \bibinfo {author}
  {\bibfnamefont {D.}~\bibnamefont {Grasso}}, \ and\ \bibinfo {author}
  {\bibfnamefont {M.~N.}\ \bibnamefont {Mazziotta}},\ }\href {\doibase
  10.1088/1475-7516/2018/07/006} {\bibfield  {journal} {\bibinfo  {journal}
  {JCAP}\ }\textbf {\bibinfo {volume} {07}},\ \bibinfo {pages} {006} (\bibinfo
  {year} {2018})},\ \Eprint {http://arxiv.org/abs/1711.09616} {arXiv:1711.09616
  [astro-ph.HE]} \BibitemShut {NoStop}%
\bibitem [{\citenamefont {Maurin}(2020)}]{Maurin:2018rmm}%
  \BibitemOpen
  \bibfield  {author} {\bibinfo {author} {\bibfnamefont {D.}~\bibnamefont
  {Maurin}},\ }\href {\doibase 10.1016/j.cpc.2019.106942} {\bibfield  {journal}
  {\bibinfo  {journal} {Comput. Phys. Commun.}\ }\textbf {\bibinfo {volume}
  {247}},\ \bibinfo {pages} {106942} (\bibinfo {year} {2020})},\ \Eprint
  {http://arxiv.org/abs/1807.02968} {arXiv:1807.02968 [astro-ph.IM]}
  \BibitemShut {NoStop}%
\bibitem [{\citenamefont {di~Mauro}\ \emph {et~al.}(2014)\citenamefont
  {di~Mauro}, \citenamefont {Donato}, \citenamefont {Goudelis},\ and\
  \citenamefont {Serpico}}]{DiMauro_2014}%
  \BibitemOpen
  \bibfield  {author} {\bibinfo {author} {\bibfnamefont {M.}~\bibnamefont
  {di~Mauro}}, \bibinfo {author} {\bibfnamefont {F.}~\bibnamefont {Donato}},
  \bibinfo {author} {\bibfnamefont {A.}~\bibnamefont {Goudelis}}, \ and\
  \bibinfo {author} {\bibfnamefont {P.~D.}\ \bibnamefont {Serpico}},\ }\href
  {\doibase 10.1103/physrevd.90.085017} {\bibfield  {journal} {\bibinfo
  {journal} {Physical Review D}\ }\textbf {\bibinfo {volume} {90}} (\bibinfo
  {year} {2014}),\ 10.1103/physrevd.90.085017}\BibitemShut {NoStop}%
\bibitem [{\citenamefont {Alt}\ \emph {et~al.}(2005)\citenamefont {Alt} \emph
  {et~al.}}]{2005_NA49}%
  \BibitemOpen
  \bibfield  {author} {\bibinfo {author} {\bibfnamefont {C.}~\bibnamefont
  {Alt}} \emph {et~al.} (\bibinfo {collaboration} {NA49}),\ }\href {\doibase
  10.1140/epjc/s2005-02391-9} {\bibfield  {journal} {\bibinfo  {journal} {Eur.
  Phys. J. C}\ }\textbf {\bibinfo {volume} {45}},\ \bibinfo {pages} {343–381}
  (\bibinfo {year} {2005})}\BibitemShut {NoStop}%
\bibitem [{\citenamefont {Aduszkiewicz}\ \emph {et~al.}(2017)\citenamefont
  {Aduszkiewicz} \emph {et~al.}}]{Aduszkiewicz:2017sei}%
  \BibitemOpen
  \bibfield  {author} {\bibinfo {author} {\bibfnamefont {A.}~\bibnamefont
  {Aduszkiewicz}} \emph {et~al.} (\bibinfo {collaboration} {NA61/SHINE}),\
  }\href {\doibase 10.1140/epjc/s10052-017-5260-4} {\bibfield  {journal}
  {\bibinfo  {journal} {Eur. Phys. J. C}\ }\textbf {\bibinfo {volume} {77}},\
  \bibinfo {pages} {671} (\bibinfo {year} {2017})},\ \Eprint
  {http://arxiv.org/abs/1705.02467} {arXiv:1705.02467 [nucl-ex]} \BibitemShut
  {NoStop}%
\bibitem [{\citenamefont {Arsene}\ \emph {et~al.}(2007)\citenamefont {Arsene},
  \citenamefont {Bearden}, \citenamefont {Beavis}, \citenamefont {Bekele},
  \citenamefont {Besliu}, \citenamefont {Budick}, \citenamefont {Bøggild},
  \citenamefont {Chasman}, \citenamefont {Christensen}, \citenamefont
  {Dalsgaard},\ and\ \citenamefont {et~al.}}]{2007_BRAHMS}%
  \BibitemOpen
  \bibfield  {author} {\bibinfo {author} {\bibfnamefont {I.}~\bibnamefont
  {Arsene}}, \bibinfo {author} {\bibfnamefont {I.~G.}\ \bibnamefont {Bearden}},
  \bibinfo {author} {\bibfnamefont {D.}~\bibnamefont {Beavis}}, \bibinfo
  {author} {\bibfnamefont {S.}~\bibnamefont {Bekele}}, \bibinfo {author}
  {\bibfnamefont {C.}~\bibnamefont {Besliu}}, \bibinfo {author} {\bibfnamefont
  {B.}~\bibnamefont {Budick}}, \bibinfo {author} {\bibfnamefont
  {H.}~\bibnamefont {Bøggild}}, \bibinfo {author} {\bibfnamefont
  {C.}~\bibnamefont {Chasman}}, \bibinfo {author} {\bibfnamefont {C.~H.}\
  \bibnamefont {Christensen}}, \bibinfo {author} {\bibfnamefont {H.~H.}\
  \bibnamefont {Dalsgaard}}, \ and\ \bibinfo {author} {\bibnamefont {et~al.}},\
  }\href {\doibase 10.1103/physrevlett.98.252001} {\bibfield  {journal}
  {\bibinfo  {journal} {Physical Review Letters}\ }\textbf {\bibinfo {volume}
  {98}} (\bibinfo {year} {2007}),\ 10.1103/physrevlett.98.252001}\BibitemShut
  {NoStop}%
\bibitem [{\citenamefont {Adare}\ \emph {et~al.}(2011)\citenamefont {Adare},
  \citenamefont {Afanasiev}, \citenamefont {Aidala}, \citenamefont {Ajitanand},
  \citenamefont {Akiba}, \citenamefont {Al-Bataineh}, \citenamefont
  {Alexander}, \citenamefont {Aoki}, \citenamefont {Aphecetche}, \citenamefont
  {Armendariz},\ and\ \citenamefont {et~al.}}]{2011_PHENIX}%
  \BibitemOpen
  \bibfield  {author} {\bibinfo {author} {\bibfnamefont {A.}~\bibnamefont
  {Adare}}, \bibinfo {author} {\bibfnamefont {S.}~\bibnamefont {Afanasiev}},
  \bibinfo {author} {\bibfnamefont {C.}~\bibnamefont {Aidala}}, \bibinfo
  {author} {\bibfnamefont {N.~N.}\ \bibnamefont {Ajitanand}}, \bibinfo {author}
  {\bibfnamefont {Y.}~\bibnamefont {Akiba}}, \bibinfo {author} {\bibfnamefont
  {H.}~\bibnamefont {Al-Bataineh}}, \bibinfo {author} {\bibfnamefont
  {J.}~\bibnamefont {Alexander}}, \bibinfo {author} {\bibfnamefont
  {K.}~\bibnamefont {Aoki}}, \bibinfo {author} {\bibfnamefont {L.}~\bibnamefont
  {Aphecetche}}, \bibinfo {author} {\bibfnamefont {R.}~\bibnamefont
  {Armendariz}}, \ and\ \bibinfo {author} {\bibnamefont {et~al.}},\ }\href
  {\doibase 10.1103/physrevc.83.064903} {\bibfield  {journal} {\bibinfo
  {journal} {Physical Review C}\ }\textbf {\bibinfo {volume} {83}} (\bibinfo
  {year} {2011}),\ 10.1103/physrevc.83.064903}\BibitemShut {NoStop}%
\bibitem [{\citenamefont {Adam}\ \emph {et~al.}(2015)\citenamefont {Adam} \emph
  {et~al.}}]{Adam:2015qaa}%
  \BibitemOpen
  \bibfield  {author} {\bibinfo {author} {\bibfnamefont {J.}~\bibnamefont
  {Adam}} \emph {et~al.} (\bibinfo {collaboration} {ALICE}),\ }\href {\doibase
  10.1140/epjc/s10052-015-3422-9} {\bibfield  {journal} {\bibinfo  {journal}
  {Eur. Phys. J. C}\ }\textbf {\bibinfo {volume} {75}},\ \bibinfo {pages} {226}
  (\bibinfo {year} {2015})},\ \Eprint {http://arxiv.org/abs/1504.00024}
  {arXiv:1504.00024 [nucl-ex]} \BibitemShut {NoStop}%
\bibitem [{\citenamefont {Sirunyan}\ \emph {et~al.}(2017)\citenamefont
  {Sirunyan}, \citenamefont {Tumasyan}, \citenamefont {Adam}, \citenamefont
  {Asilar}, \citenamefont {Bergauer}, \citenamefont {Brandstetter},
  \citenamefont {Brondolin}, \citenamefont {Dragicevic}, \citenamefont {Erö},
  \citenamefont {Flechl},\ and\ \citenamefont {et~al.}}]{2017_CMS}%
  \BibitemOpen
  \bibfield  {author} {\bibinfo {author} {\bibfnamefont {A.}~\bibnamefont
  {Sirunyan}}, \bibinfo {author} {\bibfnamefont {A.}~\bibnamefont {Tumasyan}},
  \bibinfo {author} {\bibfnamefont {W.}~\bibnamefont {Adam}}, \bibinfo {author}
  {\bibfnamefont {E.}~\bibnamefont {Asilar}}, \bibinfo {author} {\bibfnamefont
  {T.}~\bibnamefont {Bergauer}}, \bibinfo {author} {\bibfnamefont
  {J.}~\bibnamefont {Brandstetter}}, \bibinfo {author} {\bibfnamefont
  {E.}~\bibnamefont {Brondolin}}, \bibinfo {author} {\bibfnamefont
  {M.}~\bibnamefont {Dragicevic}}, \bibinfo {author} {\bibfnamefont
  {J.}~\bibnamefont {Erö}}, \bibinfo {author} {\bibfnamefont {M.}~\bibnamefont
  {Flechl}}, \ and\ \bibinfo {author} {\bibnamefont {et~al.}},\ }\href
  {\doibase 10.1103/physrevd.96.112003} {\bibfield  {journal} {\bibinfo
  {journal} {Physical Review D}\ }\textbf {\bibinfo {volume} {96}} (\bibinfo
  {year} {2017}),\ 10.1103/physrevd.96.112003}\BibitemShut {NoStop}%
\bibitem [{\citenamefont {Ferriere}(2001)}]{Ferriere:2001rg}%
  \BibitemOpen
  \bibfield  {author} {\bibinfo {author} {\bibfnamefont {K.~M.}\ \bibnamefont
  {Ferriere}},\ }\href {\doibase 10.1103/RevModPhys.73.1031} {\bibfield
  {journal} {\bibinfo  {journal} {Rev. Mod. Phys.}\ }\textbf {\bibinfo {volume}
  {73}},\ \bibinfo {pages} {1031} (\bibinfo {year} {2001})},\ \Eprint
  {http://arxiv.org/abs/astro-ph/0106359} {arXiv:astro-ph/0106359} \BibitemShut
  {NoStop}%
\bibitem [{\citenamefont {Arbuzov}(2002)}]{Arbuzov_2002}%
  \BibitemOpen
  \bibfield  {author} {\bibinfo {author} {\bibfnamefont {A.}~\bibnamefont
  {Arbuzov}},\ }\href {\doibase 10.1016/s0370-2693(01)01335-1} {\bibfield
  {journal} {\bibinfo  {journal} {Physics Letters B}\ }\textbf {\bibinfo
  {volume} {524}},\ \bibinfo {pages} {99–106} (\bibinfo {year}
  {2002})}\BibitemShut {NoStop}%
\bibitem [{\citenamefont {{Scanlon}}\ and\ \citenamefont
  {{Milford}}(1965)}]{1965ApJ...141..718S}%
  \BibitemOpen
  \bibfield  {author} {\bibinfo {author} {\bibfnamefont {J.~H.}\ \bibnamefont
  {{Scanlon}}}\ and\ \bibinfo {author} {\bibfnamefont {S.~N.}\ \bibnamefont
  {{Milford}}},\ }\href {\doibase 10.1086/148156} {\bibfield  {journal}
  {\bibinfo  {journal} {\apj}\ }\textbf {\bibinfo {volume} {141}},\ \bibinfo
  {pages} {718} (\bibinfo {year} {1965})}\BibitemShut {NoStop}%
\bibitem [{\citenamefont {Anticic}\ \emph {et~al.}(2010)\citenamefont
  {Anticic}, \citenamefont {Baatar}, \citenamefont {Bartke} \emph
  {et~al.}}]{NA49_2010}%
  \BibitemOpen
  \bibfield  {author} {\bibinfo {author} {\bibfnamefont {T.}~\bibnamefont
  {Anticic}}, \bibinfo {author} {\bibfnamefont {B.}~\bibnamefont {Baatar}},
  \bibinfo {author} {\bibfnamefont {J.}~\bibnamefont {Bartke}},  \emph
  {et~al.},\ }\href {\doibase 10.1140/epjc/s10052-010-1328-0} {\bibfield
  {journal} {\bibinfo  {journal} {The European Physical Journal C}\ }\textbf
  {\bibinfo {volume} {68}},\ \bibinfo {pages} {1–73} (\bibinfo {year}
  {2010})}\BibitemShut {NoStop}%
\bibitem [{\citenamefont {Aamodt}(2011)}]{2011_ALICE}%
  \BibitemOpen
  \bibfield  {author} {\bibinfo {author} {\bibfnamefont {K.~e.~a.}\
  \bibnamefont {Aamodt}},\ }\href {\doibase 10.1140/epjc/s10052-011-1655-9}
  {\bibfield  {journal} {\bibinfo  {journal} {The European Physical Journal C}\
  }\textbf {\bibinfo {volume} {71}} (\bibinfo {year} {2011}),\
  10.1140/epjc/s10052-011-1655-9}\BibitemShut {NoStop}%
\bibitem [{\citenamefont {Chatrchyan}\ \emph {et~al.}(2012)\citenamefont
  {Chatrchyan}, \citenamefont {Khachatryan}, \citenamefont {Sirunyan},
  \citenamefont {Tumasyan}, \citenamefont {Adam}, \citenamefont {Aguilo},
  \citenamefont {Bergauer}, \citenamefont {Dragicevic}, \citenamefont {Erö},\
  and\ \citenamefont {et~al.}}]{2012_CMS}%
  \BibitemOpen
  \bibfield  {author} {\bibinfo {author} {\bibfnamefont {S.}~\bibnamefont
  {Chatrchyan}}, \bibinfo {author} {\bibfnamefont {V.}~\bibnamefont
  {Khachatryan}}, \bibinfo {author} {\bibfnamefont {A.~M.}\ \bibnamefont
  {Sirunyan}}, \bibinfo {author} {\bibfnamefont {A.}~\bibnamefont {Tumasyan}},
  \bibinfo {author} {\bibfnamefont {W.}~\bibnamefont {Adam}}, \bibinfo {author}
  {\bibfnamefont {E.}~\bibnamefont {Aguilo}}, \bibinfo {author} {\bibfnamefont
  {T.}~\bibnamefont {Bergauer}}, \bibinfo {author} {\bibfnamefont
  {M.}~\bibnamefont {Dragicevic}}, \bibinfo {author} {\bibfnamefont
  {J.}~\bibnamefont {Erö}}, \ and\ \bibinfo {author} {\bibnamefont {et~al.}},\
  }\href {\doibase 10.1140/epjc/s10052-012-2164-1} {\bibfield  {journal}
  {\bibinfo  {journal} {The European Physical Journal C}\ }\textbf {\bibinfo
  {volume} {72}} (\bibinfo {year} {2012}),\
  10.1140/epjc/s10052-012-2164-1}\BibitemShut {NoStop}%
\bibitem [{\citenamefont {Antinucci}\ \emph {et~al.}(1973)\citenamefont
  {Antinucci}, \citenamefont {Bertin},\ and\ \citenamefont
  {Capiluppi}}]{osti_4593576}%
  \BibitemOpen
  \bibfield  {author} {\bibinfo {author} {\bibfnamefont {M.}~\bibnamefont
  {Antinucci}}, \bibinfo {author} {\bibfnamefont {A.}~\bibnamefont {Bertin}}, \
  and\ \bibinfo {author} {\bibfnamefont {P.}~\bibnamefont {Capiluppi}},\ }\href
  {\doibase 10.1007/BF02827250} {\bibfield  {journal} {\bibinfo  {journal}
  {Lett. Nuovo Cim. 6: No. 4, 121-128(27 Jan 1973).}\ } (\bibinfo {year}
  {1973}),\ 10.1007/BF02827250}\BibitemShut {NoStop}%
\bibitem [{\citenamefont {Ellis}\ and\ \citenamefont
  {Stroynowski}(1977)}]{RevModPhys.49.753}%
  \BibitemOpen
  \bibfield  {author} {\bibinfo {author} {\bibfnamefont {S.~D.}\ \bibnamefont
  {Ellis}}\ and\ \bibinfo {author} {\bibfnamefont {R.}~\bibnamefont
  {Stroynowski}},\ }\href {\doibase 10.1103/RevModPhys.49.753} {\bibfield
  {journal} {\bibinfo  {journal} {Rev. Mod. Phys.}\ }\textbf {\bibinfo {volume}
  {49}},\ \bibinfo {pages} {753} (\bibinfo {year} {1977})}\BibitemShut
  {NoStop}%
\bibitem [{\citenamefont {Alper}\ and\ \citenamefont
  {et~al.}(1975)}]{ALPER1975237}%
  \BibitemOpen
  \bibfield  {author} {\bibinfo {author} {\bibfnamefont {B.}~\bibnamefont
  {Alper}}\ and\ \bibinfo {author} {\bibnamefont {et~al.}},\ }\href {\doibase
  https://doi.org/10.1016/0550-3213(75)90618-5} {\bibfield  {journal} {\bibinfo
   {journal} {Nuclear Physics B}\ }\textbf {\bibinfo {volume} {100}},\ \bibinfo
  {pages} {237} (\bibinfo {year} {1975})}\BibitemShut {NoStop}%
\bibitem [{\citenamefont {Norbury}\ and\ \citenamefont
  {Townsend}(2007)}]{Norbury_2007}%
  \BibitemOpen
  \bibfield  {author} {\bibinfo {author} {\bibfnamefont {J.~W.}\ \bibnamefont
  {Norbury}}\ and\ \bibinfo {author} {\bibfnamefont {L.~W.}\ \bibnamefont
  {Townsend}},\ }\href {\doibase 10.1103/physrevd.75.034001} {\bibfield
  {journal} {\bibinfo  {journal} {Physical Review D}\ }\textbf {\bibinfo
  {volume} {75}} (\bibinfo {year} {2007}),\
  10.1103/physrevd.75.034001}\BibitemShut {NoStop}%
\bibitem [{\citenamefont {Feroz}\ \emph {et~al.}(2009)\citenamefont {Feroz},
  \citenamefont {Hobson},\ and\ \citenamefont {Bridges}}]{Multinest_2009}%
  \BibitemOpen
  \bibfield  {author} {\bibinfo {author} {\bibfnamefont {F.}~\bibnamefont
  {Feroz}}, \bibinfo {author} {\bibfnamefont {M.~P.}\ \bibnamefont {Hobson}}, \
  and\ \bibinfo {author} {\bibfnamefont {M.}~\bibnamefont {Bridges}},\ }\href
  {\doibase 10.1111/j.1365-2966.2009.14548.x} {\bibfield  {journal} {\bibinfo
  {journal} {Monthly Notices of the Royal Astronomical Society}\ }\textbf
  {\bibinfo {volume} {398}},\ \bibinfo {pages} {1601–1614} (\bibinfo {year}
  {2009})}\BibitemShut {NoStop}%
\bibitem [{\citenamefont {Adams}\ \emph {et~al.}(2006)\citenamefont {Adams}
  \emph {et~al.}}]{STAR:2006xud}%
  \BibitemOpen
  \bibfield  {author} {\bibinfo {author} {\bibfnamefont {J.}~\bibnamefont
  {Adams}} \emph {et~al.} (\bibinfo {collaboration} {STAR}),\ }\href {\doibase
  10.1016/j.physletb.2006.04.032} {\bibfield  {journal} {\bibinfo  {journal}
  {Phys. Lett. B}\ }\textbf {\bibinfo {volume} {637}},\ \bibinfo {pages} {161}
  (\bibinfo {year} {2006})},\ \Eprint {http://arxiv.org/abs/nucl-ex/0601033}
  {arXiv:nucl-ex/0601033} \BibitemShut {NoStop}%
\bibitem [{\citenamefont {Winkler}(2017)}]{winkler_2017}%
  \BibitemOpen
  \bibfield  {author} {\bibinfo {author} {\bibfnamefont {M.~W.}\ \bibnamefont
  {Winkler}},\ }\href {\doibase 10.1088/1475-7516/2017/02/048} {\bibfield
  {journal} {\bibinfo  {journal} {Journal of Cosmology and Astroparticle
  Physics}\ }\textbf {\bibinfo {volume} {2017}},\ \bibinfo {pages} {048–048}
  (\bibinfo {year} {2017})}\BibitemShut {NoStop}%
\bibitem [{\citenamefont {Donato}\ \emph {et~al.}(2017)\citenamefont {Donato},
  \citenamefont {Korsmeier},\ and\ \citenamefont {Di~Mauro}}]{Donato:2017ywo}%
  \BibitemOpen
  \bibfield  {author} {\bibinfo {author} {\bibfnamefont {F.}~\bibnamefont
  {Donato}}, \bibinfo {author} {\bibfnamefont {M.}~\bibnamefont {Korsmeier}}, \
  and\ \bibinfo {author} {\bibfnamefont {M.}~\bibnamefont {Di~Mauro}},\ }\href
  {\doibase 10.1103/PhysRevD.96.043007} {\bibfield  {journal} {\bibinfo
  {journal} {Phys. Rev. D}\ }\textbf {\bibinfo {volume} {96}},\ \bibinfo
  {pages} {043007} (\bibinfo {year} {2017})},\ \Eprint
  {http://arxiv.org/abs/1704.03663} {arXiv:1704.03663 [astro-ph.HE]}
  \BibitemShut {NoStop}%
\bibitem [{\citenamefont {Acharya}\ \emph {et~al.}(2022)\citenamefont {Acharya}
  \emph {et~al.}}]{NA61SHINE:2021iay}%
  \BibitemOpen
  \bibfield  {author} {\bibinfo {author} {\bibfnamefont {A.}~\bibnamefont
  {Acharya}} \emph {et~al.} (\bibinfo {collaboration} {NA61/SHINE}),\ }\href
  {\doibase 10.1140/epjc/s10052-021-09976-y} {\bibfield  {journal} {\bibinfo
  {journal} {Eur. Phys. J. C}\ }\textbf {\bibinfo {volume} {82}},\ \bibinfo
  {pages} {96} (\bibinfo {year} {2022})},\ \Eprint
  {http://arxiv.org/abs/2106.07535} {arXiv:2106.07535 [hep-ex]} \BibitemShut
  {NoStop}%
\bibitem [{\citenamefont {Aduszkiewicz}\ \emph {et~al.}(2016)\citenamefont
  {Aduszkiewicz} \emph {et~al.}}]{NA61SHINE:2015haq}%
  \BibitemOpen
  \bibfield  {author} {\bibinfo {author} {\bibfnamefont {A.}~\bibnamefont
  {Aduszkiewicz}} \emph {et~al.} (\bibinfo {collaboration} {NA61/SHINE}),\
  }\href {\doibase 10.1140/epjc/s10052-016-4003-2} {\bibfield  {journal}
  {\bibinfo  {journal} {Eur. Phys. J. C}\ }\textbf {\bibinfo {volume} {76}},\
  \bibinfo {pages} {198} (\bibinfo {year} {2016})},\ \Eprint
  {http://arxiv.org/abs/1510.03720} {arXiv:1510.03720 [hep-ex]} \BibitemShut
  {NoStop}%
\bibitem [{\citenamefont {Alt}\ and\ \citenamefont {et~al.}(2007)}]{NA49_2007}%
  \BibitemOpen
  \bibfield  {author} {\bibinfo {author} {\bibfnamefont {C.}~\bibnamefont
  {Alt}}\ and\ \bibinfo {author} {\bibnamefont {et~al.}},\ }\href {\doibase
  10.1140/epjc/s10052-006-0165-7} {\bibfield  {journal} {\bibinfo  {journal}
  {The European Physical Journal C}\ }\textbf {\bibinfo {volume} {49}},\
  \bibinfo {pages} {897–917} (\bibinfo {year} {2007})}\BibitemShut {NoStop}%
\bibitem [{\citenamefont {{Orth}}\ and\ \citenamefont
  {{Buffington}}(1976)}]{1976ApJ...206..312O}%
  \BibitemOpen
  \bibfield  {author} {\bibinfo {author} {\bibfnamefont {C.~D.}\ \bibnamefont
  {{Orth}}}\ and\ \bibinfo {author} {\bibfnamefont {A.}~\bibnamefont
  {{Buffington}}},\ }\href {\doibase 10.1086/154386} {\bibfield  {journal}
  {\bibinfo  {journal} {\apj}\ }\textbf {\bibinfo {volume} {206}},\ \bibinfo
  {pages} {312} (\bibinfo {year} {1976})}\BibitemShut {NoStop}%
\bibitem [{\citenamefont {Barr}\ \emph {et~al.}(2007)\citenamefont {Barr},
  \citenamefont {Chvala}, \citenamefont {Fischer}, \citenamefont {Kreps},
  \citenamefont {Makariev}, \citenamefont {Pattison}, \citenamefont {Rybicki},
  \citenamefont {Varga},\ and\ \citenamefont {Wenig}}]{NA49_2007_discussion}%
  \BibitemOpen
  \bibfield  {author} {\bibinfo {author} {\bibfnamefont {G.}~\bibnamefont
  {Barr}}, \bibinfo {author} {\bibfnamefont {O.}~\bibnamefont {Chvala}},
  \bibinfo {author} {\bibfnamefont {H.}~\bibnamefont {Fischer}}, \bibinfo
  {author} {\bibfnamefont {M.}~\bibnamefont {Kreps}}, \bibinfo {author}
  {\bibfnamefont {M.}~\bibnamefont {Makariev}}, \bibinfo {author}
  {\bibfnamefont {C.}~\bibnamefont {Pattison}}, \bibinfo {author}
  {\bibfnamefont {A.}~\bibnamefont {Rybicki}}, \bibinfo {author} {\bibfnamefont
  {D.}~\bibnamefont {Varga}}, \ and\ \bibinfo {author} {\bibfnamefont
  {S.}~\bibnamefont {Wenig}},\ }\href {\doibase 10.1140/epjc/s10052-006-0166-6}
  {\bibfield  {journal} {\bibinfo  {journal} {The European Physical Journal C}\
  }\textbf {\bibinfo {volume} {49}},\ \bibinfo {pages} {919–945} (\bibinfo
  {year} {2007})}\BibitemShut {NoStop}%
\bibitem [{\citenamefont {Baldin}\ \emph {et~al.}(1982)\citenamefont {Baldin}
  \emph {et~al.}}]{Baldin:1982my}%
  \BibitemOpen
  \bibfield  {author} {\bibinfo {author} {\bibfnamefont {A.~M.}\ \bibnamefont
  {Baldin}} \emph {et~al.},\ }\href@noop {} {\bibfield  {journal} {\bibinfo
  {journal} {JINR-E1-82-472}\ } (\bibinfo {year} {1982})}\BibitemShut {NoStop}%
\bibitem [{\citenamefont {Abgrall}\ and\ \citenamefont
  {et~al.}(2016)}]{NA61_2016}%
  \BibitemOpen
  \bibfield  {author} {\bibinfo {author} {\bibfnamefont {N.}~\bibnamefont
  {Abgrall}}\ and\ \bibinfo {author} {\bibnamefont {et~al.}},\ }\href {\doibase
  10.1140/epjc/s10052-016-3898-y} {\bibfield  {journal} {\bibinfo  {journal}
  {The European Physical Journal C}\ }\textbf {\bibinfo {volume} {76}}
  (\bibinfo {year} {2016}),\ 10.1140/epjc/s10052-016-3898-y}\BibitemShut
  {NoStop}%
\bibitem [{\citenamefont {Group}(2020)}]{10.1093/ptep/ptaa104}%
  \BibitemOpen
  \bibfield  {author} {\bibinfo {author} {\bibfnamefont {P.~D.}\ \bibnamefont
  {Group}},\ }\href {\doibase 10.1093/ptep/ptaa104} {\bibfield  {journal}
  {\bibinfo  {journal} {Progress of Theoretical and Experimental Physics}\
  }\textbf {\bibinfo {volume} {2020}} (\bibinfo {year} {2020}),\
  10.1093/ptep/ptaa104},\ \bibinfo {note} {083C01},\ \Eprint
  {http://arxiv.org/abs/https://academic.oup.com/ptep/article-pdf/2020/8/083C01/34673722/ptaa104.pdf}
  {https://academic.oup.com/ptep/article-pdf/2020/8/083C01/34673722/ptaa104.pdf}
  \BibitemShut {NoStop}%
\bibitem [{\citenamefont {Corke}\ and\ \citenamefont
  {Sjostrand}(2011)}]{Corke:2010yf}%
  \BibitemOpen
  \bibfield  {author} {\bibinfo {author} {\bibfnamefont {R.}~\bibnamefont
  {Corke}}\ and\ \bibinfo {author} {\bibfnamefont {T.}~\bibnamefont
  {Sjostrand}},\ }\href {\doibase 10.1007/JHEP03(2011)032} {\bibfield
  {journal} {\bibinfo  {journal} {JHEP}\ }\textbf {\bibinfo {volume} {03}},\
  \bibinfo {pages} {032} (\bibinfo {year} {2011})},\ \Eprint
  {http://arxiv.org/abs/1011.1759} {arXiv:1011.1759 [hep-ph]} \BibitemShut
  {NoStop}%
\end{thebibliography}%

\appendix

\section{Parametrization of the total inelastic cross section}
\label{app:sigma_0}

In this section we report the results for the calculation of the inelastic cross section, which appears in Eq.~\eqref{eq:main_equation}.
A new calculation of the inelastic cross section is necessary to estimate its uncertainty, that must be added to all the other uncertainties.
We take the available data from the Particle Data Group \cite{10.1093/ptep/ptaa104} for the total collision and elastic $p+p$ cross sections because very few data are available for the inelastic cross sections.
We first make a fit to the total collision ($\mathrm{\sigma_{tot}^{pp}}$) and elastic  ($\mathrm{\sigma_{el}^{pp}}$) cross section and then we derive the inelastic $\sigma_0(s)$ one as their difference.
We use the following functional form for both $\mathrm{\sigma_{tot}^{pp}}$ and $\mathrm{\sigma_{el}^{pp}}$:
\begin{eqnarray}
\mathrm{\sigma_{tot,el}^{pp}} &=& Z^{\mathrm{pp}} + B^{\mathrm{pp}}\log^2(s/s_M)  \nonumber \\ 
 &&+ Y_1^{\mathrm{pp}}(s_M/s)^{\eta_1} - Y_2^{\mathrm{pp}}(s_M/s)^{\eta_2}
\label{eq:sigmappinelparam}  
\end{eqnarray}
where $B^{pp} = \pi (\hbar c)^2/M^2$, $s_M = (2 m_p + M)^2$, all energies are given in GeV and $\mathrm{\sigma_{tot}^{pp}}$ and $\mathrm{\sigma_{el}^{pp}}$ are given in units of mb. 

We include in the fitting procedure only data referring to $E_p > 2$ GeV because at lower energies the contribution of the resonances become very important and complicate to model precisely. Moreover, for $e^\pm$ with $E>1$ GeV the contribution from protons with energies below 2 GeV is negligible. For $E_p<2$ GeV we then use the $\sigma_0(s)$ parametrization reported in Ref.~\cite{Kamae:2006bf}.

\begin{table}[b]
\begin{center} 
\vspace{1cm}
\begin{tabular}{ c c c } 
\hline \hline 
Parameter & Total & Elastic                  \\ 
\hline 
$M$        &  1.589  &  3.094               \\ 
$Z^{pp}$   &  59.58 &  21.34                 \\ 
$Y_1^{pp}$ &  0.890 &  2.667                  \\ 
$Y_2^{pp}$ &  19.35  &  14.21                \\
$\eta_1$   &  2.543 &  1.003                 \\
$\eta_2$   &  -0.0895 &  -0.0327 \\
\hline \hline
\end{tabular} \caption{Fit results for the total and elastic proton scattering cross sections according to Eq.~\eqref{eq:sigmappinelparam}.
\label{tab:protonCSfit}}
\end{center} 
\end{table}

\begin{figure}[t]
    \includegraphics[width=0.49\textwidth]{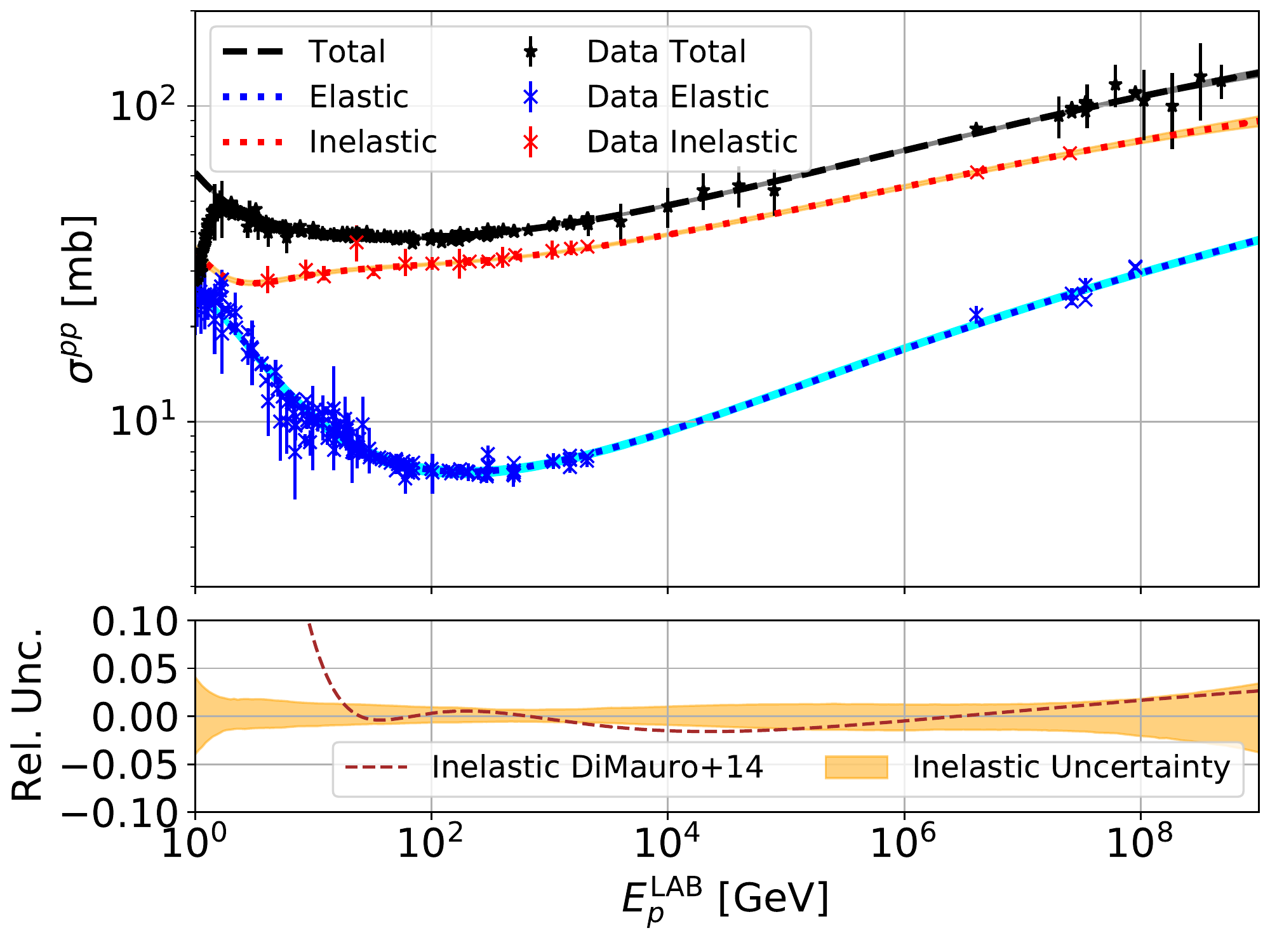}
  \caption{This plot shows the result for the fit to the total (black data and line) and elastic (blue data and line) cross section for $p+p$ collisions. We also show the inelastic cross section (red line) derived as $\mathrm{\sigma_{tot}^{pp}}-\mathrm{\sigma_{el}^{pp}}$, along with the available data. In the bottom panel we show the $1\sigma$ uncertainty band derived for the inelastic cross section and the relative difference between our best fit and the one of Ref.~\cite{DiMauro_2014} (brown dashed line).}
  \label{fig:sigmatot}
\end{figure}

The fitting parameters for both $\mathrm{\sigma_{tot}^{pp}}$ and $\mathrm{\sigma_{el}^{pp}}$ are shown in Tab.~ \ref{tab:protonCSfit}. 
The comparison between the model and the data are shown in Fig.~\ref{fig:sigmatot}.
The resulting functions provide good fits to the available data, with a $\chi^2$/d.o.f.$=0.88$ for $\mathrm{\sigma_{tot}^{pp}}$  and $2.20$ for $\mathrm{\sigma_{el}^{pp}}$, where statistical and systematic uncertainties have been added in quadrature. 
As shown in Fig.~\ref{fig:sigmatot}, the inelastic cross section $\sigma_0(s) = \mathrm{\sigma_{tot}^{pp}}-\mathrm{\sigma_{el}^{pp}}$ has a $1\sigma$ uncertainty that is on average between $2-3\%$.
\begin{figure*}[t]
  \centering {
    \includegraphics[width=0.5\textwidth]{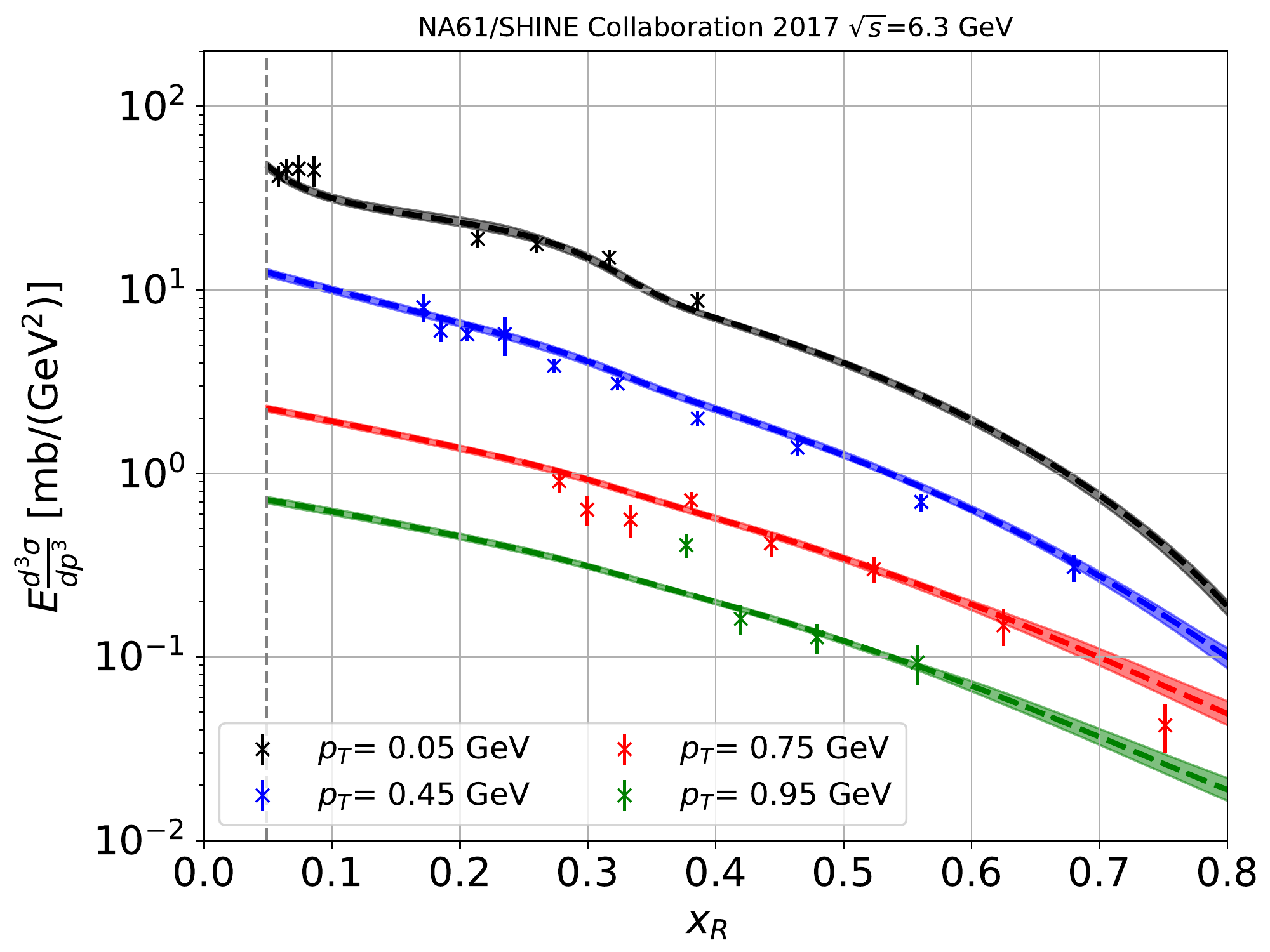}\includegraphics[width=0.5\textwidth]{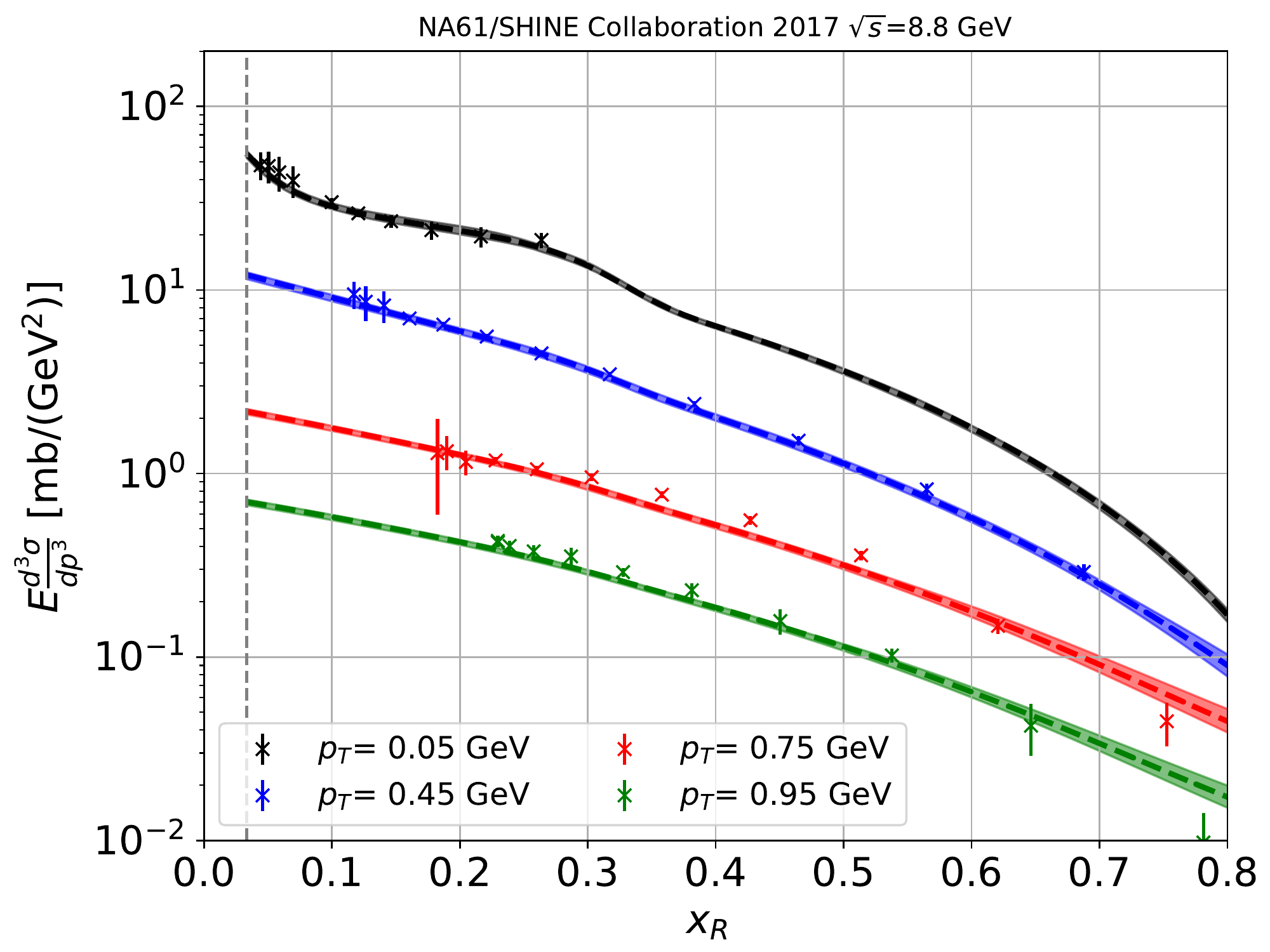}
  }
  \caption{
             Comparison between our best fit and the NA61/SHINE data \cite{Aduszkiewicz:2017sei} of Lorentz invariant cross section for the inclusive $\pi^+$ production in $p+p$ collisions at $\sS=6.3$ GeV (left panel) and $\sS=8.8$ GeV (right panel). The data is shown as a function of $x_R$ for a few representative values of $p_T$. Shaded bands show the 1$\sigma$ uncertainty. 
  }
  \label{Fig:pi-plus-NA61}
\end{figure*}
We add this uncertainty in the estimate of the source term for the secondary production of $e^{\pm}$.
Our results are in very good agreement not only with the PDG results on $\mathrm{\sigma_{tot}^{pp}}$ but also with the function for $\mathrm{\sigma_{el}^{pp}}$ and previous references such as \cite{DiMauro_2014} for the inelastic cross section.

\section{Comparison with data at small $\sqrt{s}$}
\label{app:comparison_NA61}

As discussed in Sec.~\ref{sec:piplus_differentCME} we fix the kinematic shape of $\sigmaInv(p+p\rightarrow \pi^+ +X)$
using NA49 data at $\sqrt{s}=17.3$\ GeV, while at lower energies we use the multiplicity to adjust the overall normalization of the cross section. This treatment can be cross checked by data. In particular, NA61/SHINE \cite{Aduszkiewicz:2017sei} provides measurements of the Lorentz invariant cross section for the inclusive $\pi^+$ production in $p+p$ collisions at different $x_R$ and $p_T$. In Fig.~\ref{Fig:pi-plus-NA61}, we compare our parametrization with the NA61/SHINE measurements at $\sqrt{s}=6.3$\ GeV and $\sqrt{s}=8.8$\ GeV. The invariant cross section is presented as function of $x_R$ and for a few representative values of $p_T$. Our parametrization provides a very good description of the data, especially for the more important small $p_T$ values.

\begin{table}[h]
\caption{Results from the best fit and the 1$\sigma$ error for the parameters in Eqs.~\eqref{eq:main_equation_k0s}
and \eqref{eq:main_equation_lambda}. $k_1$ and $l_1$ are in units of GeV$^{-2}$.}.
\label{tab::Fit_results_hyperons}
\begin{tabular}{ l  c c l c }
 \hline \hline
              & $K^0_S$                 & $\qquad$     &          & $\Lambda$         \\ \hline
$k_1$         &  $1.88 \pm 0.64$        &              & $l_1$    &  $(3.3 \pm 0.9) \cdot 10^{-2}$   \\
$k_2$         &  $9.23 \pm 1.47$        &              & $l_2$    &  $1.48\pm 1.04$   \\
$k_3$         &  $-6.86 \pm 1.71$       &              & $l_3$    &  $-2.21\pm 1.24$   \\
$k_4$         &  $0.20 \pm 0.08$        &              & $l_4$    &  $-0.25\pm 0.15$   \\
$k_5$         & $1.10 \pm 0.13$         &              & $l_5$    &  $0.26\pm 0.11$   \\
$k_6$         & $6.58 \pm 0.40$         &              & $l_6$    & $3.17\pm 0.28$   \\
$k_7$         & $0.89 \pm 0.04$         &              & $l_7$    &  $1.33\pm 0.08$   \\
$k_8$         &  $3.05 \pm 0.14$        &              & $l_8$    &  $2.41\pm 0.03$   \\
$k_9$         &  $0.50 \pm 1.06$       &              & $l_9$    &  $-0.44\pm 0.40$   \\
$k_{10}$      &  $0.045 \pm 0.27$       &              & $l_{10}$ &  $-0.45\pm 0.14$   \\
 \hline \hline
\end{tabular}
\end{table}

\begin{figure*}[t]
  \centering {
    \includegraphics[width=0.5\textwidth]{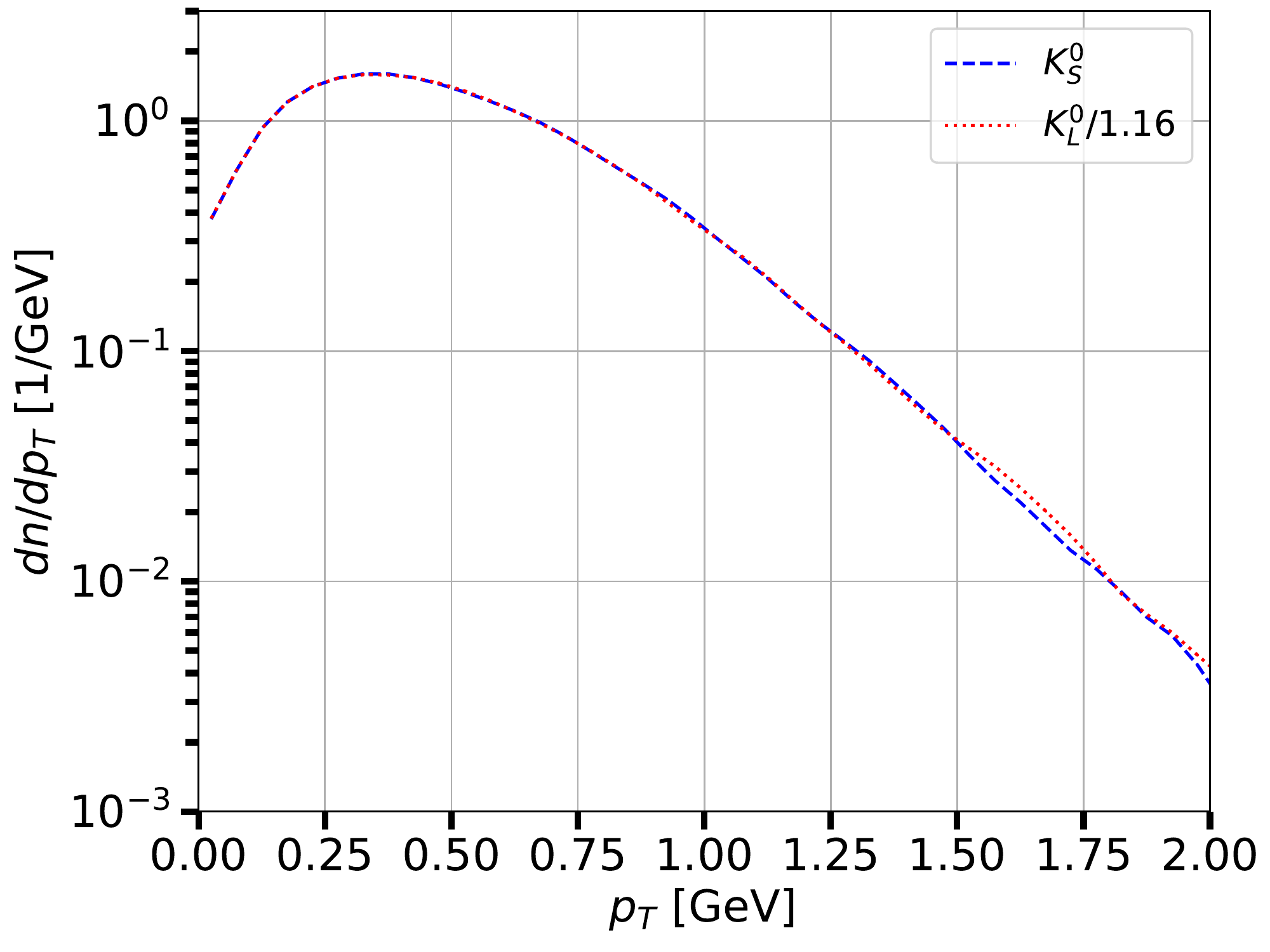}\includegraphics[width=0.5\textwidth]{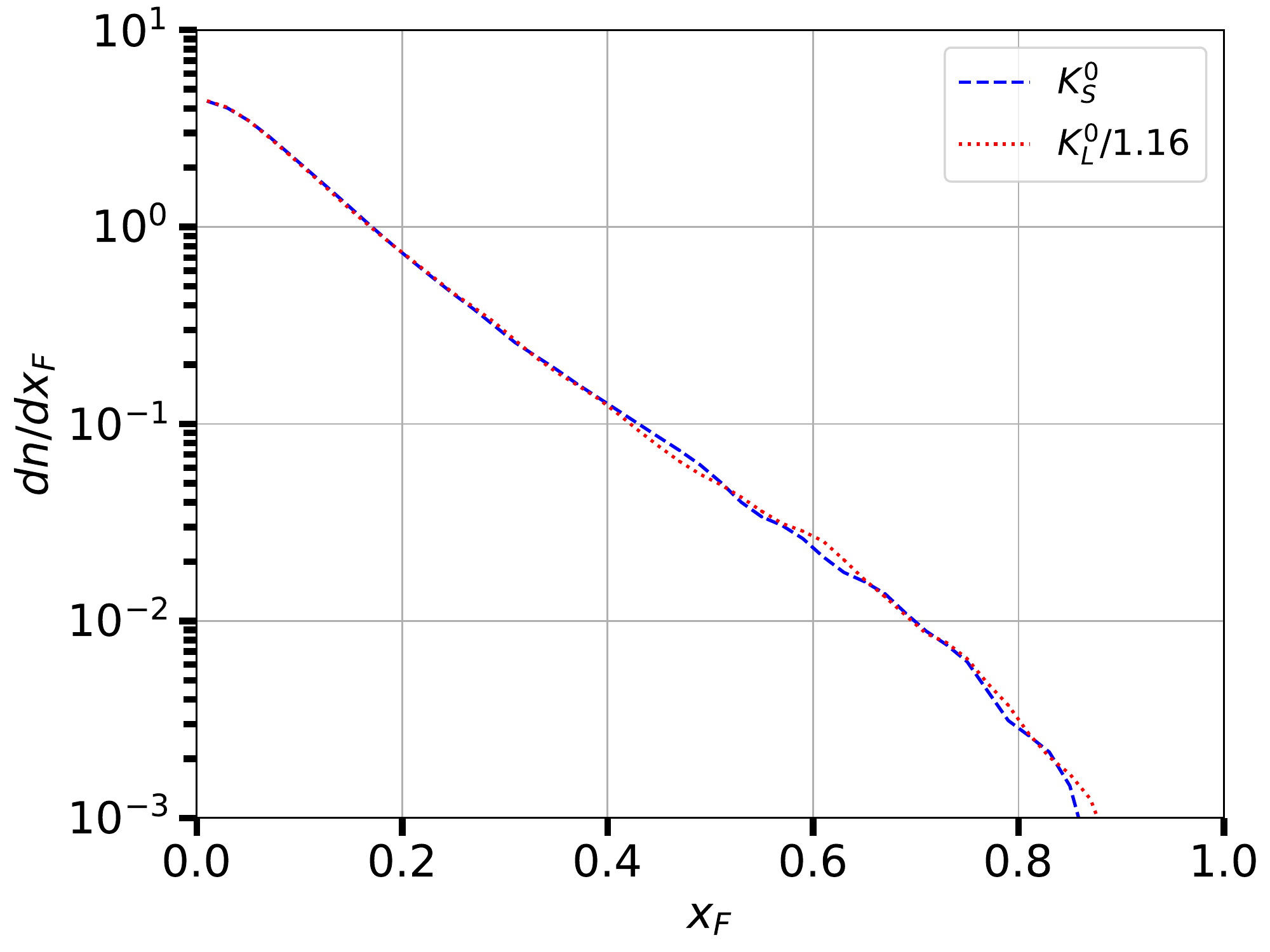}
  }
  \caption{
    The left (right) panel shows the distribution of the multiplicity with respect to the $p_T$ ($x_F$) variable for the $e^+$ produced from $K^0_L$ rescaled by a factor 1.16 (red dotted line) and $K^0_S$ (blue dashed line). The results are obtained using Pythia.
  }
  \label{fig:k0l}
\end{figure*}

\section{A few more details on the treatment of strange mesons and baryons}\label{app:details_K0_Lambda}
Here we collect a few tables and figures providing further details on our treatment of $K^0_S$, $K^0_L$, and $\Lambda$ production.  
Table~\ref{tab::Fit_results_hyperons} summarizes the best-fit parameters of the cross section as in Eq.~\eqref{eq:main_equation_k0s}
and \eqref{eq:main_equation_lambda} for the $K^0_S$ and $\Lambda$, obtained from the fit to NA61/SHINE data \cite{NA61SHINE:2021iay,NA61SHINE:2015haq}.
In Fig.~\ref{fig:k0l} we compare the $e^+$ production spectra from $K^0_S$ and $K^0_L$ using Pythia. Up to an overall normalization the distributions in $p_T$ and $x_F$ look very similar. As explained in Sec.~\ref{sec:eplus_K0l} we thus assume that the positron production cross section of $K^0_L$ is proportional to $K^0_S$.

\section{Pythia setup}
\label{sec:pythia}
Pythia is a program that is routinely used for the generation of events in high-energy collisions between elementary particles, including physics models for the evolution from a few-body hard-scattering process to complex multiparticle final state. Pythia has been designed with physics model rigorously derived from theory and others based on phenomenological models with parameters to be determined from data. We use in this paper the latest version 8.3 of Pythia \cite{Sjostrand:2014zea}.

Our benchmark setup is defined with the parameter {\tt Tune:pp} set to 4.
This is also called ``Tune 2M'' and it has been introduced with Pythia v.8.140 \cite{Corke:2010yf}. We then run the Monte Carlo also varying the tuning and choosing {\tt Tune:pp} set to $1,2,3,4,5,6,14,18$. All the other tunes are either created for very different scopes or have only slight variations with respect to the one listed above.

\begin{figure*}[t]
    \includegraphics[width=0.49\textwidth]{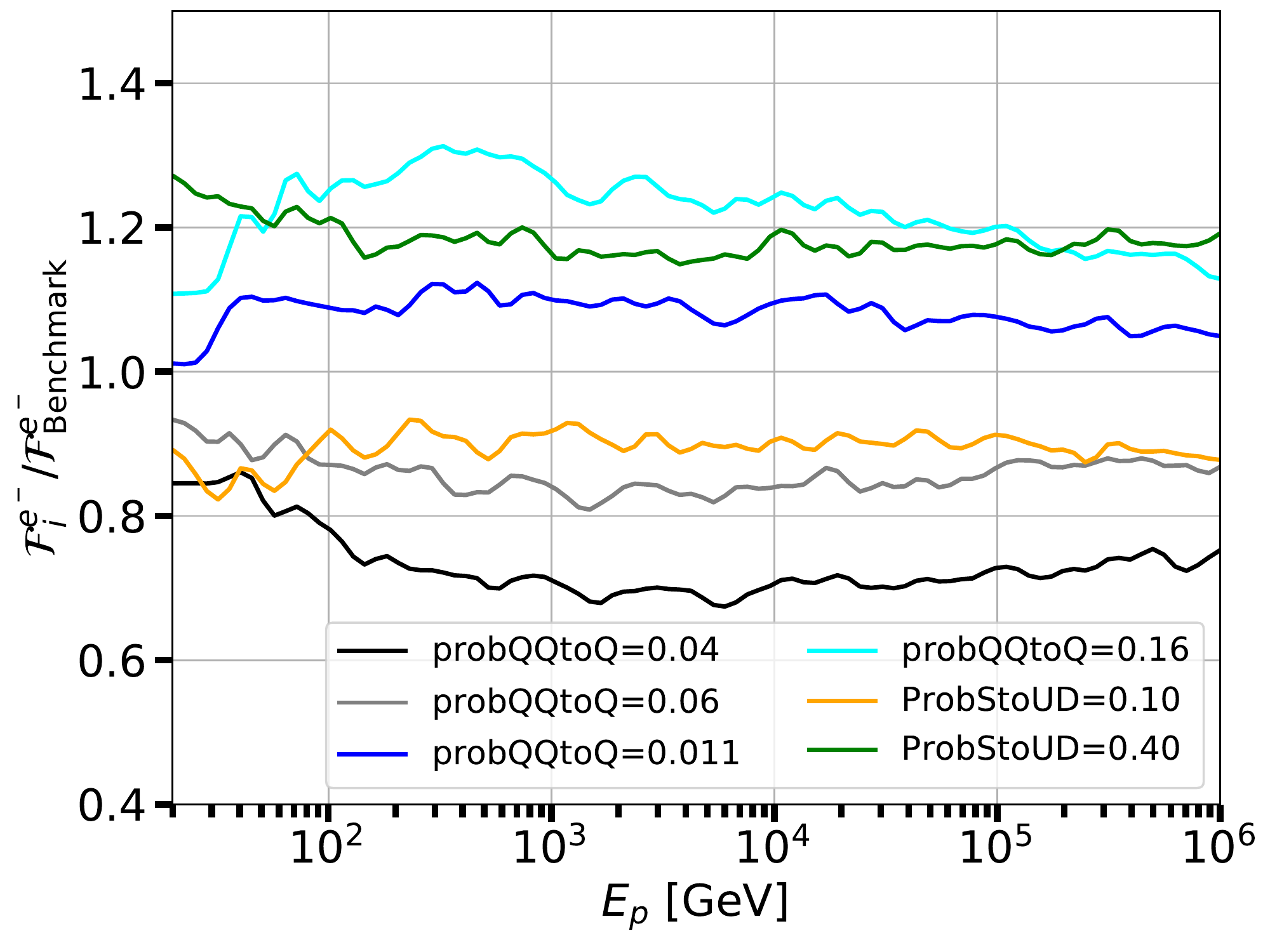}
    \includegraphics[width=0.49\textwidth]{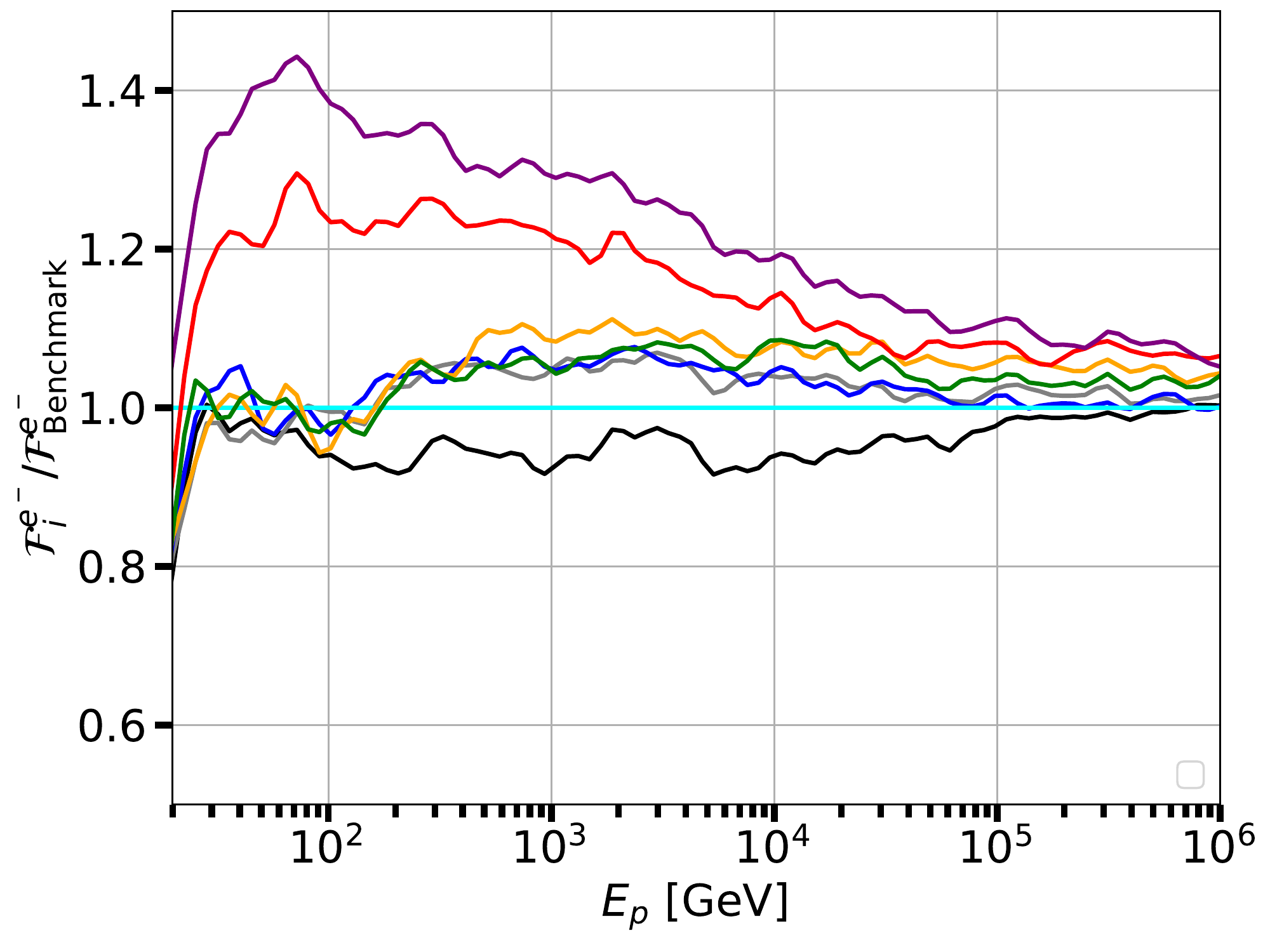}
    \includegraphics[width=0.49\textwidth]{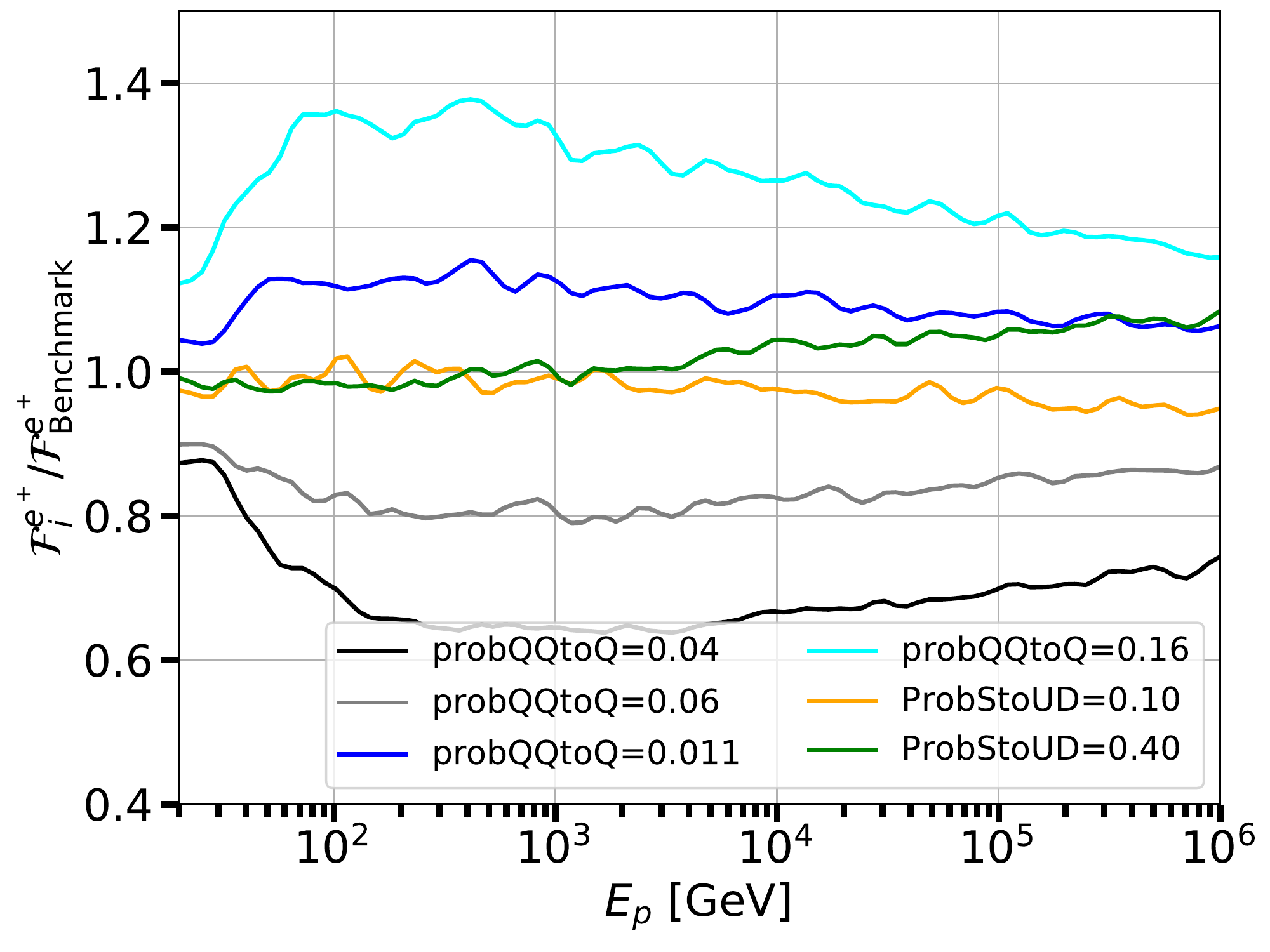}
    \includegraphics[width=0.49\textwidth]{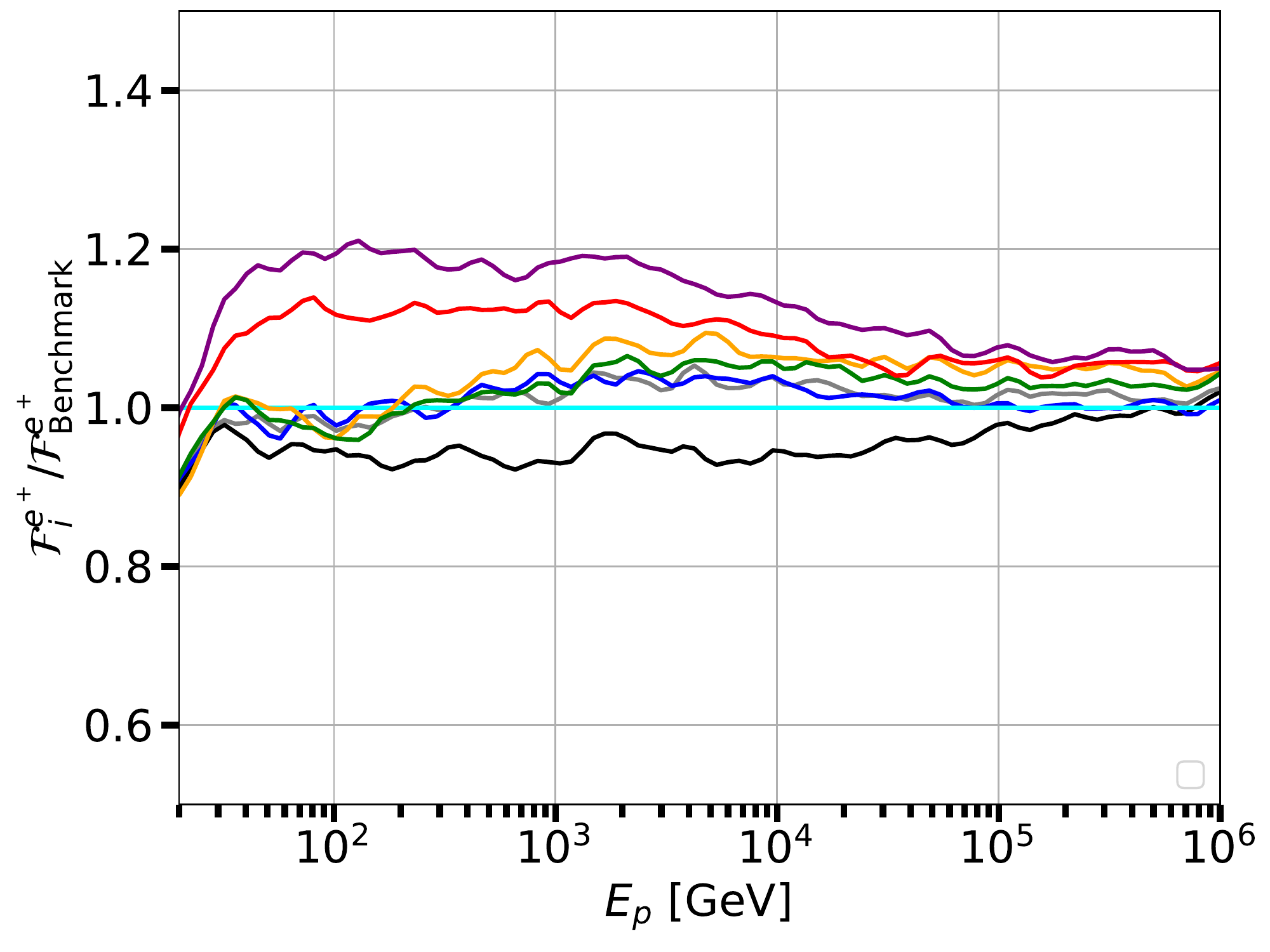}
    \caption{Ratio between the correction factor obtained with the benchmark Pythia setup, with {\tt Tune:pp} equal to 4, and the one found with all the other tunes tested (right panel) and with the variation of the parameters {\tt StringFlav:probQQtoQ} and {\tt StringFlav:ProbStoUD} (left panels). We show the results for the production of $e^+$ ($e^-$) in the top (botton) panels.
    } 
    \label{fig:pythiavar}
\end{figure*}

In addition to varying the parameter {\tt Tune:pp} we change the value of the {\tt StringFlav:probQQtoQ} and {\tt StringFlav:ProbStoUD}. The first one changes suppression of diquark production relative to quark production. In other words it changes the relative production of baryon with respect meson.
The second one modifies the suppression of s quark production relative to ordinary u or d one. We decide to perform simulations changing these two parameters by a factor of two smaller and larger with respect to the default one. Changing more than a factor of two the values of {\tt StringFlav:probQQtoQ} and {\tt StringFlav:ProbStoUD} does not have a significant impact on the results. We show in Fig.~\ref{fig:pythiavar} the ratio between the correction factor $\mathcal{F}$ obtained with our benchmark model and with the different setup explained above. These figures are equivalent for the ratio between the sum of multiplicity of the $\Sigma$ and $\Xi$ particle with respect to the on of the $\Lambda$ particle from $p+p$ collisions. We see that changing the parameter {\tt Tune:pp} varies $\mathcal{F}$ by a factor of about 20\% for electrons and 10\% for positrons. The variation of the parameters {\tt StringFlav:probQQtoQ} and {\tt StringFlav:ProbStoUD} instead produces a larger change in the correction factor that reaches 30\%. Therefore, we make the conservative choice to associate a systematic error of $40\%$ to the $\Sigma$ and $\Xi$ contribution of $e^{\pm}$.

\end{document}